\documentclass[pdflatex,sn-nature,twocolumn,oneside]{sn-jnl}
\makeatletter
\@twosidefalse
\makeatother

\usepackage{graphicx}%
\usepackage{multirow}%
\usepackage{amsmath,amssymb,amsfonts}%
\usepackage{mathtools}
\usepackage{amsthm}%
\usepackage{mathrsfs}%
\usepackage[title]{appendix}%
\usepackage{xcolor}%
\usepackage{textcomp}%
\usepackage{manyfoot}%
\usepackage{booktabs}%
\usepackage{algorithm}%
\usepackage{hyperref}
\usepackage{algorithmicx}%
\usepackage{algpseudocode}%
\usepackage{listings}%
\usepackage{physics}
\usepackage{stfloats}
\usepackage[nomarkers,nolists]{endfloat}
\usepackage{bm}
\usepackage[cal=boondoxo]{mathalpha}
\begin{document}

\title[Article Title]{Nonequilibrium Photocarrier and Phonon Dynamics from First Principles: a Unified Treatment of Carrier-Carrier, Carrier-Phonon, and Phonon-Phonon Scattering}


\author*{{Stefano} {Mocatti}$^*$}
\email{stefano.mocatti@unitn.it}

\author{Giovanni Marini}%

\author{Giulio Volpato}%

\author{Pierluigi Cudazzo}%

\author*{Matteo Calandra$^*$}%
\email{m.calandrabuonaura@unitn.it}

\affil{\orgdiv{Department of Physics}, \orgname{University of Trento}, \orgaddress{\street{Via Sommarive 14}, \city{Povo}, \postcode{38123}, \state{Trento}, \country{Italy}}}


\abstract{We develop a first-principles many-body framework to describe photocarrier and phonon dynamics in semiconductors after ultrafast excitation. The method includes explicit ab initio light–matter coupling, collision integrals for carrier–carrier, carrier–phonon, and phonon–phonon scattering, time-dependent quasiparticle and phonon-frequency renormalizations, and light-induced coherent atomic motion. The equations of motion are solved in a maximally localized Wannier basis, ensuring gauge-consistent scattering integrals and allowing for dense momentum sampling, enabling direct comparison with pump–probe experiments. The framework is computationally efficient, scalable, and can be combined with constrained density-functional theory to study longer-time light-induced structural phase transitions.

We demonstrate the method for MoS$_2$ and h-BN monolayers. In MoS$_2$, it captures photoinduced renormalizations of electronic and lattice properties, ultrafast carrier relaxation, hot-phonon dynamics, and coherent atomic motion. Including carrier–carrier scattering is essential for realistic photocarrier equilibration, while neglecting phonon–phonon scattering yields incorrect long-time lattice thermalization and overestimates the A$_{1g}$ coherent-phonon damping time by a factor of two. In h-BN, we quantify photoinduced changes in the electronic, optical, and lattice responses in quasi-equilibrium, demonstrating a fluence-dependent enhancement of screening and melting of excitonic features.}




\maketitle

\section{Introduction}

The advances in ultrafast physics and pump-probe experiments
have led to the possibility of monitoring the photoelectron and phonon dynamics in semiconductors~\cite{Maiuri2020, DelaTorre2021, Pellegrini2016} with high accuracy after laser excitation. The time evolution of physical properties
such as the electronic structure, reflectivity, optical absorption, and vibrational Raman response is now experimentally accessible as a function of the pump fluence (or photocarrier density) and from the femtosecond to the picosecond timescale. Consequently, new physical phenomena have emerged: nonthermal melting \cite{Liu1979,Shank1983,Preston1987,Siders1999}, coherent phonon dynamics ~\cite{Cho1990,Hunsche1995,Wall2012,Matsubara2016,Huang2022}, light-induced phase transitions ~\cite{Siegal1994,Kim2002,Hellmann2010,Tanimura2022,Mohr-Vorobeva2011,Dringoli2024,Hu2015,Wall2018}, tuning of ferroelectric order \cite{Nova2019}, electronic structure \cite{McIver2020} and interaction strengths ~\cite{Gerber2017,Baykusheva2022,Ron2020,Ciocys2023}, and melting of excitonic features \cite{Chernikov2015,Cunningham2017}, just to name a few. 

This rapid progress sets a high bar for theory: a predictive, versatile, computationally efficient and scalable many-body approach rooted in \textit{ab initio} theory is needed to track the coupled nonequilibrium dynamics of photocarriers (electrons and holes) and phonons. Theory should also account for coherent phonon response and damping, carrier thermalization, hot-phonon relaxation, and time and fluence evolution of electronic and optical properties. A correct description of these phenomena requires treating explicitly the laser field and the three key scattering mechanisms, namely carrier-carrier, carrier-phonon, and phonon-phonon, on an equal footing. Indeed, neglecting any one of these effects leads to nonphysical or inaccurate photocarrier and phonon dynamics.

\begin{figure}
    \centering  \includegraphics[width=0.5\linewidth]{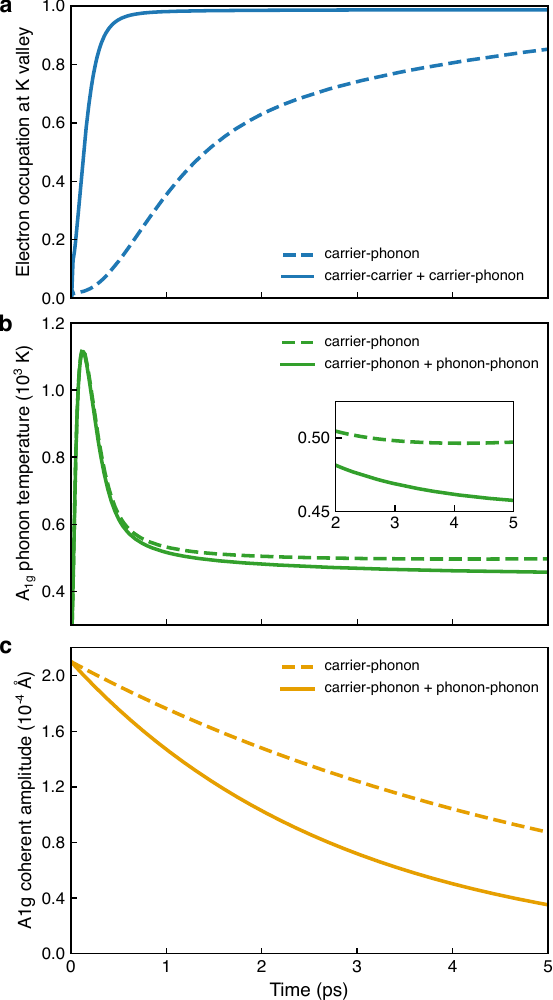}
    \caption{\textbf{Real-time dynamics of selected electronic and lattice observables in monolayer MoS$_2$ after above-gap excitation (fluence $\approx 0.85$~mJ/cm$^{2}$).}
    (a) Occupation of the lowest conduction band at $K$.
    (b) Effective temperature of the $A_{1g}$ phonon; inset: long-time cooling (2-5~ps).
    (c) Coherent amplitude of the $A_{1g}$ mode.
    Solid curves include the full interaction set: (a) carrier--carrier + carrier-phonon; (b,c) carrier-phonon + phonon-phonon, while dashed curves include only carrier-phonon.
    Discrepancies in timescales and amplitudes highlight the necessity of treating all interactions on equal footing.}
    \label{fig:sumup}
\end{figure}

To better highlight this point and the complexity of the problem, we anticipate in Fig.~\ref{fig:sumup} three representative real-time observables calculated with our {\it ab initio} many-body technique, including all relevant scattering effects for photoexcited monolayer MoS$_2$ at a fluence of $\approx 0.85$~mJ/cm$^{2}$ and an initial temperature of $300$~K.
Panel (a) shows that neglecting carrier-carrier scattering 
results in electron equilibration times
that are one order of magnitude too large.
Panels (b) and (c) demonstrate that
neglecting phonon-phonon scattering
leads to an incorrect long-time cooling of the $A_{1g}$ hot-phonon mode and an underestimation of the decay time of coherent $A_{1g}$ oscillations. It is worth mentioning that single-layer MoS$_2$ is a weakly anharmonic material and phonon-phonon scattering can have even more disruptive effects in the case of strong anharmonicity \cite{Mocatti2023, Furci2024}.
In short, Fig.~\ref{fig:sumup} 
highlights that neglecting any scattering mechanisms leads to inaccurate photocarrier, phonon, and ionic nonequilibrium dynamics. 

A range of methods has been developed to describe the nonequilibrium dynamics of photocarriers and phonons. However, none of these approaches includes all three scattering mechanisms. Nonequilibrium Green’s functions (NEGF) approaches based on the Kadanoff-Baym ansatz~\cite{Lipavsky1986, Kadanoff2018} and density-matrix formulations~\cite{Rossi2002} have captured photocarrier multiplication and exciton melting in graphene~\cite{Karlsson2021, Pavlyukh2022, Pavlyukh_Perfetto2022, Perfetto_Pavlyukh2022} and coherent/incoherent dynamics in MoS$_2$ when coupled to Ehrenfest dynamics~\cite{Perfetto2023}. While these approaches include both electron-electron and electron-phonon effects, a fully \textit{ab initio} treatment addressing material-specific coupling strengths and quantum anharmonicity remains out of reach.

Markovian reduction yields the semiconductor Bloch equations \cite{Schäfer2002,Haug2009,Haug2010,Kira2011} and their electron-phonon generalization ~\cite{Stefanucci2024}. In this context, photocarrier-induced changes in electronic and optical \cite{Steinhoff2014, Schmidt2016, Erben2018} and lattice dynamics \cite{Girotto_Novko2023} at quasi-equilibrium and photoexcited carrier dynamics~\cite{Butscher2007, Malic2011, Breusing2011, Steinhoff2016} have been partly addressed.  

Further simplifications lead to semiclassical Boltzmann equations~\cite{Marini2013,Bernardi2014,Sangalli2015,Sadasivam2017,OMahony2019,Tong2021,Chen2022,Maliyov2024,Sjakste2025} and the two-temperature model~\cite{Caruso2022}. In practice, however, most Boltzmann-based studies retain only a subset of scattering channels, often omitting explicit carrier-carrier~\cite{Tong2021, Caruso2021, Caruso2022, Emeis2024} and phonon-phonon~\cite{OMahony2019, Bernardi2014} collisions.

Time-dependent density-functional theory and \textit{ab initio} molecular dynamics have been widely applied to ultrafast phenomena, including ultrafast demagnetization~\cite{Krieger2015, Zhang2023}, Hubbard-$U$ renormalization~\cite{Tancogne-Dejean2018}, photoinduced structural dynamics~\cite{Lian2020, Wen-Hao2022, Chen2018, Liu2022}, and hot-carrier relaxation \cite{Zheng2023, Lively2024}, while constrained density-functional theory (cDFT) and density-functional perturbation theory have been used to describe quasi-equilibrium structural responses~\cite{Tangney1999, Tangney2002, Murray2007, Marini2021, Murray2015}, light-induced transitions~\cite{Mocatti2023, Furci2024}, and nonthermal melting~\cite{Corradini2025}. Yet a unified, first-principles framework that \emph{simultaneously} treats electron-electron, electron-phonon, and phonon-phonon interactions in real time, without neglecting any one of these effects, has remained elusive.

In this work, we develop a unified theoretical and computational framework for the real-time nonequilibrium dynamics of photocarriers and phonons in semiconductors that
(i) treats electron-electron, electron-phonon, and phonon-phonon interactions on the same footing in the quantum equations of motion;
(ii) captures time-dependent quasiparticle and phonon-frequency renormalizations tied to the evolving photocarriers and phonon distributions; 
(iii) addresses changes in the electronic and optical properties as functions of fluence, by properly including the single-particle bandgap shrinking and renormalization of exciton energies and spectral widths, 
(iv) accounts for the time evolution of screening due to the formation of an electron-hole plasma,
(v) can be coupled with cDFT and cDFPT \cite{Marini2021} to describe light-induced structural transitions at long times, and, finally,
(vi) is implemented in a Wannier representation with a direct-interpolation scheme~\cite{Volpato2025} allowing for ultradense Brillouin-zone sampling, long propagation times, and realistic pump-probe conditions with a high level of accuracy.

Our formulation builds on top of the semiconducting Bloch equations \cite{Rossi2002} and NEGF \cite{Stefanucci2023} by adding \textit{ab initio} phonon-phonon scattering and by enforcing a Bloch gauge-consistent \cite{Volpato2025} evaluation of the scattering integrals, thereby linking, within a single predictive calculation, coherent polarization, carrier relaxation, hot-phonon dynamics, coherent atomic motion, photocarrier-induced enhancement of screening, and linear optical absorption in the transient state. 
From a technical point of view, our implementation is scalable and efficient, enabling direct, parameter-free comparison to pump-probe measurements at experimental fluences and temperatures.

We apply our approach to single-layer MoS$_2$ and h-BN. In the first case, we address photoinduced renormalizations of electronic and lattice properties, ultrafast carrier relaxation, hot-phonon lifetimes, and displacive coherent phonon dynamics. In the case of single-layer h-BN, we investigate changes in electronic, optical, and lattice dynamics properties at long times in a quasi-equilibrium configuration, namely, we study the fluence-induced electronic screening renormalization and melting of excitonic features.
Where possible, we benchmark forces and long-time trends against cDFPT and available experiments, and we delineate regimes in which simplified descriptions fail qualitatively.

The paper is organized as follows. In \nameref{sec2} we introduce the theoretical formalism, including the \textit{ab initio} Hamiltonian and the derivation of quasiparticle renormalizations, collision integrals, and the Ehrenfest equation. Then, we detail the computational implementation, including workflow, Wannier framework, time propagation, and performance. In \nameref{sec3} we presents applications to monolayer h-BN and MoS$_2$. Conclusions are given in \nameref{sec: Conclusion and perspectives}.   

\section{Methods}\label{sec2}

\subsection{Equilibrium structure in the absence of an external drive}

We start from the full quantum Hamiltonian of electrons and nuclei. At equilibrium, we employ the Born-Oppenheimer (BO) factorization of the total wavefunction,
\begin{equation}
\Phi(\vb{r},\vb{R},t)=\Psi(\vb{r};\vb{R},t)\,\chi(\vb{R},t),
\end{equation}
where $\vb{r}$ and $\vb{R}$ are collective electronic and nuclear coordinates, respectively. $\Psi(\vb{r};\vb{R},t)$ is the electronic many-body state at fixed nuclear configuration, and $\chi(\vb{R},t)$ is the nuclear many-body state.

We focus on crystalline solids in which nuclei oscillate about equilibrium positions $\vb{R}_i^0$ repeated periodically. The composite index $i\!\equiv\!(I,a)$ labels unit cell $I$ and atom $a$ within the cell, so that $\vb{R}_i^0=\vb{R}_I+\boldsymbol{\tau}_a$, with $\vb{R}_I$ a Bravais lattice vector and $\boldsymbol{\tau}_a$ the basis position. The number of atoms per cell is $N_{\text{at}}$.

Equilibrium positions are defined by vanishing forces on the BO energy surface $E(\vb{R})$:
\begin{equation}
\label{eq:zero_force_condition}
\vb{F}_i=-\left.\nabla_{\vb{R}_i}E(\vb{R})\right|_{\vb{R}=\vb{R}^0}=0.
\end{equation}
We write the instantaneous nuclear coordinates as displacements from equilibrium,
\begin{equation}
\vb{R}_i=\vb{R}_i^0+\vb{Q}_i,
\end{equation}
with $\vb{Q}_i$ the ionic displacement. We impose Born-von Kármán boundary conditions over $N$ unit cells of volume $\Omega$, obtaining a total volume $V=N\Omega$. From now on, unless otherwise stated, we work in Rydberg atomic units (Ry a.u.), i.e., $\hbar=2m_e=e^2/2=4\pi\epsilon_0=1$.

\subsection{Harmonic phonon Hamiltonian at equilibrium}
\label{sec:Harmonic phonon Hamiltonian at equilibrium}

To describe nuclear motion at equilibrium, we adopt the harmonic approximation, expanding $E(\vb{R})$ to second order in ionic displacements. The resulting lattice dynamics is set by the force constant matrix in reciprocal space,
\begin{equation}
\label{eq:force_constant_matrix}
C_{a\alpha b\beta}(\vb{q})=\sum_{I}e^{-i\vb{q}\cdot\vb{R}_I}
\left.\frac{\partial^2 E(\vb{R})}{\partial R_{Ia\alpha}\,\partial R_{0b\beta}}\right|_{\vb{R}=\vb{R}^0},
\end{equation}
where Greek indices denote cartesian components.

For each wavevector $\vb{q}$, the $N_m=3N_{\text{at}}$ normal mode frequencies $\omega^0_{\vb{q}\nu}$ and eigenvectors $e^{a\alpha}_{\vb{q}\nu}$ follow from the dynamical matrix
\begin{equation}
\label{eq:dynamical_matrix_equilibrium}
D_{a\alpha b\beta}(\vb{q})=\frac{C_{a\alpha b\beta}(\vb{q})}{\sqrt{M_a M_b}},
\end{equation}
with $M_a$ the mass of atom $a$, via diagonalization:
\begin{equation}
\label{eq:dynamical_matrix_normal_modes}
D_{\vb{q}\nu\nu'}=\sum_{a,b,\alpha,\beta}
\big(e^{a\alpha}_{\vb{q}\nu}\big)^{*} D_{a\alpha b\beta}(\vb{q})\,e^{b\beta}_{\vb{q}\nu'}
= \delta_{\nu\nu'}\big(\omega^0_{\vb{q}\nu}\big)^2.
\end{equation}
The normal modes define phonon creation and annihilation operators, $\hat{a}^\dagger_{\vb{q}\nu}$ and $\hat{a}_{\vb{q}\nu}$, satisfying canonical commutation relations (see Supplementary Section 1). From these, the nuclear displacement and momentum operators can be defined:
\begin{align}
\hat{Q}_{Ia\alpha}
&=\sqrt{\frac{1}{M_a N}}
\sum_{\vb{q},\nu}\frac{e^{a\alpha}_{\vb{q}\nu}}{\sqrt{\omega^0_{\vb{q}\nu}}}
e^{i\vb{q}\cdot\vb{R}_I}\,
\frac{\hat{a}^\dagger_{-\vb{q}\nu} + \hat{a}_{\vb{q}\nu}}{\sqrt{2}},
\label{eq:at_displ_fourier}\\
\hat{P}_{Ia\alpha}
&= i\sqrt{\frac{M_a}{N}}
\sum_{\vb{q},\nu}\sqrt{\omega^0_{\vb{q}\nu}}\,
e^{a\alpha}_{\vb{q}\nu}e^{i\vb{q}\cdot\vb{R}_I}\,
\frac{\hat{a}^\dagger_{-\vb{q}\nu}-\hat{a}_{\vb{q}\nu}}{\sqrt{2}}.
\label{eq:at_mom_fourier}
\end{align}
It is convenient to introduce reciprocal-space displacement and momentum operators,
\begin{align}
\hat{Q}_{\vb{q}\nu}&=\frac{\hat{a}^\dagger_{-\vb{q}\nu} + \hat{a}_{\vb{q}\nu}}{\sqrt{2}},
\label{eq:displacement_operator_reciprocal_space}\\
\hat{P}_{\vb{q}\nu}&= i\,\frac{\hat{a}^\dagger_{-\vb{q}\nu}-\hat{a}_{\vb{q}\nu}}{\sqrt{2}},
\label{eq:momentum_operator_reciprocal_space}
\end{align}
which obey canonical commutation relations, moreover their Hermitian conjugates satisfy $\hat{Q}^\dagger_{\vb{q}\nu}=\hat{Q}_{-\vb{q}\nu}$ and $\hat{P}^\dagger_{\vb{q}\nu}=\hat{P}_{-\vb{q}\nu}$.

In this basis, the harmonic nuclear Hamiltonian becomes
\begin{align}
\label{eq:harmonic_phonon_equilibrium1}
\hat{H}_{\text{BO}}
&=\frac{1}{2}\sum_{\vb{q},\nu,\nu'}
\Bigg[
\delta_{\nu\nu'}\,\omega^0_{\vb{q}\nu}\,\hat{P}^\dagger_{\vb{q}\nu}\hat{P}_{\vb{q}\nu'}
+\hat{Q}^\dagger_{\vb{q}\nu}
\frac{D_{\vb{q}\nu\nu'}}{\sqrt{\omega^0_{\vb{q}\nu}\omega^0_{\vb{q}\nu'}}}
\hat{Q}_{\vb{q}\nu'}
\Bigg]
\\
\label{eq:harmonic_phonon_equilibrium2}
&=\sum_{\vb{q},\nu}\omega^0_{\vb{q}\nu}\left(\hat{a}^\dagger_{\vb{q}\nu}\hat{a}_{\vb{q}\nu}+\frac{1}{2}\right)
=\sum_{\vb{q},\nu}\omega^0_{\vb{q}\nu}\left(\hat{n}_{\vb{q}\nu}+\frac{1}{2}\right),
\end{align}
with $\hat{n}_{\vb{q}\nu}=\hat{a}^\dagger_{\vb{q}\nu}\hat{a}_{\vb{q}\nu}$ the phonon number operator.

The harmonic frequencies $\omega^0_{\vb{q}\nu}$ are computed within density-functional perturbation theory (DFPT)~\cite{Baroni2001} using a semilocal exchange-correlation functional, and already include the adiabatic renormalization due to electron-phonon interactions~\cite{Calandra2010}.

If the harmonic Hamiltonian is dynamically unstable, one may construct an auxiliary, positive-definite harmonic Hamiltonian through a full quantum anharmonic minimization, e.g., by employing the stochastic self-consistent harmonic approximation (SSCHA)~\cite{Monacelli2021}, which then replaces the original harmonic Hamiltonian. The role of anharmonicity in nonequilibrium dynamics will be addressed later.

\subsection{Electronic Hamiltonian at fixed nuclei}
\label{sec:electronic_hamiltonian}

We begin by choosing an appropriate single-particle basis for the electrons. Since our starting point is a density-functional theory (DFT) calculation, we use the Kohn-Sham (KS) states $\psi_{\vb{k} n}(\vb{r},\sigma)$ and band energies $\varepsilon_{\vb{k} n}^{\text{KS}}$, where $\vb{k}$ is the crystal momentum and $n$ the band index. This choice allows us to define second-quantized operator $\hat{c}_{\vb{k}n}$ ($\hat{c}^\dagger_{\vb{k}n}$), obeying the canonical anticommutation relations (see Supplementary Section 1), that annihilates (creates) an electron in the $n$th Kohn-Sham orbital with momentum $\vb{k}$.

In this basis, the auxiliary Kohn-Sham Hamiltonian is diagonal:
\begin{equation}
\label{eq:el_KS}
\begin{gathered}
    \hat{H}_{\text{KS}} = \sum_{\vb{k},n} \varepsilon^{\text{KS}}_{\vb{k}n}\hat{c}^\dagger_{\vb{k}n} \hat{c}_{\vb{k}n},
\end{gathered}
\end{equation}
with
\begin{eqnarray}
 \label{eq:KS_eigenvalues}
  \varepsilon^{\text{KS}}_{\vb{k}n} &=& \langle\psi_{\vb{k}n}|\hat{\vb{p}}^2 +  \hat{V}_{\text{H}xc} + \hat{V}_{ext} + \hat{V}_{ion}|\psi_{\vb{k}n}\rangle,
\end{eqnarray}
and where $\hat{V}_{\text{H}xc}$ is the (semilocal) Hartree plus exchange-correlation potential. The external potential is
\begin{equation}
    \hat{V}_{ext}({\vb{r}}) = - \sum_{i=1}^{N_n} Z_{i} v(|\hat{\vb{r}} - \vb{R}^0_{i}|)
\end{equation}
and the ionic potential is
\begin{equation}
    \hat{V}_{ion} = \dfrac{1}{2}\sum_{i\neq j}^{N_n} Z_{i}Z_{j} v(|\vb{R}^0_{i} - \vb{R}^0_{j}|). 
\end{equation}
Because we employ a pseudopotential framework, $Z_i$ is the pseudo-ionic atomic number for the $i$th atomic species, and $N_n$ is the number of ions in the crystal.
The potential $v(|\vb{r}|)=\frac{2}{|\vb{r}|}$ is the bare Coulomb potential, and $\vb{R}^0_i$ denote the ionic equilibrium positions, determined by the zero-force condition in the absence of an external drive (see Eq.~\eqref{eq:zero_force_condition}).

We discuss the equilibrium electronic Hamiltonian before the action of the external drive ($t<0$). To define the bare single particle Hamiltonian matrix from the Kohn-Sham eigenvalues, we subtract the Hartree, exchange-correlation potential, i.e.,
\begin{equation}
\label{eq:bare_electronic_hamiltonian2}
\bar{h}_{\vb{k}nm} = \delta_{nm}\varepsilon_{\vb{k} n}^{\text{KS}}-\langle \psi_{\vb{k} n}| \hat{V}_{\text{H}xc}|\psi_{\vb{k} m}\rangle
\end{equation}
and obtain the bare electron single-particle Hamiltonian as
\begin{equation}
\label{eq:bare_electronic_hamiltonian}
\hat{H}_{sp}=\sum_{\vb{k},n,m}  \bar{h}_{\vb{k}nm}\hat{c}^\dagger_{\vb{k}n} \hat{c}_{\vb{k}m}.
\end{equation}
The \textit{ab initio} many-body Hamiltonian of the electronic system at equilibrium is composed of three terms:
\begin{equation}
 \hat{H}_e =  \hat{H}_{sp} + \hat{H}_{ee} + \hat{H}_{ep}.
 \label{eq:clamped}
 \end{equation}
The electron-electron interaction part is
\begin{equation}
\begin{gathered}
\hat{H}_{ee} = \dfrac{1}{2N} \sum_{\mathclap{\substack{\vb{k,k',q} \\ n,n',m,m' }}} v_{\vb{kk'k-qk'+q}}^{nn'mm'} \hat{c}^\dagger_{\vb{k}n}\hat{c}^\dagger_{\vb{k'}n'}\hat{c}_{\vb{k'+q}m'}\hat{c}_{\vb{k-q}m}, 
\end{gathered}
\end{equation}
where the matrix element of the bare Coulomb potential is defined as
\begin{equation}
    v_{\vb{kk'k-qk'+q}}^{nn'mm'} = \mel{\psi_{\vb{k}n}\psi_{\vb{k'}n'}}{v(|\hat{\vb{r}}-\hat{\vb{r}}'|)}{{\psi_{\vb{k-q}m}\psi_{\vb{k'+q}m'}}}.
\end{equation}
The electron-phonon interaction part is
\begin{equation}
\label{eq:elph_interaction_hamiltonian}
\begin{aligned}
\hat{H}_{ep} =\dfrac{1}{\sqrt{N}}\sum_{\mathclap{\substack{\vb{k,q} \\ n,m,\nu}}} & \bar{g}_{mn}^{\nu}(\vb{k},\vb{q})  \hat{c}^\dagger_{\vb{k+q}m}\hat{c}_{\vb{k}n}(\hat{a}_{\mathbf{q}\nu}+\hat{a}^{\dagger}_{-\mathbf{q}\nu}),
\end{aligned}
\end{equation}
involving the bare electron-phonon matrix element defined as 
\begin{equation}
\label{eq:bare_elph_matrix}
\bar{g}_{mn}^\nu(\vb{k},\vb{q}) =
\sum_{a,\alpha}\dfrac{{e}_{\vb{q}\nu}^{a\alpha}}{{\sqrt{2M_a\omega_{\vb{q}\nu}^{0}}}}\bar{{d}}^{a\alpha}_{mn}(\vb{k,q}),   
\end{equation}
where the bare deformation potential is obtained from
\begin{equation}
\label{eq:bare_deformation_potential_reciprocal_space}
\bar{{d}}^{a\alpha}_{mn}(\vb{k,q}) = \sum_{I} e^{-i\vb{q}\cdot \vb{R}_I} \mel{\psi_{\vb{k+q}m}}{\bar{d}_{Ia\alpha}(\hat{\vb{r}})}{\psi_{\vb{k}n}} 
\end{equation}
with
\begin{equation}
\label{eq:bare_deformation_potential_real_space}
    \bar{d}_{Ia\alpha}(\vb{r}) = \dfrac{\partial V_{ext}({\vb{r}})}{\partial Q_{Ia\alpha}}.
\end{equation}
We assume that the external drive couples directly only to the electronic degrees of freedom. This is the case for a time-dependent uniform external electric field $\vb{E}(t)$ with photon energies in the eV range, inducing dipolar electronic excitations. The ions are perturbed only indirectly, via the field-induced electronic response. We thus first consider the electrons at clamped nuclei and analyze their dynamics. The dynamics of phonons and ions triggered by the drive-induced change in the electronic state is addressed later.

The perturbed nonequilibrium electronic Hamiltonian is obtained by adding a time-dependent light-matter interaction term $\hat{H}_{drive}(t)$. This term follows from the minimal-coupling substitution. In the Coulomb gauge and within the dipole approximation, the second-quantized light-matter interaction reads
\begin{equation}
\label{eq:electronic_drive_hamiltonian}
    \hat{H}_{drive}(t) = \sum_{\mathclap{\vb{k},n,m}} \bar{\Omega}_{\vb{k}nm}(t) \hat{c}^\dagger_{\vb{k}n} \hat{c}_{\vb{k}m},
\end{equation} 
where we introduce the bare Rabi frequency matrix
\begin{equation}
    \bar{\Omega}_{\vb{k}nm}(t) = \sqrt{2}\vb{E}(t) \cdot \mathbfcal{D}_{\vb{k}nm},
\end{equation}
and the electric-dipole matrix elements
\begin{equation}
    \mathbfcal{D}_{\vb{k}nm}= \mel{\psi_{\vb{k}n}}{\hat{\vb{r}}}{\psi_{\vb{k}m}}.
\end{equation}
The full nonequilibrium electronic Hamiltonian at clamped nuclei is therefore
\begin{equation}
    \hat{H}_{el}(t) = \hat{H}_{sp} + \hat{H}_{ee} + \hat{H}_{ep} + \hat{H}_{drive}(t).
    \label{eq:total_el_hamiltonian}
\end{equation}

\subsection{Electronic Bloch equations}
\label{sec:electronic_bloch equations}

The time evolution of observables is governed by quantum thermal averages computed from the system’s density matrix. We introduce the time-dependent nonequilibrium many-body density-matrix operator $\hat{\rho}(t)$, which describes the state of electrons and phonons. Its time dependence arises from the time-dependent Hamiltonian. In the Schrödinger picture, the nonequilibrium average of an operator $\hat{\mathcal{O}}$ is defined as
\begin{equation}
\label{eq:thermal_average}
  \langle \hat{\cal O} \rangle_t = \mathrm{Tr}[\hat{\rho}(t) \hat{\cal O}].
\end{equation}
Its time derivative satisfies:
\begin{eqnarray}
    i \dfrac{d \langle \hat{\cal O} \rangle_t}{dt} &=& 
    \langle [\hat{\cal O}, \hat{H}] \rangle_t. \label{eq:Heis0}
\end{eqnarray}
To track the dynamics of the electronic subsystem, we evaluate the nonequilibrium electronic occupations $f_{\vb{k}n}(t)$ and microscopic dipole polarizations $p_{\vb{k}nm}(t)$, defined as
\begin{eqnarray}
f_{\vb{k} n}(t) &=& \langle \hat{c}_{\vb{k} n}^\dagger \hat{c}_{\vb{k} n} \rangle_t, \label{eq:occ_0} \\
p_{\vb{k} nm}(t) &=& \langle \hat{c}_{\vb{k} m}^\dagger \hat{c}_{\vb{k} n} \rangle_t \quad {\rm for} \quad n \ne m. \label{eq:pol_0}
\end{eqnarray}
Using Eqs.~\eqref{eq:total_el_hamiltonian}--\eqref{eq:pol_0}, the time evolution of these quantities reads
\begin{eqnarray}
i \frac{d f_{\vb{k} n}(t)}{dt} &=& \langle [\hat{c}_{\vb{k} n}^\dagger \hat{c}_{\vb{k} n}, \hat{H}_{sp} + \hat{H}_{drive}] \rangle_t +\nonumber \\
&& \langle [\hat{c}_{\vb{k} n}^\dagger \hat{c}_{\vb{k} n}, \hat{H}_{ee} + \hat{H}_{ep}] \rangle_t, \label{eq:df_nk/dt_commutators} \\
i \dfrac{d p_{\vb{k}nm}(t)}{dt} &=& \langle [\hat{c}_{\vb{k} m}^\dagger \hat{c}_{\vb{k} n}, \hat{H}_{sp} + \hat{H}_{drive}] \rangle_t +\nonumber \\
&& \langle [\hat{c}_{\vb{k} m}^\dagger \hat{c}_{\vb{k} n}, \hat{H}_{ee} + \hat{H}_{ep}] \rangle_t. \label{eq:dp_nmk/dt_commutators}
\end{eqnarray}
To derive the equations of motion (EOMs), we evaluate the equal-time commutators in the expressions above. These fall into two categories: single-particle commutators, arising from the quadratic Hamiltonians $\hat{H}_{sp}$ and $\hat{H}_{drive}(t)$, which can be computed analytically; and many-body commutators, involving $\hat{H}_{ee}$ and $\hat{H}_{ep}$, which are more complex to evaluate.

The single-particle commutators can be evaluated directly using the equal-time canonical anticommutation relations (see Eq.(S1) in the Supplementary Information). This yields
\begin{align}
    \hspace{-5pt} \langle [\hat{c}^\dagger_{\vb{k}m}\hat{c}_{\vb{k}n}, \hat{H}_{sp}]\rangle_t &= \bar{h}_{\vb{k}nm}[f_{\vb{k}m}(t) - f_{\vb{k}n}(t)] + \nonumber\\
    & \hspace{-80pt} \sum_{\mathclap{m'\neq m}} \bar{h}_{\vb{k}nm'}\, p_{\vb{k}m'm}(t) - 
    \sum_{\mathclap{n'\neq n}} \bar{h}_{\vb{k}n'm}\, p_{\vb{k}nn'}(t).  \label{eq:comm_cc_Hsp} \\
    \hspace{-5pt} \langle [\hat{c}^\dagger_{\vb{k}m}\hat{c}_{\vb{k}n}, \hat{H}_{drive}]\rangle_t &= \bar{\Omega}_{\vb{k}nm}(t)\left[f_{\vb{k}m}(t) - f_{\vb{k}n}(t)\right] + \nonumber \\
    & \hspace{-80pt} \sum_{\mathclap{m'\neq m}} \bar{\Omega}_{\vb{k}nm'}(t)\, p_{\vb{k}m'm}(t) - 
    \sum_{\mathclap{n'\neq n}} \bar{\Omega}_{\vb{k}n'm}(t)\, p_{\vb{k}nn'}(t). \label{eq:comm_cc_Hdrive} 
\end{align}
The single-particle contributions to the equation of motion for the electronic occupations are obtained by setting $n = m$ in Eqs.~\eqref{eq:comm_cc_Hsp} and \eqref{eq:comm_cc_Hdrive}. All terms derived from the single-particle commutators involve the bare single-particle Hamiltonian matrix eigenvalues and Rabi frequencies. However, physically meaningful equations must depend on dressed electronic degrees of freedom. Therefore, Eqs.~\eqref{eq:comm_cc_Hsp} and \eqref{eq:comm_cc_Hdrive} alone are insufficient to fully capture the system’s dynamics. The required dressing is provided by many-body effects due to electron-electron and electron-phonon interactions.

The evaluation of the many-body commutators in Eqs.~\eqref{eq:df_nk/dt_commutators} and \eqref{eq:dp_nmk/dt_commutators} is considerably more involved than that of the quadratic terms in the Hamiltonian. A direct application of the canonical anticommutation relations leads to an infinite Bogoliubov-Born-Green-Kirkwood-Yvon (BBGKY) hierarchy \cite{Kadanoff2018, Balescu1975}, involving operators of increasing complexity.

To truncate this hierarchy, two main strategies are typically employed. The first is to evaluate the commutators explicitly and apply a suitable truncation to the electron-electron and electron-phonon terms \cite{Wyld1963, Kuhn1998, Rossi2002}.

\begin{figure}
    \centering
    \includegraphics[width=0.95\linewidth]{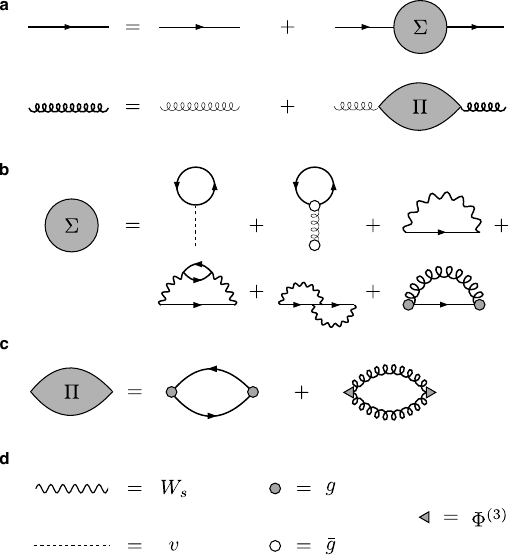}
    \caption{\textbf{Electron and phonon Dyson equations and self-energy approximations.} (a) Dyson equations for the interacting electron (straight line) and phonon (curly line) propagators. Bold lines denote interacting propagators; thin lines, noninteracting ones. (b,c) Electron and phonon self-energy diagrams included in the present approximation. (d) Conventions for the bare and screened Coulomb interactions $v$ and $W_s$, electron-phonon couplings $\bar{g}$ and $g$, and third-order anharmonic coupling $\Phi^{(3)}$.}
    \label{fig:diagrams}
\end{figure}

The second approach relies on the nonequilibrium Green’s function (NEGF) formalism \cite{Kadanoff2018, Stefanucci2023, Stefanucci2024}, which introduces lesser and greater Green’s functions and self-energies. These enter a nonequilibrium Dyson equation used to determine dressed propagators from a closed expression for the self-energy, as illustrated in Fig.~\ref{fig:diagrams}(a). A Feynman diagrammatic expansion is then employed, together with the Generalized Kadanoff-Baym ansatz (GKBA) \cite{Lipavsky1986,Karlsson2021,Schäfer2002, Haug2010} or its mirrored counterpart \cite{Stefanucci2024} (MGKBA) and the Markov approximation, to obtain a computationally tractable scheme (see Supplementary Section 2) that naturally accounts for both quasiparticle renormalization and interaction-vertex screening. 

The net result of the NEGF formalism is that the many-body commutator can be decomposed into two contributions:
\begin{gather}
\label{eq:commutator_elel_elph}
    \langle [\hat{c}_{\vb{k} m}^\dagger \hat{c}_{\vb{k} n}, \hat{H}_{ee} + \hat{H}_{ep}] \rangle_t = \Delta_{\vb{k}nm}(t) + i\mathcal{I}_{\vb{k}nm}(t).
\end{gather}
The first term, $\Delta_{\vb{k}nm}(t)$, represents the time-dependent quasiparticle renormalization due to many-body interactions. The second term, $\mathcal{I}_{\vb{k}nm}(t)$, is a collision integral that captures scattering-induced changes in occupations and polarizations arising from carrier-carrier and carrier-phonon processes.

The specific form of these two contributions depends on the approximation scheme used to truncate the BBGKY hierarchy. In general, both terms can be separated into electron-electron and electron-phonon components:
\begin{align}
    \mathcal{I}_{\vb{k}nm}(t) &= \mathcal{I}^{ee}_{\vb{k}nm}(t) + \mathcal{I}^{ep}_{\vb{k}nm}(t), \label{eq:scattering_integral_separation} \\
    \Delta_{\vb{k}nm}(t) &= \Delta^{ee}_{\vb{k}nm}(t) + \Delta^{ep}_{\vb{k}nm}(t).
\end{align}
In what follows, we denote diagonal terms using a single band index for notational simplicity.

The renormalization term originates from the singular part of the electron self-energy~\cite{Stefanucci2023, Stefanucci2024}, which collects all time-local contributions to quasiparticle properties. In its exact form, this singular component includes the Hartree-Fock term, the electron-phonon tadpole, and the Debye-Waller contributions~\cite{Allen1976, Marini2015, Stefanucci2023}. Additional static terms may also appear, depending on the chosen quasiparticle approximation, as shown in Eqs.(S17) and (S18) in the Supplementary Information.

The renormalization term can be written as a commutator with a time-dependent effective single-particle Hamiltonian, in close analogy with Eqs.~\eqref{eq:comm_cc_Hsp} and \eqref{eq:comm_cc_Hdrive}; see also Supplementary Section 2. Its explicit expression reads
\begin{align}
\label{eq:renormalization_term}
    \Delta_{\vb{k}nm}(t) = \langle [\hat{c}^\dagger_{\vb{k}m} \hat{c}_{\vb{k}n}, \hat{\Delta}] \rangle_t,
\end{align}
where
\begin{equation}
    \hat{\Delta} = \sum_{\vb{k},n,m} \Sigma_{\vb{k}nm}(t)\, \hat{c}^\dagger_{\vb{k}n} \hat{c}_{\vb{k}m}
\end{equation}
is a single-particle operator constructed from the time-local, singular self-energy $\Sigma(t)$.

Given the structure of the commutators in Eqs.~\eqref{eq:comm_cc_Hsp} and \eqref{eq:comm_cc_Hdrive}, the net effect of the renormalization term is to dress the single-particle part of the electronic Hamiltonian. In other words, the quadratic part of the Hamiltonian becomes
\begin{equation}
\label{eq:dressing_sp_matrix}
    \bar{h}_{\vb{k}nm} + \bar{\Omega}_{\vb{k}nm}(t) \;\to\;
    \bar{h}_{\vb{k}nm} + \bar{\Omega}_{\vb{k}nm}(t) + \Sigma_{\vb{k}nm}(t).
\end{equation}
Consider now this expression at equilibrium in the absence of the external perturbation, i.e., at $t=0$. Under these assumptions, the bare Rabi frequency vanishes and the renormalization term dresses the bare single-particle Hamiltonian matrix. Within the quasiparticle approximation on the Kohn-Sham basis, the dressed single-particle Hamiltonian becomes diagonal:
\begin{equation}
    h_{\vb{k}nm} = \bar{h}_{\vb{k}nm} + \Sigma_{\vb{k}nm}(0) = \delta_{nm} \varepsilon_{\vb{k}n}^0,
\end{equation}
where we defined the equilibrium dressed single-particle electronic eigenvalues $\varepsilon_{\vb{k}n}^0$. 
Out of equilibrium ($t > 0$), the renormalization dresses both the eigenvalues and the Rabi frequency through the diagonal and off-diagonal components of the self-energy variation, respectively. We define the time-dependent change of the self-energy as
\begin{equation}
\label{eq:self_energy_variation}
    \Delta \Sigma_{\vb{k}nm}(t) = \Sigma_{\vb{k}nm}(t) - \Sigma_{\vb{k}nm}(0).
\end{equation}
Using Eqs.~\eqref{eq:dressing_sp_matrix}-\eqref{eq:self_energy_variation}, we obtain the following expressions for the dressed time-dependent electronic quantities:
\begin{align}
    \varepsilon_{\vb{k}n}(t) &= \varepsilon_{\vb{k}n}^0 + \Delta \Sigma_{\vb{k}n}(t), \label{eq:renorm_eig2} \\
    \Omega_{\vb{k}nm}(t) &= \bar{\Omega}_{\vb{k}nm}(t) + \Delta \Sigma_{\vb{k}nm}(t). \label{eq:renorm_Rabi_freq}
\end{align}
Thus, in the single-particle commutators of Eqs.~\eqref{eq:comm_cc_Hsp} and \eqref{eq:comm_cc_Hdrive}, combined with the many-body renormalization term in Eq.~\eqref{eq:renormalization_term}, renormalization effects can be included by replacing the bare Hamiltonian matrix and Rabi frequencies with their time-dependent dressed counterparts given in Eqs.~\eqref{eq:renorm_eig2} and \eqref{eq:renorm_Rabi_freq}. Thus, we obtain:
\begin{align}
    & \langle [\hat{c}^\dagger_{\vb{k}m}\hat{c}_{\vb{k}n}, \hat{H}_{sp} + \hat{H}_{drive} + \hat{\Delta}]\rangle_t = \sum_{\mathclap{m'\neq m}} {\Omega}_{\vb{k}nm'}(t)p_{\vb{k}m'm}(t) \nonumber \\  
    & - \sum_{\mathclap{n'\neq n}} {\Omega}_{\vb{k}n'm}(t)\, p_{\vb{k}nn'}(t) + 
    \left[\varepsilon_{\vb{k}n}(t) - \varepsilon_{\vb{k}m}(t)\right]p_{\vb{k}nm}(t) \nonumber \\
    & + {\Omega}_{\vb{k}nm}(t)\left[f_{\vb{k}m}(t) - f_{\vb{k}n}(t)\right] \label{eq:comm_cc_Hdrive_H_sp_dressed} 
\end{align}
The single-particle contribution to the electronic occupations is found by setting $n=m$. In this case, the only non-vanishing terms are the first two on the right-hand side, leading to
\begin{equation}
    \langle [\hat{c}^\dagger_{\vb{k}n}\hat{c}_{\vb{k}n}, \hat{H}_{sp} + \hat{H}_{drive} + \hat{\Delta}]\rangle_t = 2\Im\sum_{\mathclap{m'\neq m}} {\Omega}_{\vb{k}nm'}(t)p_{\vb{k}m'm}(t)
\end{equation}
In the following, we detail the approximations used to construct the time-dependent quasiparticle self-energy variation.

We focus on the electron-electron contribution to the self-energy. This is treated within the GW approximation \cite{Hedin1965, Aryasetiawan1998}, where the bare Coulomb interaction is screened by a frequency-dependent dielectric function computed within the random-phase approximation (RPA).

In our framework, the time-local component of the self-energy is identified with the time-dependent Hartree term plus the static GW contribution, namely, the screened-exchange and Coulomb-hole terms known collectively as the COHSEX approximation \cite{Hedin1965, Bruneval2006}. We denote these as $\Delta \Sigma^{\text{H}}(t)$ and $\Delta \Sigma^{\text{GW}}(t)$, respectively. 

These contributions correspond to the first and third diagrams in Fig.~\ref{fig:diagrams}(b). Their time dependence arises from the evolution of electronic occupations and polarizations, which modify both the charge density and the static screening. The explicit expressions for $\Delta \Sigma^{\text{H}}(t)$ and $\Delta \Sigma^{\text{GW}}(t)$ are given in Eqs.(S23), (S28) and (S30) in the Supplementary Information.

We emphasize that the Hartree self-energy includes only inhomogeneous ($\vb{G} \neq 0$) contributions, as the homogeneous ($\vb{G} = 0$) component vanishes due to overall charge neutrality. Consequently, the Hartree variation contributes only in the presence of local-field effects.

To describe the time evolution of the screened Coulomb interaction, we employ a time-dependent static-screening approximation, namely
\begin{equation}
\label{eq:time_dependent_static_screening_guess_W}
    W(t,t') = W_s(t)\, \delta(t - t'),
\end{equation}
where $W_s(t)$ is the statically screened interaction evaluated at time $t$. In real space, it is given by
\begin{equation}
\label{eq:screened_time_dependent_W}
    W_s(\vb{r}, \vb{r}', t) = \int d\vb{r}_1\, v(|\vb{r} - \vb{r}_1|)\, \epsilon^{-1}_s(\vb{r}_1, \vb{r}', t),
\end{equation}
where $\epsilon^{-1}_s(\vb{r}, \vb{r}', t)$ is the time-dependent static inverse dielectric function evaluated within the RPA. In the present implementation, its time dependence is determined by the instantaneous nonequilibrium occupations, i.e., by the diagonal part of the electronic density matrix, while the contribution of microscopic interband polarizations is neglected, see Supplementary Section 4 for details on the time-dependent RPA dielectric function modeling.

This approximation is appropriate for the plasma-dominated regime considered here, where strong above-gap excitation rapidly generates incoherent carriers and the transient screening is mainly controlled by their occupations, less so in the earliest coherent stage after a near-resonant excitation of a bright exciton at low photoexcited density. In that case, including the full density matrix (occupations and polarizations) in the screening can suppress the long-wavelength screening response \cite{Perfetto_Marini_Stefanucci_2020}, reflecting the fact that a bound neutral electron--hole pair screens much less efficiently than an unbound electron--hole plasma.

Under the excitation conditions studied in this work, this coherence-driven correction is expected to be short-lived or subdominant, so that the occupation-based update captures the leading transient effect. Extending the present framework to a coherence-inclusive dielectric response is nevertheless straightforward.

We approximate the electron-phonon contribution to the electron self-energy using the Fan-Migdal (FM) approximation \cite{Fan1951, Migdal1958, Giustino2017}. Within this framework, the bare electron-phonon coupling matrix in Eq.~\eqref{eq:bare_elph_matrix} is replaced by the statically screened electron-phonon vertex
\begin{equation}
\label{eq:dress_elph_matrix}
{g}_{mn}^\nu(\vb{k},\vb{q}) =
 \sum_{a,\alpha}\dfrac{{e}_{\vb{q}\nu}^{a\alpha}}{{\sqrt{2M_a\omega^{0}_{\vb{q}\nu}}}}d^{a\alpha}_{mn}(\vb{k,q}),
\end{equation}
where the dressed deformation potential in reciprocal space is defined as
\begin{equation}
\label{eq:dressed_deformation_potential_reciprocal_space}
{d}_{mn}^{a\alpha}(\vb{k},\vb{q}) =
\mel{\psi_{\vb{k+q}m}}{ \sum_{I}e^{-i\vb{q}\cdot\vb{R}_{I}}d_{Ia\alpha}(\vb{\hat{r}})}{{\psi_{\vb{k}n}}},
\end{equation}
and the real-space deformation potential reads
\begin{equation}
\label{eq:screened_deformation_potential}
    d_{Ia\alpha}(\vb{r}) = \int d\vb{r'} \, \bar{d}_{Ia\alpha}(\vb{r'}) \, \epsilon^{-1}(\vb{r}, \vb{r'}).
\end{equation}
The dressed deformation potential is typically computed within DFPT \cite{Baroni2001} using a semilocal exchange-correlation kernel. As a result, the underlying static dielectric matrix includes both RPA screening and DFT exchange-correlation effects. The static dielectric matrix employed for the electron-electron self-energy is instead computed within the RPA alone; see Eq.~\eqref{eq:screened_time_dependent_W}.

This difference stems from practical considerations: the screened electron-phonon matrix elements are conventionally extracted from the self-consistent Kohn-Sham potential in DFPT. While, in principle, one could go beyond this level by incorporating GW corrections to the electron-phonon vertex \cite{Li2019}, such treatments remain computationally demanding and are not explored here.

A further distinction between Eqs.~\eqref{eq:screened_time_dependent_W} and \eqref{eq:screened_deformation_potential} concerns the treatment of time dependence. Whereas the statically screened Coulomb interaction $W_s(t)$ accounts for time-dependent screening effects arising from evolving occupations, the deformation potential is assumed to be time independent. In principle, this time dependence could be included by evaluating the dielectric matrix at each time step using instantaneous time-dependent DFT states; however, we leave this extension for future work.

The time-local electron-phonon self-energy is thus approximated by a time-dependent FM self-energy in the quasiparticle approximation, denoted $\Delta \Sigma^{\text{FM}}(t)$. This contribution corresponds to the sixth diagram in Fig.~\ref{fig:diagrams}(b). We also include the electron-phonon tadpole term, represented by the second diagram in Fig.~\ref{fig:diagrams}(b), which accounts for renormalizations induced by coherent atomic displacements. This term, denoted $\Delta \Sigma^{\text{AM}}(t)$, is discussed later..

An additional time-local singular self-energy from electron-phonon interaction is the Debye-Waller term. Although this term should, in principle, be included, we neglect it for practical reasons.

The explicit expressions for the tadpole and FM self-energies are given in Eqs.(S26) and (S29) in the Supplementary Information.

The collision term in Eq.~\eqref{eq:commutator_elel_elph} arises from all self-energy contributions beyond the singular part. It accounts for quasiparticle scattering and incoherent processes in the dynamics of the electronic state, thereby inducing the decay of quasiparticle populations and the dephasing of the system’s polarization.

Within the NEGF formalism \cite{Kadanoff2018,Marini2013, Stefanucci2023}, the scattering term is expressed as a time convolution between the greater/lesser components of the self-energy and the corresponding Green’s functions, as given in Eq.(S16) in the Supplementary Information. While this expression is formally exact, it has limited practical utility, as it depends on the full two-time structure of the Green’s functions and therefore does not yield a closed equation for the single-time electronic observables of interest.

To obtain a tractable description, we adopt the GKBA \cite{Lipavsky1986} in combination with the quasiparticle approximation for the electron propagator. This procedure yields a closed equation of motion, given a suitable form of the self-energy~\cite{Marini2013, Stefanucci2024}. However, the resulting scattering integral retains memory effects, as its evaluation requires access to the system’s history at all preceding times.

A further simplification comes from the Markov approximation, which assumes that scattering at a given time depends only on the system's instantaneous state. This approximation eliminates memory effects and yields time-local collision integrals \cite{Marini2013, Stefanucci2024}.

By combining the GKBA with the Markov approximation and specifying a closed form for the self-energy, we derive explicit expressions for the scattering integrals. In the following, we present the self-energy approximations adopted in this work and the corresponding expressions for the scattering terms.

To approximate the collision integrals arising from electron-electron interactions, we adopt the time-local form of the screened Coulomb interaction introduced in Eq.~\eqref{eq:time_dependent_static_screening_guess_W} and express the self-energy in terms of the statically screened, time-dependent interaction. Specifically, we employ the second-order Born approximation \cite{Haug2010}, wherein the bare Coulomb interaction is replaced by its statically screened counterpart. This approach corresponds to evaluating the fourth and fifth diagrams in Fig.~\ref{fig:diagrams}(b). Together with the third diagram, these terms represent the leading-order contributions to the electron self-energy in a perturbative expansion in powers of $W_s$. The first-order term is purely singular and yields only quasiparticle energy renormalizations, whereas the second-order terms incorporate retardation effects and contribute to scattering processes via the collision integrals.

Within these approximations, the diagonal electron-electron collision integral takes the form \cite{Marini2013, Steinhoff2016, Stefanucci2024}:
\begin{equation}
\label{eq:elel_collision_int}
    \begin{aligned}
    \mathcal{I}_{\vb{k}n}^{ee}(t) & = \dfrac{\pi}{N^2} \sum_{\mathclap{\substack{\vb{k'},\vb{q} \\ n',m,m'}}} \big|W_{\vb{kk'k-qk'+q}}^{nn'mm'}(t) - W_{\vb{kk'k'+qk-q}}^{nn'm'm}(t)\big|^2 \\
    &  \times \big[f_{\vb{k-q}m}(t)f_{\vb{k'+q}m'}(t)(1-f_{\vb{k}n}(t))(1-f_{\vb{k'}n'}(t)) - \\
     &
     f_{\vb{k}n}(t)f_{\vb{k'}n'}(t)(1-f_{\vb{k-q}m}(t))(1-f_{\vb{k'+q}m'}(t))\big] \times \\
    &\delta(\varepsilon_{\vb{k}n}(t) + \varepsilon_{\vb{k'}n'}(t) - \varepsilon_{\vb{k'+q}m'}(t) - \varepsilon_{\vb{k-q}m}(t)),
    \end{aligned}
  \end{equation}
where we introduce the time-dependent screened Coulomb matrix element,
\begin{equation}
    W_{\vb{kk'k''k'''}}^{nn'mm'}(t) = \mel{\psi_{\vb{k}n}\psi_{\vb{k'}n'}}{W_s(\hat{\vb{r}},\hat{\vb{r}}',t)}{{\psi_{\vb{k''}m}\psi_{\vb{k'''}m'}}},
\end{equation}
with $\vb{k} + \vb{k'} - \vb{k''} - \vb{k'''} = \vb{G}$, where $\vb{G}$ is a reciprocal-lattice vector.

The off-diagonal collision integral is approximated using a relaxation-time ansatz (RTA):
\begin{equation}
\label{eq:rta_polarization}
    \mathcal{I}_{\vb{k}nm}(t) \simeq - \Gamma_{\vb{k}nm}(t)p_{\vb{k}nm}(t),
\end{equation}
where $\Gamma_{\vb{k}nm}(t)$ denotes the time-dependent dephasing rate. These rates are often treated as phenomenological quantities and related to temperature-dependent excitonic linewidths \cite{Marini2008, Selig2016}. Here, we compute them within the quasiparticle approximation \cite{Stefanucci2024}, whereby
\begin{equation}
\label{eq:RTA_quasiparticle_approximation}
    \Gamma_{\vb{k}nm}(t) \simeq \Gamma_{\vb{k}n}(t) + \Gamma_{\vb{k}m}(t),
\end{equation}
with $\Gamma_{\vb{k}n}(t)$ denoting the relaxation rate of an electron in the $n$th Kohn-Sham orbital with momentum $\vb{k}$.

Under these assumptions, the electron-electron relaxation rate reads:
\begin{equation}
\label{eq:elel_dephasing_rate}
    \begin{aligned}
        \Gamma_{\vb{k}n}^{ee}(t) = & \dfrac{\pi}{2N^2} \sum_{\mathclap{\substack{\vb{k'},\vb{q} \\ m,n',m'}}} \big|W_{\vb{kk'k-qk'+q}}^{nn'mm'}(t) - W_{\vb{kk'k'+qk-q}}^{nn'm'm}(t)\big|^2 \\
        & \times \big[f_{\vb{k'}n'}(t)(1-f_{\vb{k-q}m}(t))(1-f_{\vb{k'+q}m'}(t)) + \\
     & f_{\vb{k-q}m}(t)f_{\vb{k'+q}m'}(t)(1-f_{\vb{k'}n'}(t))\big] \times
     \\
     & \delta(\varepsilon_{\vb{k}n}(t) + \varepsilon_{\vb{k'}n'}(t) - \varepsilon_{\vb{k'+q}m'}(t) - \varepsilon_{\vb{k-q}m}(t)).
    \end{aligned}
\end{equation}    
A link between this quantity, the diagonal collision integral, and the imaginary part of the electron self-energy is provided in Supplementary Section 5.

The explicit form of the electron-phonon collision term is derived from the full FM self-energy. Owing to the statically screened electron-phonon vertex adopted in this work, the collision integral follows directly without additional vertex approximation. Within this approximation, the diagonal contribution to the collision integral reads \cite{Marini2013, OMahony2019, Stefanucci2024}:

\begin{equation}
\label{eq:elph_collision_int}
\begin{aligned}
& \mathcal{I}^{ep}_{\vb{k}n}(t) = \dfrac{2\pi}{N} \sum_{\mathclap{\substack{\vb{q} \\ m,\nu}}} |g_{mn}^\nu(\vb{k,q})|^2 \Big\{  \\
& \big[f_{\vb{k+q}m}(t)(1 - f_{\vb{k}n}(t)) - n_{\vb{q}\nu}(t)(f_{\vb{k}n}(t) - f_{\vb{k+q}m}(t))\big] \times \\ 
&\delta({\varepsilon}_{\vb{k+q}m}(t) - {\varepsilon}_{\vb{k}n}(t) - \omega_{\vb{q}\nu}(t)) + \\
&\big[n_{\vb{q}\nu}(t)(f_{\vb{k+q}m}(t) - f_{\vb{k}n}(t)) - f_{\vb{k}n}(t)(1-f_{\vb{k+q}m}(t))\big] \times
\\
&\delta({\varepsilon}_{\vb{k+q}m}(t) - {\varepsilon}_{\vb{k}n}(t) + \omega_{\vb{q}\nu}(t)) \Big\}.
\end{aligned}
\end{equation}
In Eq.~\eqref{eq:elph_collision_int}, $n_{\vb{q}\nu}(t)$ and $\omega_{\vb{q}\nu}(t)$ denote the time-dependent phonon occupation and frequency of mode $\nu$ with momentum $\vb{q}$; their dynamics will be addressed later.

The off-diagonal component of the collision integral is obtained analogously to the electron-electron case by combining Eqs.~\eqref{eq:rta_polarization} and \eqref{eq:RTA_quasiparticle_approximation}. The corresponding quasiparticle decay rate due to electron-phonon interaction reads:
\begin{equation}
\begin{aligned}            
&{\Gamma}^{ep}_{\vb{k}n}(t) = \dfrac{\pi}{N}\sum_{\mathclap{\substack{\vb{q} \\ m,\nu}}} |g_{mn}^\nu(\mathbf{k},\mathbf{q})|^2 \Big\{ 
\\
&\big[n_{\vb{q}\nu}(t) + 1-f_{\vb{k+q}m}(t)\big]\delta({\varepsilon}_{\vb{k+q}m}(t) - {\varepsilon}_{\vb{k}n}(t) + {\omega}_{\vb{q}\nu}(t)) \\
& + \big[n_{\vb{q}\nu}(t) + f_{\vb{k+q}m}(t)\big]\delta({\varepsilon}_{\vb{k+q}m}(t) - {\varepsilon}_{\vb{k}n}(t) - {\omega}_{\vb{q}\nu}(t))\Big\}.
\end{aligned}
    \label{eq:elph_dephasing_rate}
\end{equation}
As in the electron-electron case, the connection between the decay rate, diagonal collision integral, and imaginary part of the FM self-energy is discussed in Supplementary Section 5.

We showed that the many-body commutator in Eq.~\eqref{eq:commutator_elel_elph} introduces two main effects: quasiparticle renormalization and scattering. The former leads to a time-dependent dressing of the electronic eigenvalues and Rabi frequencies, as described by Eqs.~\eqref{eq:renorm_eig2} and \eqref{eq:renorm_Rabi_freq}. The latter accounts for the decay of occupations and the dephasing of interband polarizations.

The full set of dynamical equations for electronic occupations and polarizations follows from Eqs.~\eqref{eq:df_nk/dt_commutators} and \eqref{eq:dp_nmk/dt_commutators}, including the single-particle contributions from Eqs.~\eqref{eq:comm_cc_Hsp} and \eqref{eq:comm_cc_Hdrive} and the many-body terms in Eq.~\eqref{eq:commutator_elel_elph}.

The resulting electronic EOMs read:
\begin{align}
\frac{d f_{\vb{k}n}(t)}{dt}
&= 2\,\Im\!\left[\sum_{\mathclap{m\neq n}} \Omega_{\vb{k}nm}(t)\, p_{\vb{k}mn}(t)\right] \nonumber \\
&
 \quad + \mathcal{I}_{\vb{k}n}^{ee}(t) + \mathcal{I}_{\vb{k}n}^{ep}(t),
\label{eq:EOM_el_occupations} \\
\frac{d p_{\vb{k}nm}(t)}{dt}
&= -\Big[\Gamma^{ee}_{\vb{k}n}(t)+\Gamma^{ee}_{\vb{k}m}(t)
       +\Gamma^{ep}_{\vb{k}n}(t)+\Gamma^{ep}_{\vb{k}m}(t) \nonumber \\
&\hspace{-43pt} - i\big(\varepsilon_{\vb{k}n}(t)-\varepsilon_{\vb{k}m}(t)\big)\Big]
      p_{\vb{k}nm}(t)
 - i\sum_{\mathclap{m'\neq m}}\Omega_{\vb{k}nm'}(t)p_{\vb{k}m'm}(t) \nonumber
 \\
 & \hspace{-43pt} + i\sum_{\mathclap{n'\neq n}}\Omega_{\vb{k}n'm}(t)\,p_{\vb{k}nn'}(t) - i\big(f_{\vb{k}m}(t)-f_{\vb{k}n}(t)\big)\,\Omega_{\vb{k}nm}(t)
\label{eq:EOM_el_polarizations}
\end{align}
The collision integrals entering the equations above are given in Eqs.~\eqref{eq:elel_collision_int} and \eqref{eq:RTA_quasiparticle_approximation}--\eqref{eq:elph_dephasing_rate}, while the renormalized electronic eigenvalues and Rabi frequencies are defined in Eqs.~\eqref{eq:renorm_eig2} and \eqref{eq:renorm_Rabi_freq}.

\subsection{Phonon Hamiltonian}
\label{subsec:Full phonon Hamiltonian}

Earlier, we introduced the harmonic phonon Hamiltonian at equilibrium within the BO approximation. This allowed us to define creation and annihilation operators for the normal modes of vibration from the Hessian of the BO energy surface; see Eqs.~\eqref{eq:dynamical_matrix_equilibrium}--\eqref{eq:harmonic_phonon_equilibrium2}. However, to account for electron-phonon coupling, we must instead start from the bare phonon Hamiltonian.

The first step is to derive a quadratic bare phonon Hamiltonian starting from Eq.~\eqref{eq:harmonic_phonon_equilibrium2}. In principle, this can be achieved by removing from the force-constant matrix in Eq.~\eqref{eq:force_constant_matrix} the electron-phonon self-energy evaluated within the adiabatic approximation~\cite{Calandra2010, Giustino2017}.

Diagonalizing the resulting matrix, namely the bare crystal elastic tensor $C_0$, yields the bare phonon frequencies. Here, however, we adopt a different but equivalent approach. Specifically, we descreen the force-constant matrix by using the static screened-screened approximation for the self-energy~\cite{Calandra2010, Berges2023, Caldarelli2025, Stefanucci_Perfetto2025}:
\begin{align}
\label{eq:screened_screened_ph_self_energy_cartesian_basis}
    \widetilde{\Pi}_{a\alpha b\beta}(\vb{q}) = \dfrac{1}{N}\sum_{\vb{k},n,m} & d^{a\alpha}_{mn}(\vb{k,q}) d^{b\beta}_{mn}(\vb{k,q}) \times \nonumber \\
    & \Re\left[\dfrac{f_{\vb{k}n}^0 - f_{\vb{k+q}m}^0}{\varepsilon^{\text{KS}}_{\vb{k}n} - \varepsilon^{\text{KS}}_{\vb{k+q}m}+i\eta}\right],
\end{align}
where $\eta \to 0^+$. We then define the bare force-constant matrix as
\begin{equation}
\label{eq:bare_force_constant_matrix}
    \bar{C}_{a\alpha b\beta}(\vb{q}) = C_{a\alpha b\beta}(\vb{q}) - \widetilde{\Pi}_{a\alpha b\beta}(\vb{q}).
\end{equation}
This expression can be computed directly from a density-functional calculation, thereby connecting the unscreened force-constant matrix with DFPT, in analogy with the bare electronic single-particle matrix; see Eq.~\eqref{eq:bare_electronic_hamiltonian2}. Note, however, that using this definition of the bare force-constant matrix is equivalent to using the bare crystal elastic tensor, since it amounts to a different partitioning between the reference propagator and a static phonon self-energy contribution, while leaving the equilibrium dressed dynamical matrix unchanged. We adopt this choice solely for transparency and to connect more directly with the linear-response density-functional framework.

The bare phonon Hamiltonian is then obtained by replacing the BO dynamical matrix in Eq.~\eqref{eq:harmonic_phonon_equilibrium1} with the bare one, as defined through Eqs.~\eqref{eq:dynamical_matrix_equilibrium}, \eqref{eq:dynamical_matrix_normal_modes}, and \eqref{eq:bare_force_constant_matrix}. The resulting expression reads
\begin{equation}
\label{eq:bare_ph_sp_hamiltonian}
\hat{H}_{h} = 
\dfrac{1}{2}\sum_{\mathclap{\substack{\vb{q} \\ \nu,\nu'}}}\Bigg[ \delta_{\nu\nu'}\omega^0_{\vb{q}\nu}\hat{P}^\dagger_{\vb{q}\nu}\hat{P}_{\vb{q}\nu'} + \hat{Q}^\dagger_{\vb{q}\nu}\dfrac{\bar{D}_{\vb{q}\nu\nu'}}{\sqrt{\omega^0_{\vb{q}\nu}\omega^0_{\vb{q}\nu'}}}\hat{Q}_{\vb{q}\nu'}\Bigg].
\end{equation}
A key difference relative to the BO harmonic Hamiltonian in Eq.~\eqref{eq:harmonic_phonon_equilibrium2} is that the bare phonon Hamiltonian is not diagonal in the normal-mode indices and includes quadratic terms with two creation or two annihilation operators. The full phonon Hamiltonian is obtained by combining Eqs.~\eqref{eq:elph_interaction_hamiltonian} and \eqref{eq:bare_ph_sp_hamiltonian}, resulting in
\begin{equation}
\label{eq:full_ph_hamiltonian}
    \hat{H}_{ph} = \hat{H}_{h} + \hat{H}_{ep},
\end{equation}
where $\hat{H}_{ep}$ describes the coupling between electrons and phonons.

Anharmonic phonon-phonon interactions can also be incorporated into the formalism, as discussed later.

\subsection{Phonon Bloch equations}
\label{subsec:Phonon Bloch equations}

We employ the same density-matrix formalism used previously for the electronic case. We track the time evolution of phonon degrees of freedom via the nonequilibrium phonon occupations $n_{\vb{q}\nu}(t)$, defined as
\begin{equation}
\label{eq:phonon_occupations_definition}
    n_{\vb{q}\nu}(t) = \langle \hat{a}^\dagger_{\vb{q}\nu}\hat{a}_{\vb{q}\nu} \rangle_t.
\end{equation}
The EOMs for the phonon occupations follow from Eqs.~\eqref{eq:Heis0}, \eqref{eq:full_ph_hamiltonian}, and \eqref{eq:phonon_occupations_definition}, and take the form
\begin{align}
\label{eq:comm_ph_occupations}
    i\dfrac{dn_{\vb{q}\nu}(t)}{dt} = \langle [\hat{a}^\dagger_{\vb{q}\nu}\hat{a}_{\vb{q}\nu}, \hat{H}_h] \rangle_t + \langle [\hat{a}^\dagger_{\vb{q}\nu}\hat{a}_{\vb{q}\nu}, \hat{H}_{ep}] \rangle_t.
\end{align}
The first commutator yields quadratic terms in second-quantized operators. Hence, we obtain nonequilibrium averages of the type $\langle \hat{a}^\dagger_{\vb{q}\nu}\hat{a}_{\vb{q}\nu'}\rangle_t$, $\langle \hat{a}_{\vb{q}\nu}\hat{a}^\dagger_{\vb{q}\nu'}\rangle_t$, and $\langle \hat{a}_{\vb{q}\nu}\hat{a}_{\vb{-q}\nu'}\rangle_t$. This indicates that Eq.~\eqref{eq:comm_ph_occupations} alone is not sufficient to determine the EOMs for phonon occupations and should be coupled to the corresponding equations for these additional averages. 

The second commutator, involving the many-body electron-phonon Hamiltonian, leads to a BBGKY hierarchy, analogous to the electronic case. This many-body problem can be addressed using Green’s-function techniques~\cite{Giustino2017, Stefanucci2023}, which yield the EOMs for the phonon occupations~\cite{Stefanucci2024}.

In principle, one can derive EOMs for averages involving two creation or two annihilation operators, which give rise to effects such as phonon squeezing~\cite{Johnson2009, Benatti2017, Lakehal2020} and time-dependent oscillations in thermal diffuse scattering~\cite{Teitelbaum2018, Fahy2016}. However, since we are primarily interested in phonon relaxation, the impact of such correlations can be neglected. Therefore, we restrict our treatment to the evolution of the phonon occupation alone.

Here we summarize the main outcomes of the many-body treatment of Eq.~\eqref{eq:comm_ph_occupations}. The commutator with the electron-phonon interaction Hamiltonian can be split, similarly to Eq.~\eqref{eq:commutator_elel_elph}, into two contributions:
\begin{equation}
\label{eq:elph_commutator_phonon}
    \langle [\hat{a}^\dagger_{\vb{q}\nu} \hat{a}_{\vb{q}\nu},\hat{H}_{ep}]\rangle_t = \Delta^{pe}_{\vb{q}\nu}(t) + i\mathcal{I}^{pe}_{\vb{q}\nu}(t),
\end{equation}
where the first term corresponds to a many-body renormalization and the second term is a collision integral. The explicit forms of $\Delta_{\vb{q}\nu}(t)$ and $\mathcal{I}_{\vb{q}\nu}(t)$ depend on the self-energy approximation adopted to handle the many-body hierarchy.

The renormalization term, as in the electronic case, originates from the singular (time-local) part of the phonon self-energy~\cite{Stefanucci2023, Stefanucci2024}. It can be expressed as a commutator between the phonon-occupation operator and an effective Hamiltonian contribution:
\begin{equation}
    \Delta_{\vb{q}\nu}^{pe}(t) = \langle [\hat{a}^\dagger_{\vb{q}\nu}\hat{a}_{\vb{q}\nu},\hat{\Delta}] \rangle,
\end{equation}
with
\begin{equation}
    \hat{\Delta} = \sum_{\vb{q},\nu,\nu'} \hat{Q}^\dagger_{\vb{q}\nu} \Pi_{\vb{q}\nu\nu'}(t) \hat{Q}_{\vb{q}\nu'}.
\end{equation}
Here, \(\Pi(t)\) denotes the time-local part of the phonon self-energy. By combining this renormalization term with the commutator involving the harmonic Hamiltonian in Eq.~\eqref{eq:comm_ph_occupations}, it is readily shown that the renormalization effectively modifies the harmonic force constant matrix. This leads to a dressed, time-dependent dynamical matrix
\begin{equation}
\label{eq:dressed_dynamical_matrix}
    D_{\vb{q}\nu\nu'}(t) = \bar{D}_{\vb{q}\nu\nu'} + 2\sqrt{\omega^0_{\vb{q}\nu}\omega^0_{\vb{q}\nu'}}\,\Pi_{\vb{q}\nu\nu'}(t).
\end{equation}
An explicit form for the time-dependent dynamical matrix follows once an approximation for the phonon self-energy is specified. We treat it via the screened-screened electron-hole bubble approximation~\cite{Calandra2010, Berges2023, Caldarelli2025,
Stefanucci_Perfetto2025}, the corresponding diagram is the first contribution in Fig.~\ref{fig:diagrams}(c).

Within this approximation, the time-local phonon self-energy is given by the adiabatic electron-hole bubble \(\Pi^{\text{EH}}(t)\), whose full expression is provided in Eq.(S32) in the Supplementary Information. At equilibrium (\(t=0\)), this self-energy satisfies
\begin{align}
    \Pi_{\vb{q}\nu\nu'}(0) = \sum_{a,b,\alpha,\beta} \dfrac{(e^{a\alpha}_{\vb{q}\nu})^*}{\sqrt{2 M_a \omega^0_{\vb{q}\nu}}} \widetilde{\Pi}_{a\alpha b\beta}(\vb{q}) \dfrac{e^{b\beta}_{\vb{q}\nu'}}{\sqrt{2 M_b \omega^0_{\vb{q}\nu'}}},
\end{align}
where \(\widetilde{\Pi}(\vb{q})\) is defined by Eq.~\eqref{eq:screened_screened_ph_self_energy_cartesian_basis}. Combining this expression with the definitions of the bare force constant matrix (Eq.~\eqref{eq:bare_force_constant_matrix}), the equilibrium dynamical matrix (Eq.~\eqref{eq:dynamical_matrix_equilibrium}), and Eq.~\eqref{eq:dressed_dynamical_matrix} evaluated at \(t=0\), we recover the BO dynamical matrix:
\begin{equation}
    D_{\vb{q}\nu\nu'}(0) = \delta_{\nu\nu'}\left[\omega_{\vb{q}\nu}^0\right]^2.
\end{equation}
This result implies that the time-dependent dynamical matrix can be written as
\begin{align}
    D_{\vb{q}\nu\nu'}(t) = \delta_{\nu\nu'}[\omega_{\vb{q}\nu}^0]^2 + 2\sqrt{\omega_{\vb{q}\nu}^0\omega_{\vb{q}\nu'}^0}\, \Delta \Pi^{\text{EH}}_{\vb{q}\nu\nu'}(t),
\end{align}
where \(\Delta \Pi(t) = \Pi(t) - \Pi(0)\) denotes the nonequilibrium correction to the self-energy. The time-dependent phonon frequencies are then obtained by diagonalizing \(D_{\vb{q}\nu\nu'}(t)\), which generally yields phonon eigenvectors different from the equilibrium ones.

To avoid this complication, we adopt the no mixing mode approximation, in which the normal-mode basis is fixed and only the diagonal part of the self-energy is retained. The time-dependent phonon frequencies thus reduce to
\begin{align}
\label{eq:renormalized_phonon_frequencies}
    [\omega_{\vb{q}\nu}(t)]^2 = \left[\omega_{\vb{q}\nu}^0\right]^2 + 2\omega_{\vb{q}\nu}^0\, \Delta\Pi^{\text{EH}}_{\vb{q}\nu}(t),
\end{align}
namely, the dressed dynamical matrix remains diagonal at all times. Hence, by neglecting the time evolution of quadratic phonon averages other than the occupations, the commutators involving the harmonic Hamiltonian combined with the renormalization term in Eq.~\eqref{eq:comm_ph_occupations} vanish. As a result, the only remaining contribution to the equation of motion for the phonon occupations is the scattering term.

The scattering integral originates from all phonon self-energy contributions other than the singular part. As in the electronic case, an explicit expression can be derived within the NEGF approach~\cite{Stefanucci2024}. Once again, a closed form is obtained via the phonon-GKBA~\cite{Karlsson2021} and the Markov approximation, thus neglecting memory effects.

We approximate the phonon self-energy with the electron-hole bubble with two statically screened vertices~\cite{Calandra2010, Caldarelli2025, Stefanucci_Perfetto2025}. Within the present approximation, the phonon-carrier scattering term coincides with the expression found in semiclassical phonon-transport theory~\cite{Ziman2001}:
\begin{align}
\label{eq:phel_collision_int}
&\mathcal{I}^{pe}_{\vb{q}\nu}(t) = -\dfrac{2\pi}{N}\sum_{\vb{k},m,n} |g_{mn}^\nu(\vb{k},\vb{q})|^2 \times \nonumber \\
&
\big[ n_{\vb{q}\nu}(t)(f_{\vb{k}n}(t) - f_{\vb{k+q}m}(t)) - f_{\vb{k+q}m}(t)(1 - f_{\vb{k}n}(t)) \big] \times \nonumber
\\
&\delta(\varepsilon_{\vb{k+q}m}(t) - \varepsilon_{\vb{k}n}(t) - \omega_{\vb{q}\nu}(t)),
\end{align}
where the screened electron-phonon matrix elements \(g_{mn}^\nu(\vb{k},\vb{q})\) are defined in Eq.~\eqref{eq:dress_elph_matrix}. As in the electronic case, this scattering integral can be linked to the imaginary part of the electron-hole bubble self-energy through the RTA; see Supplementary Section 5.

\subsection{Anharmonic effects}
\label{subsec:Anharmonic effects}

Up to now, we considered a harmonic phonon Hamiltonian coupled to electrons through Eq.~\eqref{eq:full_ph_hamiltonian}. While the harmonic approximation provides a solid foundation for lattice dynamics, real materials inevitably exhibit anharmonic effects. These are key to understanding many phonon-related phenomena, especially phonon transport, thermal expansion, and phase transitions \cite{Ziman2001, Fugallo2013, Bianco2017}.

For nonequilibrium phonon dynamics, anharmonicity plays a crucial role. It provides the microscopic mechanism behind intrinsic phonon-phonon scattering, which is essential for bringing the phonon subsystem to thermal equilibrium. This is particularly relevant for the relaxation of nonequilibrium phonon distributions generated by electron-phonon interactions. For instance, after an ultrafast electronic excitation, the energy transferred to the lattice initially populates specific phonon modes via carrier-phonon scattering \cite{OMahony2019}, and the subsequent redistribution of this energy among all phonon modes occurs primarily through anharmonic phonon-phonon scattering processes \cite{Caruso2022}.

To include anharmonicity, we expand the BO energy up to fourth order in the atomic displacement operators \cite{Mendez1984, Maradudin1962, Paulatto2015}, obtaining the effective anharmonic Hamiltonian:
\begin{equation}
\begin{aligned}
\label{eq:anharmonic_hamiltonian}
\hat{H}_{pp} = \hat{H}^{(3)} + \hat{H}^{(4)},
\end{aligned}    
\end{equation}
where 
\begin{equation}
        \hat{H}^{(n)} = \dfrac{2^{n/2} N}{n! N^{n/2}} \sum_{\mathclap{\substack{\vb{q}_1,..,\vb{q}_n \\ \nu_1,..,\nu_n}}} \Phi^{(n)}_{\nu_1..\nu_n}(\vb{q}_1,..,\vb{q}_n) \hat{Q}_{\vb{q}_1\nu_1}..\hat{Q}_{\vb{q}_n\nu_n}.
\end{equation}
Here, we introduce the third- and fourth-order anharmonic matrices in the normal-mode basis~\cite{Mendez1984, Maradudin1962, Paulatto2015}. The total phonon Hamiltonian is thus obtained by combining Eqs.~\eqref{eq:harmonic_phonon_equilibrium1} and \eqref{eq:anharmonic_hamiltonian}. This clarifies why we consider anharmonic effects as an extension of the results obtained earlier. While the electron-phonon coupling problem is solved from the bare harmonic dynamical matrix and deformation potential, the anharmonic part is defined from a BO energy expansion to fourth order, based on the equilibrium quadratic Hamiltonian in Eq.~\eqref{eq:harmonic_phonon_equilibrium1}.

At this point, we distinguish two regimes: perturbative and nonperturbative. Which case applies depends on the anharmonicity of the lattice potential, the temperature range considered, and the strength of quantum ionic effects; this is generally system-specific.

In the perturbative case, anharmonic effects are relatively small. The harmonic dynamical matrix is positive definite, so the BO energy can be expanded directly in terms of atomic displacements. The anharmonic matrices can be obtained either via DFPT \cite{Paulatto2013} or finite-difference methods \cite{Togo2015}. From these, one can estimate phonon linewidths and shifts using the anharmonic self-energy \cite{Maradudin1962, Paulatto2015}.

In the nonperturbative case, anharmonicity is strong enough to render the harmonic dynamical matrix unstable, and a direct expansion is no longer possible. Here, one can employ the SSCHA \cite{Monacelli2021} and perform a full quantum anharmonic minimization. This yields a new set of equilibrium positions and renormalized phonon frequencies (the SSCHA frequencies). The anharmonic matrices are then extracted via stochastic sampling of the free energy using an effective quadratic density matrix. This approach allows computation of nonperturbative linewidths and shifts as well \cite{Monacelli2021}.

In this work, we focus on the perturbative case for simplicity and defer the nonperturbative extension to future work. Moreover, for practical reasons, we retain only the third-order terms in Eq.~\eqref{eq:anharmonic_hamiltonian}. Even though fourth-order terms can yield significant phonon shifts, phonon linewidths are entirely determined by third-order terms \cite{Maradudin1962, Paulatto2015}.

Within this framework, the relevant self-energy is the phonon bubble (PB), see Fig.~\ref{fig:diagrams}(c), which gives a time-dependent correction to the BO phonon frequencies via the real part of $\Pi^{\text{PB}}(t)$ \cite{Maradudin1962, Paulatto2015, Monacelli2021}; see Eq.(S33) in the Supplementary Information. This correction is included in the quasiparticle and no mixing mode approximations. The resulting time-dependent phonon frequencies are then obtained from Eq.~\eqref{eq:renormalized_phonon_frequencies} by adding $\Pi^{\text{PB}}(t)$ to $\Delta \Pi^{\text{EH}}(t)$, namely
\begin{equation}
\label{eq:renormalized_phonon_frequencies2}
    \big[\omega_{\vb{q}\nu}(t)\big]^2 = \big[\omega_{\vb{q}\nu}^0\big]^2 + 2\omega_{\vb{q}\nu}^0[\Delta \Pi^{\text{EH}}_{\vb{q}\nu}(t) + \Pi^{\text{PB}}_{\vb{q}\nu}(t)].
\end{equation}
Anharmonicity also introduces phonon-phonon scattering, leading to an additional collision integral. This term, derived from Fermi’s golden rule for three-phonon processes, reads \cite{Ziman2001}:
\begin{equation}
\label{eq:phph_collision_int}
    \begin{aligned}        
    & \mathcal{I}_{\vb{q}\nu}^{pp}(t) = -\dfrac{\pi}{N} \sum_{\mathclap{\substack{\vb{q'},\vb{q''} \\ \nu',\nu''}}} \big|\Phi^{(3)}_{\nu\nu'\nu''}(\vb{q},\vb{q'},\vb{q''})\big|^2 \bigg\{ \\
    &\big[n_{\vb{q}\nu}(t)(n_{\vb{q}'\nu'}(t)+1)+ n_{\vb{q}''\nu''}(t)(n_{\vb{q}\nu}(t) - n_{\vb{q}'\nu'}(t))\big] \times \\ 
         & \delta({\omega}_{\vb{q}\nu}(t) - {\omega}_{\vb{q}'\nu'}(t) -{\omega}_{\vb{q}''\nu''}(t)) + 2\times
         \\
         &
         \big[n_{\vb{q}\nu}(t)(n_{\vb{q'}\nu'}(t) - n_{\vb{q''}\nu''}(t)) - n_{\vb{q''}\nu''}(t)(1 + n_{\vb{q'}\nu'}(t))\big] \times \\
         & \delta({\omega}_{\vb{q}\nu}(t) + {\omega}_{\vb{q'}\nu'}(t) - {\omega}_{\vb{q''}\nu''}(t))\bigg\}.
    \end{aligned}
\end{equation}
Again, this scattering term can be linked to the imaginary part of the phonon bubble self-energy via the RTA, as discussed in Supplementary Section 5. With this information, the EOMs for phonon occupations become 
\begin{equation}
\label{eq:EOM_ph_occupations}
    \dfrac{d n_{\vb{q}\nu}(t)}{dt} = \mathcal{I}_{\vb{q}\nu}^{pe}(t) + \mathcal{I}_{\vb{q}\nu}^{pp}(t),
\end{equation}
where the explicit expressions for the collision integrals are given in Eqs.~\eqref{eq:phel_collision_int} and \eqref{eq:phph_collision_int}, while the time-dependent phonon frequencies are obtained from Eq.~\eqref{eq:renormalized_phonon_frequencies2}.

\subsection{Coherent atomic motion}
\label{subsec:Coherent atomic motion}

In the previous sections, we derived the nonequilibrium EOMs for the coupled electron-phonon system, which in principle allow us to track its time evolution under the discussed approximations. At that stage, we neglected the possibility of light-induced coherent atomic motion. In other words, we assumed that the nonequilibrium thermal averages of ionic displacements and momenta, defined in Eqs.~\eqref{eq:at_displ_fourier} and \eqref{eq:at_mom_fourier}, vanish. However, photoexcited carriers can generate nonzero forces on the ions, inducing a net atomic displacement. This process is known as displacive excitation of coherent phonons~\cite{Cheng1991, Zeiger1992}, and it manifests through light-induced oscillations in Bragg peaks, reflectivity, and electronic bands~\cite{Teitelbaum_Shin2018, Huang2022, Emeis2024}.

To describe the time evolution of atomic displacements and momenta, we start from Eqs.~\eqref{eq:at_displ_fourier} and \eqref{eq:at_mom_fourier}. The ion dynamics is determined by the following nonequilibrium averages:
\begin{align} 
Q_{Ia\alpha}(t) &= \langle \hat{Q}_{Ia\alpha} \rangle_t, \\ 
P_{Ia\alpha}(t) &= \langle \hat{P}_{Ia\alpha} \rangle_t. 
\end{align}
We restrict our analysis to situations where the crystal periodicity is preserved, assuming $Q_{Ia\alpha}$ and $P_{Ia\alpha}$ independent of $I$. The EOMs for atomic displacements can be derived from the \textit{ab initio} electron-phonon Hamiltonian, as discussed in Refs.~\cite{Stefanucci2023,Stefanucci_Perfetto2025_2,Pan_Emeis_Juaernik_Bauer_Caruso2025}, and read:
\begin{align} 
\label{eq:d2Qdt2_real_space} 
M_a \dfrac{d^2 Q_{Ia\alpha}(t)}{dt^2} = & -\sum_{J,b,\beta} C^0_{Ia\alpha Jb\beta} Q_{Jb\beta}(t) \nonumber \\
& - \int d\vb{r} \bar{d}_{Ia\alpha}(\vb{r}) \Delta n(\vb{r},t), 
\end{align}
where the bare crystal elastic tensor is defined as
\begin{align} 
C^0_{Ia\alpha Jb\beta} =& \left. \dfrac{\partial^2 E_{nn}}{\partial Q_{Ia\alpha}\partial Q_{Jb\beta}} \right|_{\vb{R} = \vb{R}^0} \nonumber \\ 
& + \int d\vb{r} \ n^0(\vb{r}) \left. \dfrac{\partial^2 V_{ext}(\vb{r})}{\partial Q_{Ia\alpha}\partial Q_{Jb\beta}} \right|_{\vb{R} = \vb{R}^0}, 
\end{align}
with $E_{nn}$ denoting the nuclear-nuclear interaction energy and $n^0(\vb{r})$ the equilibrium electronic density. The bare deformation potential $\bar{d}_{Ia\alpha}(\vb{r})$ is defined in Eq.~\eqref{eq:bare_deformation_potential_real_space}, and the electronic-density variation from equilibrium is $\Delta n(\vb{r},t) = n(\vb{r},t) - n^0(\vb{r})$. The right-hand side of Eq.~\eqref{eq:d2Qdt2_real_space} represents the total force acting on the $a$th ion in the $I$th unit cell, including both harmonic elastic forces and the contribution from the nonequilibrium electronic density. This expression is consistent with the Ehrenfest force theorem derived from the harmonic crystal Hamiltonian~\cite{Baroni2001, Calandra2010}. Note that in Eq.~\eqref{eq:d2Qdt2_real_space} we neglect the force contribution arising from the direct coupling of the external field to the nuclear charges \cite{Stefanucci_Perfetto2025_2}, since we are interested in scenarios where the field frequency far exceeds any phonon frequency (non-resonant condition). 

We now analyze the nonequilibrium density variation $\Delta n(\vb{r},t)$ induced by the time-dependent light pulse, which generally contains both linear and nonlinear contributions in the electric field \cite{Stefanucci_Perfetto2025_2}. The linear terms, which give rise to dynamical Born-effective charges, are neglected in the non-resonant scenario studied here. We separate the density variation into two contributions:
\begin{equation}
\label{eq:density_variation_separation}
\Delta n(\vb{r},t) = \Delta n_{1}(\vb{r},t) + \Delta n_{2}(\vb{r},t),
\end{equation}
where $\Delta n_1(\vb{r},t)$ arises from the direct coupling between the electrons and the electric field, resulting in a variation of the electronic state. This term yields a nonequilibrium ionic force and thus an ionic displacement through Eq.~\eqref{eq:d2Qdt2_real_space}.
The term $\Delta n_2(\vb{r},t)$ is the electronic density variation induced by the ionic coherent motion initiated by the aforementioned force. Namely,
when ions start moving, after carrier thermalization, they induce a variation in the electron-hole plasma density, labeled $\Delta n_2(\vb{r},t)$,
which yields an additional force on top of the driving force related to $\Delta n_1(\vb{r},t)$.
We then assume that the variation $\Delta n_2(\vb{r},t)$ can be treated within linear response theory in the induced phonon displacement. This assumption is justified as long as the displacement involved in the coherent motion is small. Finally, we underline that, although the variations $\Delta n_1$ and $\Delta n_2$ are nonlinear in the pump electric field, if the nonequilibrium density variations $\Delta n_1$ and $\Delta n_2$ are small, one can still treat their effects within linear response in the variation of the electronic occupations and phonon displacement, as explained below. 

We evaluate the ionic force arising from $\Delta n_1(\vb{r},t)$, that is, the second term on the right-hand side of Eq.~\eqref{eq:d2Qdt2_real_space} due to $\Delta n_1(\vb{r},t)$. We note that the corresponding ionic force has the same form as that obtained in (time-dependent) DFPT with vanishing ionic displacements~\cite{Baroni2001,Calandra2010}. For this reason, we work within a linear-response DFPT framework and derive an expression for the nonequilibrium force at first order in the nonequilibrium electronic population change, i.e., $\Delta f_{\vb{k}n}$ . 

To this end, we start from the photoinduced force term, see Eqs.~\eqref{eq:d2Qdt2_real_space} and \eqref{eq:density_variation_separation}:
\begin{equation}
\label{eq:force_nonequilibrium_density_real_space}
    F_{Ia\alpha}(t) = - \int d\vb{r}\, \bar{d}_{Ia\alpha}(\vb{r}) \,\Delta n_{1}(\vb{r},t).
\end{equation}
For simplicity, and to highlight the connection with a time-independent DFPT framework, we assume a quasistationary electronic density variation. This assumption is well justified for times larger than tens of femtoseconds, since the density varies rapidly during the pulse and only slowly after carrier thermalization. This means that the force expression we obtain will accurately describe the ionic motion on its characteristic timescale (hundreds of femtoseconds) after carrier thermalization.

We relabel the time-independent density variation after carrier thermalization as $\Delta n(\vb{r})=\Delta n_{1}(\vb{r},t)$. We express the density variation in the KS basis as
\begin{align}
\label{eq:tot_density_variation}
    \Delta n(\vb{r}) \simeq \dfrac{1}{N} \sum_{\mathclap{\vb{k},n}} \Big[ & \Delta f_{\vb{k}n} |\psi_{\vb{k}n}(\vb{r})|^2 + f_{\vb{k}n}^0\Delta \psi_{\vb{k}n}^*(\vb{r})\psi_{\vb{k}n}(\vb{r}) \nonumber \\
    & + f_{\vb{k}n}^0\psi_{\vb{k}n}^*(\vb{r})\Delta\psi_{\vb{k}n} (\vb{r}) \Big],
\end{align}
accounting for both the change in electronic occupations and in the single-particle orbitals. This differs significantly from standard many-body treatments, where the single-particle orbitals are kept fixed and only changes in the occupations are considered. Here, instead, we also include the change in the single-particle orbitals induced by an out-of-equilibrium electronic occupation set.

We now express the change of the electronic orbitals using linear-response theory~\cite{Baroni2001}, obtaining 
\begin{align}
\label{eq:second_third_density_variation}
    \Delta n(\vb{r}) & \simeq \dfrac{1}{N} \sum_{\mathclap{\vb{k},n}}  \Delta f_{\vb{k}n} |\psi_{\vb{k}n}(\vb{r})|^2 \nonumber \\
    & + \frac{1}{N}\int d\vb{r}^{\prime} \chi_0(\mathbf{r},\mathbf{r}') \Delta V_{\text{KS}}(\vb{r}'),
\end{align}
where we used the definition of the static irreducible density-density response function
\begin{align}
    \chi_0(\vb{r,r'}) = \sum_{\substack{\vb{k,k'} \\ n,n'}} & \dfrac{f^0_{\vb{k}n} - f^0_{\vb{k'}n'}}{\varepsilon_{\vb{k}n}^{\text{KS}} - \varepsilon_{\vb{k'}n'}^{\text{KS}}} \psi_{\vb{k}n}^*(\vb{r})\psi_{\vb{k'}n'}(\vb{r}) \nonumber \\
    & \times \psi_{\vb{k'}n'}^*(\vb{r'})\psi_{\vb{k}n}(\vb{r'}).
\end{align}
We write the density variation to first order in the variation of the single-particle occupations, namely
\begin{equation}
\label{eq:density_variation_in_terms_of_occupations}
    \Delta n(\vb{r}) \simeq \dfrac{1}{N}\sum_{\vb{k},n} \dfrac{\delta n(\vb{r})}{\delta f_{\vb{k}n}} \Delta f_{\vb{k}n}.
\end{equation}
By employing Eqs.~\eqref{eq:tot_density_variation} and~\eqref{eq:second_third_density_variation}, we obtain the derivative of the density with respect to the single-particle occupations:
\begin{align}
\label{eq:derivative_density_wrt_occupations}
    \dfrac{\delta n(\vb{r})}{\delta f_{\vb{k}n}} =  \int d\vb{r'} \chi_0(\mathbf{r},\mathbf{r}') \frac{\delta V_{\text{KS}}(\vb{r})}{\delta f_{\vb{k}n}} 
    + |\psi_{\vb{k}n}(\vb{r})|^2.
\end{align}
The derivative of the Kohn-Sham potential can be written as
\begin{align}
\label{eq:derivative_KS_wrt_occupations}
    \dfrac{\delta V_{\text{KS}}(\vb{r})}{\delta f_{\vb{k}n}} = & \dfrac{\delta V_{ext}(\vb{r})}{\delta f_{\vb{k}n}} + \dfrac{\delta V_{\text{H}xc}(\vb{r})}{\delta f_{\vb{k}n}} = \dfrac{\delta V_{\text{H}xc}(\vb{r})}{\delta f_{\vb{k}n}} \nonumber \\
    = & \int d\vb{r'} K_{\text{H}xc}(\vb{r,r'}) \dfrac{\delta n(\vb{r}')}{\delta f_{\vb{k}n}},
\end{align}
where we used the fact that the external potential does not depend on the electronic occupations and employed the definition of the Hartree plus exchange-correlation kernel, $K_{\text{H}xc}(\vb{r,r'}) = \delta V_{\text{H}xc}(\vb{r})/\delta n(\vb{r'})$.

We plug the expression in Eq.~\eqref{eq:derivative_KS_wrt_occupations} into Eq.~\eqref{eq:derivative_density_wrt_occupations} and obtain the following self-consistent equation for the derivative of the density:
\begin{align}
\label{eq:derivative_density_wrt_occupations1}
    \dfrac{\delta n(\vb{r})}{\delta f_{\vb{k}n}} = & \int d\vb{r'}\int d\vb{r''} \chi_0(\vb{r,r'}) K_{\text{H}xc}(\vb{r',r''})\dfrac{\delta n(\vb{r''})}{\delta f_{\vb{k}n}} \nonumber \\
    & + |\psi_{\vb{k}n}(\vb{r})|^2.
\end{align}
We employ the standard definition of the ground-state static dielectric function
\begin{equation}
    \epsilon(\vb{r,r'}) = \delta(\vb{r-r'}) - \int d\vb{r''} \chi_0^R(\vb{r,r''})K_{\text{H}xc}(\vb{r'',r'})
\end{equation}
and rearrange Eq.~\eqref{eq:derivative_density_wrt_occupations1} to express the derivative of the density with respect to occupations, obtaining
\begin{equation}
    \int d\vb{r'} \epsilon(\vb{r,r'}) \dfrac{\delta n(\vb{r'})}{\delta f_{\vb{k}n}} = |\psi_{\vb{k}n}(\vb{r})|^2,
\end{equation}
or equivalently 
\begin{equation}
\label{eq:derivative_density_wrt_occupations2}
    \dfrac{\delta n(\vb{r})}{\delta f_{\vb{k}n}} = \int d\vb{r'} \epsilon^{-1}(\vb{r,r'}) |\psi_{\vb{k}n}(\vb{r'})|^2.
\end{equation}
We use Eqs.~\eqref{eq:derivative_density_wrt_occupations2} and~\eqref{eq:density_variation_in_terms_of_occupations} to rewrite Eq.~\eqref{eq:force_nonequilibrium_density_real_space} as
\begin{align}
\label{eq:ionic_force_real_space_final}
    F_{Ia\alpha} = & -\dfrac{1}{N}\sum_{\vb{k},n} \Delta f_{\vb{k}n} \mel{\psi_{\vb{k}n}}{\int d\vb{r'} \epsilon^{-1}(\hat{\vb{r}}',\hat{\vb{r}})\bar{d}_{Ia\alpha}(\vb{r'})}{\psi_{\vb{k}n}} \nonumber \\
    = & -\dfrac{1}{N}\sum_{\vb{k},n} \Delta f_{\vb{k}n} \mel{\psi_{\vb{k}n}}{d_{Ia\alpha}(\hat{\vb{r}})}{\psi_{\vb{k}n}},
\end{align}
where we employed the definition of the dressed deformation potential (Eq.~\eqref{eq:screened_deformation_potential}). This expression has the same form as the forces used in Refs.~\cite{OMahony2019, Liu2022, Emeis2024}. In Supplementary Section 6 we present an alternative derivation of the forces induced by a variation of the electronic occupations based on Janak’s theorem \cite{Janak1978}.

We underline that the real-time forces are expressed in terms of the \textit{screened}, rather than the bare, electron-phonon vertex, as is usually assumed in the Ehrenfest approximation. As shown above, this vertex dressing arises from the change in the single-particle orbitals induced by the nonequilibrium electronic distribution. For this reason, our quasi-equilibrium treatment goes beyond state-of-the-art Ehrenfest dynamics, in which the orbitals are kept fixed. It is worth noting that vertex screening yields quantitative differences in the real-time ionic forces, with important consequences for the phonon dynamics.

It is worth comparing our screened electron-phonon coupling with the one entering the corresponding equation of motion derived in Ref.~\cite{Stefanucci_Perfetto2025}. In our density-functional framework, the dressing of the vertex arises directly from the self-consistent response of the single-particle problem: the photoinduced change in the electronic occupations modifies the Kohn--Sham potential and, therefore, the orbital basis used to evaluate the electron-phonon matrix elements. In the Green's function formulation of Ref.~\cite{Stefanucci_Perfetto2025}, by contrast, the electronic basis is kept fixed, and the corresponding vertex renormalization is generated through self-energy corrections. As a consequence, the screened coupling entering the force differs in its explicit form, even though the two prescriptions coincide in the electron-gas limit, as discussed in Ref.~\cite{Stefanucci_Perfetto2025}. Establishing a more general equivalence represents an interesting future direction of research.

The result in Eq.~\eqref{eq:ionic_force_real_space_final} has been derived under the assumption that the density deviation from equilibrium depends only weakly on time. This is similar in spirit to the adiabatic approximation and represents a significant simplification, as it enables the use of time-independent perturbation theory at quasi-equilibrium. However, light-induced forces are generally time dependent. Here, we introduce the time dependence naturally through the evolution of the electronic occupations. This implies that the forces will be most accurate when the occupations vary slowly in time.

On timescales longer than those associated with carrier relaxation and the electric-field pulse duration, this is a good approximation, as the system reaches a quasi-equilibrium electronic distribution~\cite{Tangney2002, Marini2021}. By contrast, on the femtosecond timescale this is generally not the case, and the forces in Eq.~\eqref{eq:ionic_force_real_space_final} may deviate from the actual ones. However, since atomic displacements occur on a much longer timescale, this has little effect on the ionic dynamics, as the dominant contribution comes from the force at long times. For these reasons, in the following we use Eq.~\eqref{eq:ionic_force_real_space_final} together with time-dependent electronic occupations.

In this part, we focus on the density variation induced by a nonzero ionic displacement, initiated by changes in the carrier state. To account for the density response to the displacement of the $I$th ion, we use linear-response theory under the assumptions of quasi-equilibrium and a monochromatic perturbation~\cite{Baroni2001, Calandra2010}. The density variation due to a displacement in the $I$th cell reads
\begin{equation} 
\label{eq:monochromatic_density_response} 
\Delta n^I_2(\vb{r},\omega) = \int d\vb{r}' \chi_0(\vb{r},\vb{r}',\omega) \sum_{a,\alpha} d_{Ia\alpha}(\vb{r}',\omega) Q_{Ia\alpha}(\omega), 
\end{equation}
where $\chi_0(\vb{r},\vb{r}',\omega)$ is the irreducible density-density response function, $\omega$ is the perturbation frequency, and the frequency-dependent screened deformation potential is
\begin{equation} 
d_{Ia\alpha}(\vb{r},\omega) = \bar{d}_{Ia\alpha}(\vb{r}) + \int d\vb{r}' K_{\text{H}xc}(\vb{r},\vb{r}',\omega) \Delta n^I_2(\vb{r}',\omega), 
\end{equation}
with $K_{\text{H}xc}(\vb{r},\vb{r}',\omega)$ the Hartree plus exchange-correlation kernel. Summing Eq.~\eqref{eq:monochromatic_density_response} over all $I$ and inserting it into Eq.~\eqref{eq:d2Qdt2_real_space}, we obtain
\begin{align} 
\label{eq:d2Qdt2_real_space2} 
M_a \dfrac{d^2 Q_{Ia\alpha}(t)}{dt^2} = & -\sum_{J,b,\beta} \int dt' C_{Ia\alpha Jb\beta}(t-t') Q_{Jb\beta}(t') \nonumber \\
& + F_{Ia\alpha}(t),
\end{align}
where $F_{Ia\alpha}(t)$ is the force contribution due to nonequilibrium electronic occupations in Eq.~\eqref{eq:ionic_force_real_space_final}, and the Fourier transform of the nonadiabatic force constant matrix is defined as~\cite{Calandra2010}
\begin{equation}
\label{eq:non_adiabatic_force_constant_matrix} 
\begin{aligned} 
C_{Ia\alpha Jb\beta}(\omega) = & \int d\vb{r} d\vb{r}' \bar{d}_{Ia\alpha}(\vb{r}) \chi_0(\vb{r},\vb{r}',\omega) d_{Jb\beta}(\vb{r}',\omega)\\ 
& + C^0_{Ia\alpha Jb\beta}.
\end{aligned}
\end{equation}
Assuming a monochromatic ionic response, the force constant matrix in Eq.~\eqref{eq:non_adiabatic_force_constant_matrix} must be evaluated at the response frequency, which is itself determined by solving Eq.~\eqref{eq:d2Qdt2_real_space2}. Moreover, the nonadiabatic force constants are generally complex, implying that the harmonic response decays exponentially with a certain damping rate. To determine the ionic response, we Fourier transform Eq.~\eqref{eq:non_adiabatic_force_constant_matrix}, exploit crystal periodicity, and obtain the force constants in reciprocal space, $C_{a\alpha b\beta}(\vb{q},\omega)$. The nonadiabatic phonon frequencies $\widetilde{\omega}_{\vb{q}\nu}$, eigenvectors $\widetilde{\vb{e}}^a_{\vb{q}\nu}$, and half-width-at-half-maximum (HWHM) damping constants $\Gamma_{\vb{q}\nu}$ are then found from the following self-consistent equations~\cite{Calandra2010}:
\begin{gather}
    \det\left|\Re\left[\dfrac{C_{a\alpha b\beta}(\vb{q},\widetilde{\omega}_{\vb{q}\nu})}{\sqrt{M_a M_b}}\right] - \widetilde{\omega}^2_{\vb{q}\nu}\right| = 0, 
    \\
    \Gamma_{\vb{q}\nu} = \dfrac{1}{2\widetilde{\omega}_{\vb{q}\nu}} \sum_{\substack{a,b \\
    \alpha,\beta}} \dfrac{(\widetilde{{e}}^{a\alpha}_{\vb{q}\nu})^*}{\sqrt{M_a}} \Im\left[C_{a\alpha b\beta}(\vb{q},\widetilde{\omega}_{\vb{q}\nu})\right] \dfrac{\widetilde{{e}}^{b\beta}_{\vb{q}\nu}}{\sqrt{M_b}}.
\end{gather}
We assume that the BO eigenvectors, obtained by diagonalizing the equilibrium dynamical matrix (see Eq.~\eqref{eq:dynamical_matrix_normal_modes}), provide a good approximation to the nonadiabatic ones. We further assume that the nonadiabatic phonon frequencies deviate only slightly from the BO frequencies, so that we set $\widetilde{\omega}_{\vb{q}\nu} = \omega_{\vb{q}\nu}^0$. This is generally a good approximation in semiconductors. As for the imaginary part of the nonadiabatic force-constant matrix, which determines the damping constant of the nuclear motion, we adopt the framework of Ref.~\cite{Calandra2010}, in which the force constant is replaced by a stationary functional of the density response containing two statically screened electron-phonon vertices. This differs qualitatively from the structure of the nonadiabatic matrix in Eq.~\eqref{eq:non_adiabatic_force_constant_matrix}, which instead contains one bare and one dynamically screened electron-phonon vertex. Note that a recently proposed formulation within the NEGF framework has shown that the same conclusion regarding the vertex structure can be obtained from the static limit of an exact rewriting of the imaginary part of the electron-phonon self-energy~\cite{Stefanucci_Perfetto2025}.

Under the above assumptions, the real part of the dynamical matrix is diagonal in the BO basis. We therefore insert the expansion of the atomic displacements in the BO basis, Eq.~\eqref{eq:at_displ_fourier}, into Eq.~\eqref{eq:d2Qdt2_real_space2} and express the equations of motion in terms of atomic displacements in reciprocal space, namely the thermal average of the operator in Eq.~\eqref{eq:displacement_operator_reciprocal_space}.

If crystal periodicity is preserved, the only nonzero average occurs at $\vb{q}=0$. Moreover, we neglect the time evolution of the acoustic modes. The corresponding equation reads
\begin{align}
\label{eq:d2Qnudt2_electron_phonon}
    \dfrac{d^2 Q_{\nu}(t)}{dt^2} = & -\left({\omega}^0_{\vb{0}\nu}\right)^2 Q_{\nu}(t) - 2\Gamma_{\vb{0}\nu}^{ep}\dfrac{dQ_{\nu}(t)}{dt} + F_{\nu}^{ep}(t),
\end{align}
where the force induced by the photocarriers is
\begin{equation}
\label{eq:ep_ionic_force_normal_modes}
    F^{ep}_\nu(t) = -\omega_{\vb{0}\nu}^{0} \sqrt{\dfrac{2}{N}} \sum_{\vb{k},n} \Delta f_{\vb{k}n}(t) g_{nn}^\nu(\vb{k,0}).
\end{equation}
Because of the photoexcited electrons, both the phonon frequencies and the damping constants become time dependent and deviate from their equilibrium values. Formally, this amounts to evaluating the change in the density due to a nuclear displacement using the instantaneous electronic and phononic state as the reference. To estimate the nonequilibrium frequency, we use a time-dependent adiabatic approximation for the dynamical matrix. The phonon frequencies are then obtained using Eq.~\eqref{eq:renormalized_phonon_frequencies}, and we set ${\omega}_{\vb{0}\nu}^0 \to \omega_{\vb{0}\nu}(t)$. For the HWHM phonon damping, the screened-screened vertex structure of the approximate electron-phonon self-energy~\cite{Calandra2010,Stefanucci_Perfetto2025} yields the Allen formula~\cite{Allen1972}:
\begin{align}
\label{eq:gamma_nu_elph}
    \Gamma_{\vb{q}\nu}^{ep}(t) = &\dfrac{\pi}{N}\sum_{\mathclap{\vb{k},n,m}} \big|g_{nm}^\nu(\vb{k,q})\big|^2 \big[f_{\vb{k}n}(t) - f_{\vb{k+q}m}(t)\big] \nonumber \\
    & \times \delta\!\big(\varepsilon_{\vb{k}n}(t) - \varepsilon_{\vb{k+q}m}(t) + \omega_{\vb{q}\nu}(t)\big),
\end{align}
where we include the time dependence of both the electronic and phononic degrees of freedom.

Using the same approach as for the anharmonic effects discussed above, we include anharmonic effects on top of Eq.~\eqref{eq:d2Qnudt2_electron_phonon}. The effect of anharmonicity is to renormalize the phonon frequencies and decay rate through the time-dependent phonon bubble self-energy. The time-dependent phonon frequencies thus obey Eq.~\eqref{eq:renormalized_phonon_frequencies2}, while the total decay rates acquire an anharmonic contribution~\cite{Maradudin1962,Paulatto2015} that reads:
\begin{align}
\label{eq:gamma_nu_phph}
    &\Gamma_{\vb{q}\nu}^{pp} = \dfrac{\pi}{2N} \sum_{\substack{\vb{q}_1,\vb{q}_2 \\ \nu_1,\nu_2}} \big|\Phi^{(3)}_{\nu\nu_1\nu_2}(\vb{q},\vb{q}_1,\vb{q}_2)\big|^2 \times\Big\{ \nonumber
    \\
    &[1+n_{\vb{q}_1\nu_1}(t) + n_{\vb{q}_2\nu_2}(t)]\delta(\omega_{\vb{q}\nu}(t) - \omega_{\vb{q}_1\nu_1}(t) - \omega_{\vb{q}_2\nu_2}(t)) \nonumber
    \\
    &2[n_{\vb{q}_1\nu_1}(t) - n_{\vb{q}_2\nu_2}(t)]\delta(\omega_{\vb{q}\nu}(t) + \omega_{\vb{q}_1\nu_1}(t) - \omega_{\vb{q}_2\nu_2}(t))\Big\}.
\end{align}
Another effect of anharmonicity is an additional atomic-force contribution \cite{Caruso2023}, which reads
\begin{align}
\label{eq:anharmonic_force_component}
    F^{pp}_\nu(t) = -\omega^0_{\vb{0}\nu} \sqrt{\dfrac{1}{2N}} \sum_{\mathclap{\vb{q},\nu_1}} \Phi^{(3)}_{\nu\nu_1\nu_1}(\vb{0},\vb{q},\vb{-q}) \times \nonumber \\
    \big(1 + n_{\vb{q}\nu_1}(t) + n_{\vb{-q}\nu_1}(t)\big).
\end{align}
In Supplementary Section 7, we show how the time-dependent phonon frequency renormalization, decay, and force arise in the coherent atomic motion equation from a third-order anharmonic Hamiltonian \cite{Wenschuh1995, Pan_Emeis_Juaernik_Bauer_Caruso2025}; see Eqs.~\eqref{eq:harmonic_phonon_equilibrium1} and~\eqref{eq:anharmonic_hamiltonian}.

The final equations for the coherent atomic motion read
\begin{align}
    \label{eq:EOM_atomic_displacement}
    \dfrac{d^2 Q_{\nu}(t)}{dt^2} = -\omega_{\vb{0}\nu}^2(t) Q_{\nu}(t) - 2\Gamma_{\vb{0}\nu}(t) \dfrac{dQ_{\nu}(t)}{dt} + F_\nu(t),
\end{align}
where the time-dependent phonon frequencies obey Eq.~\eqref{eq:renormalized_phonon_frequencies2} and
\begin{align}
    \Gamma_{\vb{0}\nu}(t) &= \Gamma_{\vb{0}\nu}^{ep}(t) + \Gamma_{\vb{0}\nu}^{pp}(t), \\
    F_{\nu}(t) &= F_{\nu}^{ep}(t) + F_{\nu}^{pp}(t).
\end{align}
Coherent atomic motion also changes the electronic states. We associate a time-dependent self-energy contribution $\Delta \Sigma^{\text{AM}}_{\vb{k}n}(t)$ due to ionic motion, which renormalizes the electronic structure via Eq.~\eqref{eq:renorm_eig2}. This self-energy corresponds to the electron-phonon tadpole diagram, shown as the second contribution in Fig.~\ref{fig:diagrams}(b).

The expression of this self-energy can be derived either from the definition of the screened electron-phonon coupling matrix in a DFPT framework or from many-body Green’s-function theory~\cite{Marini2015, Stefanucci2023}, and can be used to describe time-dependent ARPES oscillations in pump-probe experiments~\cite{Emeis2024}. The corresponding expression is reported in Eq.(S26) in the Supplementary Information.

\subsection{Final equations}
\label{sec:Final equations}

\begin{figure}[t]
    \centering
    \includegraphics[width=0.95\linewidth]{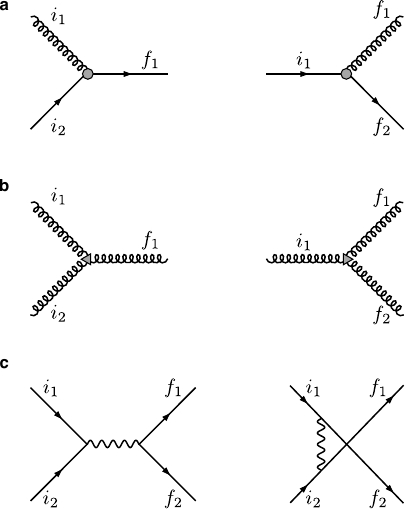}
    \caption{
    \textbf{Pictorial representation of the collisions considered in the dynamics.} The labels $i_{n}$ and $f_{n}$ denote the initial and final states, respectively. (a) Carrier-phonon processes where a phonon is absorbed or emitted by an incoming electron. (b) Phonon-phonon processes where a phonon is absorbed or emitted by an incoming phonon. (c) Carrier-carrier processes involving two-electron collisions.}
    \label{fig:collisions}
\end{figure}

In the preceding discussion, we derived the EOMs for electrons, phonons, and coherent atomic motion. These equations arise from a many-body hierarchy that is fully determined by the chosen self-energy approximations.

\begin{figure*}[t]
\includegraphics[width=0.95\textwidth]{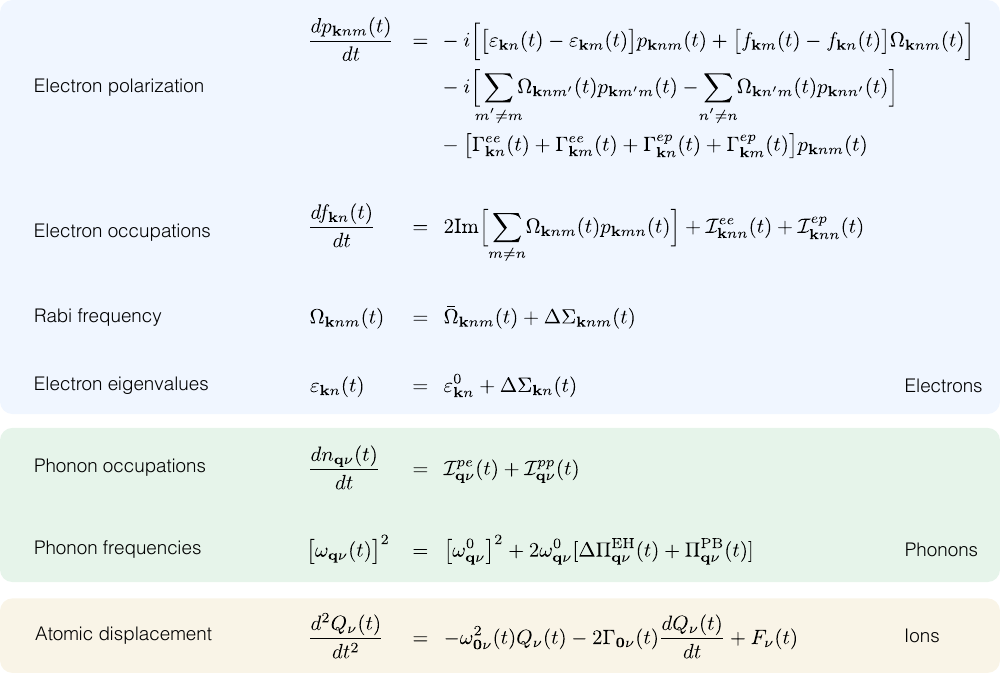}
\caption{\textbf{Full set of dynamical EOMs of the coupled electron-phonon system.} This is naturally divided into three main blocks, each one describing the coupled electron, phonon, and ion dynamics.}
\label{fig:EOM}
\end{figure*}

The role of the electron and phonon self-energies is twofold. On one hand, they lead to quasiparticle renormalizations of energies and Rabi frequencies (see Eqs.~\eqref{eq:renorm_Rabi_freq},~\eqref{eq:renorm_eig2}, and~\eqref{eq:renormalized_phonon_frequencies2}). On the other hand, they determine the form of the collision integrals due to many-body interactions (see Eqs.~\eqref{eq:elel_collision_int},~\eqref{eq:elph_collision_int},~\eqref{eq:phel_collision_int}, and~\eqref{eq:phph_collision_int}). Many-body renormalizations are linked to the real part of the quasiparticle self-energies, while the collision integrals are connected to their imaginary part via the RTA; see Supplementary Section 5.

Figure~\ref{fig:diagrams}(b,c) shows the diagrams corresponding to the chosen self-energy approximations. In the following, we briefly summarize the underlying approximations and their implications in terms of quasiparticle renormalizations and scattering. In particular, all the collision integrals in Eqs.~\eqref{eq:elel_collision_int},~\eqref{eq:elph_collision_int},~\eqref{eq:phel_collision_int}, and~\eqref{eq:phph_collision_int} correspond to semiclassical scattering processes, analogous to those found in transport theory~\cite{Ziman2001}, and can be obtained by means of Fermi’s golden rule.

We first focus on the electron self-energy shown in Fig.~\ref{fig:diagrams}(b). The first, third, fourth, and fifth diagrams represent the Hartree plus electron-electron contributions, expressed in terms of the time-dependent statically screened Coulomb interaction $W_s$. The first and third diagrams correspond to the time-dependent Hartree and COHSEX~\cite{Hedin1965}. While they do not contribute to the collision integrals, they are responsible for the many-body renormalization of the electronic energies and Rabi frequencies (see Eqs.(S23),(S28), and (S30) in the Supplementary Information). By contrast, the fourth and fifth diagrams contribute only to the collision integrals (see Eq.~\eqref{eq:elel_collision_int}) and not to quasiparticle renormalizations.

The electron-electron collision integral is associated with two distinct scattering processes, illustrated in Fig.~\ref{fig:collisions}(c). These are two-body collisions between carriers that conserve particle number but allow energy- and momentum-exchange between the initial and final states. The two types of collisions are topologically different, as the momentum conservation at each node imposes different constraints on the final momenta. Carrier-carrier scattering is efficient at redistributing energy, as there are many possible combinations of initial and final states that satisfy energy conservation. However, the scattering probability, proportional to $|W_s|^2$, decreases with increasing momentum transfer $\vb{q}$. Thus, carrier-carrier scattering is efficient for energy redistribution, but gets weaker at large momentum transfer.

The second and sixth diagrams in the electron self-energy represent the tadpole \cite{Marini2015} and Fan-Migdal approximations~\cite{Fan1951, Migdal1958, Giustino2017}, which capture the electron-phonon interaction. The electron-phonon tadpole diagram is responsible for quasiparticle renormalizations due to coherent atomic motion; see Eq.(S26) in the Supplementary Information. As for the Fan-Migdal contribution, it leads to both many-body renormalizations and collision integrals; see Eqs.(S29) in the Supplementary Information and Eq.~\eqref{eq:elph_collision_int}. The associated scattering processes, shown in Fig.~\ref{fig:collisions}(a), involve the absorption or emission of a phonon by a carrier. These processes conserve the number of electrons but not the number of phonon quasiparticles. The energy exchanged during these collisions is of the order of the phonon energy, while the momentum exchange can span the entire Brillouin zone (BZ). As a result, carrier-phonon scattering is effective at redistributing crystal momentum but less efficient in redistributing energy.

Regarding the phonon self-energy, the first diagram in Fig.~\ref{fig:diagrams}(c) is the screened-screened electron-hole bubble~\cite{Calandra2010, Stefanucci_Perfetto2025}, which contributes to both phonon-frequency renormalization and collision integrals for carrier-phonon scattering; see Eqs.(S32) in the Supplementary Information and ~\eqref{eq:phel_collision_int}. The corresponding quasiparticle processes are analogous to those discussed for electrons in Fig.~\ref{fig:collisions}(a), involving energy and momentum exchange between electron and phonon degrees of freedom. These scattering events occur with a probability proportional to $|g|^2$, implying that the most active phonon modes are those with large electron-phonon coupling matrix elements. In principle, this mechanism drives the phonon subsystem toward a thermalized state, albeit we underline that energy and momentum redistribution among phonons can occur only indirectly if phonon-phonon scattering is neglected, leading to a less efficient thermalization.

The second diagram in Fig.~\ref{fig:diagrams}(c) represents the phonon bubble~\cite{Maradudin1962, Paulatto2015, Monacelli2021}, which accounts for anharmonic phonon-frequency renormalization and phonon-phonon scattering; see Eqs.(S33) in the Supplementary Information and~\eqref{eq:phph_collision_int}. The corresponding semiclassical scattering events are shown in Fig.~\ref{fig:collisions}(b), involving phonon creation or annihilation mediated by another phonon. These processes enable direct energy and momentum exchange among phonons, providing an efficient mechanism for lattice thermalization after mode-specific activation via electron-phonon interactions.

The full set of dynamical equations, yielding the time evolution of electronic and phononic degrees of freedom and atomic displacements, constitute the EOMs for the nonequilibrium system. These equations,  reported in Fig.~\ref{fig:EOM}, together with initial equilibrium conditions, form a coupled set of nontrivial differential equations. Solving them yields the nonequilibrium dynamics and enables the computation of time-resolved observables.

A suitable computational strategy must interface with electronic-structure codes to import essential input data, such as electronic eigenvalues, normal-mode frequencies and eigenvectors, and \textit{ab initio} matrix elements required for the collision integrals and self-energy corrections. Additionally, it must handle matrix-element manipulation, support flexible momentum-grid choices, and be optimized for time integration, parallelism, and scalability.

\begin{figure*}
\includegraphics[width=0.95\textwidth]{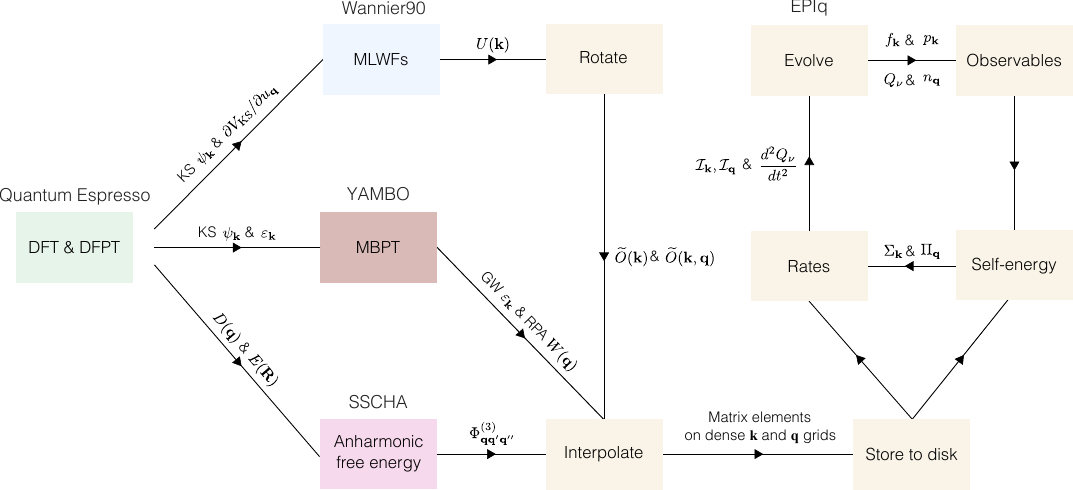}
\caption{\textbf{The workflow comprises four main stages.} (i) \textit{Ab initio} DFT and DFPT calculations on coarse $\vb{k}$ and $\vb{q}$ grids. These yield Kohn-Sham orbitals and eigenvalues (from DFT), along with the dynamical matrix, deformation potential, and third-order anharmonic matrix (from DFPT). (ii) Maximally localized Wannier functions (MLWF) construction, GW correction within many-body perturbation theory (MBPT), and (optional) anharmonic SSCHA calculation. Kohn-Sham orbitals are employed to compute MLWFs, static screening, and GW eigenvalues, while harmonic dynamical-matrix and total-energy calculations are used to estimate the nonperturbative anharmonic matrix and SSCHA frequencies. (iii) Preparation of matrix elements on dense $\vb{k}$ and $\vb{q}$ grids. MLWF matrices and phonon eigenvalues are employed to transform the matrix elements into the optimally smooth subspace and normal-mode basis, followed by interpolation onto dense grids. (iv) Explicit time integration of the EOMs. The time derivatives of occupations, polarization, atomic displacements, and momenta are evaluated through the EOMs (see Fig.~\ref{fig:EOM}). The dynamical variables are evolved and observables computed at each step. The cycle continues until quasi-equilibrium is achieved. The steps (iii) and (iv), represented by light yellow blocks in the workflow, are all performed within the \textsc{epi}q code}
\label{fig:workflow}
\end{figure*}

To this end, we implement the approach within the \textsc{epi}q suite~\cite{Marini2024}. The computational workflow is sketched in Fig.~\ref{fig:workflow}. In the following, we outline the implementation, focusing on its interface with electronic-structure codes, matrix-element handling, EOMs integration, and performance optimization.

\subsection{Interface with \textit{ab initio} codes}

The \textsc{epi}q suite, and thus this implementation, serves as a post-processing tool for \textit{ab initio} DFT, DFPT, GW, and anharmonic calculations. It is natively interfaced with \textsc{Quantum ESPRESSO} \cite{Giannozzi2009,Giannozzi2017}, which provides KS eigenvalues $\varepsilon_{\vb{k}n}^{\text{KS}}$ and orbitals $\psi_{\vb{k}n}$ for $N_b$ bands. The orbitals are employed to compute the Hamiltonian operator, yielding the KS band structure and velocity operator, needed to construct the dipole matrix, and the oscillator-strength matrix, used to build the screened Coulomb interaction. These matrix elements are evaluated on uniform $\vb{k}$- and $\vb{q}$-point grids and read:

\begin{align}
\label{eq:hamiltonian_matrix}
{H}^{\text{KS}}_{mn}(\vb{k}) = & \ \mel{\psi_{\vb{k}m}}{\hat{H}_{\text{KS}}}{\psi_{\vb{k}n}}_V,
\\
\label{eq:velocity_operator}
\vb{v}_{mn}(\vb{k}) = & \ i\mel{\psi_{\vb{k}m}}{[\hat{H}_{\text{KS}},\hat{\vb{r}}]}{\psi_{\vb{k}n}}_V,
\\
\label{eq:oscillator_strength}
\rho_{mn}^{\vb{G}}(\vb{k,q}) = & \ \mel{\psi_{\vb{k+q}m}}{e^{i(\vb{q+G}) \cdot \hat{\vb{r}}}}{\psi_{\vb{k}n}}_{V},
\end{align}
where the integration is performed over the whole volume $V$.

Phonon-related quantities are computed via DFPT \cite{Baroni2001}. The dynamical matrix reads:
\begin{equation}
\label{eq:dynamical_matrix}
D_{a\alpha b\beta}(\vb{q}) = \sum_{I} \frac{e^{i\vb{q} \cdot \vb{R}_{I}}}{\sqrt{M_a M_b}} \frac{\partial^2 E(\vb{R})}{\partial R_{Ia\alpha} \partial R_{0b\beta}} \bigg|_{\vb{R}^0},
\end{equation}
where \( E(\vb{R}) \) is the total energy at clamped nuclei. The deformation potential is given by
\begin{equation}
\label{eq:deformation_potential}
d^{a\alpha}_{mn}(\vb{k,q}) = \mel{\psi_{\vb{k+q}m}}{\sum_{I}e^{i\vb{q} \cdot \vb{R}_I} \frac{\partial \hat{V}_{\text{KS}}}{\partial Q_{Ia\alpha}}}{\psi_{\vb{k}n}}_V,
\end{equation}
where \( V_{\text{KS}} \) is the self-consistent Kohn-Sham potential. The third-order anharmonic matrix is
\begin{align}
\label{eq:thirdorder_matrix_cartesian}
\Phi^{(3)}_{a\alpha b\beta c\gamma}(\vb{q}_1, \vb{q}_2, \vb{q}_3) = \sum_{\vb{G},I,J} & \delta_{\vb{q}_1 + \vb{q}_2 + \vb{q}_3, \vb{G}} \frac{e^{i(\vb{q}_1 \cdot \vb{R}_I + \vb{q}_2 \cdot \vb{R}_J)}}{\sqrt{M_a M_b M_c}} \nonumber \\
& \times \frac{\partial^3 E(\vb{R})}{\partial R_{Ia\alpha} \partial R_{Jb\beta} \partial R_{0c\gamma}} \bigg|_{\vb{R}^0},
\end{align}
computed using either the \textsc{D3}q code \cite{Paulatto2013} or the nonperturbative SSCHA \cite{Monacelli2021}.

The same $\vb{k}$-grid, denoted by $N^{\text{pw}}_{k1}\times N^{\text{pw}}_{k2} \times N^{\text{pw}}_{k3}$,  is used for the eigenvalues, orbitals, velocity operator, oscillator strength, and deformation potential. The $\vb{q}$-grid, denoted $N^{\text{ph}}_{q1}\times N^{\text{ph}}_{q2}\times N^{\text{ph}}_{q3}$, must be equal to or a subgrid of the $\vb{k}$-grid and must be consistent for oscillator strength, dynamical matrix, and deformation potential. The flexibility of different $\vb{k}$ and $\vb{q}$ grids stems from a newly developed interpolation scheme that avoids real-space transformations \cite{Volpato2025}. The anharmonic matrix can instead be computed on an independent $\vb{q}$-grid, $N^{\text{anh}}_{q1}\times N^{\text{anh}}_{q2}\times N^{\text{anh}}_{q3}$. 

Kohn-Sham eigenvalues and orbitals also enable the calculation of the static screened Coulomb interaction and GW quasiparticle corrections via many-body perturbation theory. We use \textsc{yambo} \cite{Marini2009,Sangalli2019} to compute the static RPA screened interaction:
\begin{equation}
\label{eq:static_screened_RPA}
W_{\vb{GG'}}(\vb{q}) = (\delta_{\vb{GG'}} - v_{\vb{G}}(\vb{q})\chi^{0}_{\vb{GG'}}(\vb{q},0))^{-1}v_{\vb{G'}}(\vb{q}),
\end{equation}
where $v_{\vb{G}}(\vb{q})$ is the Fourier transform of the Coulomb interaction, see Eq.(S25) in the Supplementary Information, and $\chi^0_{\vb{GG'}}$ the irreducible density response. This is evaluated on a separate $\vb{q}$-grid $N_{q1}^{\text{w}}\times N_{q2}^{\text{w}}\times N_{q3}^{\text{w}}$, while GW-corrected energies are computed on the same $\vb{k}$-grid as the Kohn-Sham values, yielding a $\vb{k}$-dependent scissor operator:
\begin{equation}
\label{eq:scissor_operator}
    \Delta_{\vb{k}n} = \varepsilon_{\vb{k}n}^{\text{GW}} - \varepsilon_{\vb{k}n}^{\text{KS}}.
\end{equation}

If the electronic wavefunctions are kept fixed in the GW calculation, i.e., we stick to the G$_0$W$_0$ approximation, we can obtain the GW Hamiltonian from a scissor operator acting on the matrix in Eq.~\eqref{eq:hamiltonian_matrix}.

That is, we define the GW Hamiltonian as
\begin{equation}
    H^{\text{GW}}_{mn}(\vb{k}) = H_{mn}^{\text{KS}}(\vb{k}) + \delta_{mn} \Delta_{\vb{k}n} = \delta_{mn} \varepsilon^{\text{GW}}_{\vb{k}n}.
\end{equation}
Kohn-Sham orbitals are employed to construct maximally localized Wannier functions (MLWFs) \cite{Marzari2012} via the transformation
\begin{equation}
\label{eq:wannier_functions}
w_{\vb{R}n}(\vb{r}) = \frac{1}{\sqrt{N}} \sum_{\vb{k},m} e^{-i\vb{k} \cdot \vb{R}} U_{mn}(\vb{k}) \psi_{\vb{k}m}(\vb{r}),
\end{equation}
where $U_{mn}(\vb{k})$ are unitary matrices. The Wannier functions are thus defined up to a unitary transformation. This freedom, known as \textit{gauge freedom}, affects matrix elements such as Eqs.~\eqref{eq:hamiltonian_matrix}--\eqref{eq:oscillator_strength} and \eqref{eq:deformation_potential}. Consistent use of $\psi_{\vb{k}n}$ and $U_{mn}(\vb{k})$, i.e., gauge fixing, is therefore essential when combining these quantities \cite{Marini_Calandra_Cudazzo2024}.

To fix the gauge, we adopt the maximal-localization criterion, minimizing the spread functional \cite{Marzari1997}
\begin{equation}
\Omega = \sum_{n} \left[\mel{w_{\vb{0}n}}{\hat{\vb{r}}^2}{w_{\vb{0}n}} - \mel{w_{\vb{0}n}}{\hat{\vb{r}}}{w_{\vb{0}n}}^2\right].
\end{equation}
This procedure yields $N_w \le N_b$ Wannier functions from the KS bands, with disentanglement applied when necessary \cite{Souza2001}. We use \textsc{wannier90} \cite{Pizzi2020}, fully interfaced with \textsc{epi}q, to determine the $U_{mn}(\vb{k})$ matrices on the coarse $\vb{k}$-grid.

\subsection{Preparation of matrix elements and equilibrium conditions}

The \textit{ab initio} quantities are read and preprocessed before time-propagation. This includes rotation and interpolation of matrix elements, and initialization of equilibrium-state conditions.

The first step is the rotation of matrix elements appearing in Eqs.~\eqref{eq:hamiltonian_matrix}--\eqref{eq:oscillator_strength} and \eqref{eq:deformation_potential} using the unitary matrices $U_{mn}(\vb{k})$ obtained from the MLWFs. This optimally smooth subspace (OSS) transformation ensures smoothness in reciprocal space, essential for accurate interpolation. Gauge-independent quantities, such as the dynamical matrix, third-order anharmonic matrix, and statically screened Coulomb interaction, do not require an OSS transformation. This transformation reads:
\begin{gather}
    \widetilde{\mathcal{O}}_{mn}(\vb{k,q}) = \sum_{n',m'} U^*_{mm'}(\vb{k}) \mathcal{O}_{m'n'}(\vb{k,q}) U_{n'n}(\vb{k+q}). \label{eq:OSS_transf2}
\end{gather}
Here, $\mathcal{O}$ and $\widetilde{\mathcal{O}}$ denote the matrix elements in the original and OSS gauges. Momentum-diagonal ($\vb{q} = 0$) quantities include $H^{\text{KS}}_{mn}(\vb{k})$, $H^{\text{GW}}_{mn}(\vb{k})$, and $\vb{v}_{mn}(\vb{k})$, while momentum-nondiagonal ones include $\rho_{mn}^{\vb{G}}(\vb{k,q})$ and $d^{a\alpha}_{mn}(\vb{k,q})$.

After rotation, the matrix elements are interpolated onto dense uniform $\vb{k}$- and $\vb{q}$-grids with $N_1 \times N_2 \times N_3$ points; the total number of points being $N$. Quantities are interpolated by using the spline-based scheme of Ref.~\cite{Volpato2025}. On the dense grid, unitary matrices are obtained by diagonalizing the interpolated Hamiltonian and are then used to rotate gauge-dependent quantities back to the original gauge via the inverse OSS transformation. Additional manipulations yield all matrix elements required for EOMs propagation.

The equilibrium quasiparticle energies are evaluated as
\begin{equation}
    \varepsilon_{\vb{k}n}^{0} = \varepsilon_{\vb{k}n}^{\text{GW}} + \Sigma^{\text{FM}}_{\vb{k}n}(0),
\end{equation}
where $\Sigma^{\text{FM}}$ is the Fan-Migdal self-energy from Eq.(S29) in the Supplementary Information. Since $\Sigma^{\text{FM}}$ depends on $\varepsilon_{\vb{k}n}^{0}$, this equation is solved self-consistently.

Phonon frequencies and eigenvectors ($N_m = 3N_\text{at}$) are obtained by diagonalizing the interpolated dynamical matrix:
\begin{equation}
\label{eq:diagonalization_dynamical_matrix}
    \sum_{b,\beta} D_{a\alpha b\beta}(\vb{q}){e}^{b\beta}_{\vb{q}\nu} = \big(\omega_{\vb{q}\nu}^0\big)^2{e}^{a\alpha}_{\vb{q}\nu}.
\end{equation}
The dipole matrix is computed from the velocity operator as
\begin{equation}
    \mathbfcal{D}_{mn}(\vb{k}) = \dfrac{\vb{v}_{mn}(\vb{k})}{\varepsilon_{\vb{k}m}^{\text{KS}} - \varepsilon_{\vb{k}n}^{\text{KS}}}.
\end{equation}
The statically screened Coulomb interaction is written in terms of oscillator strength and dielectric matrix as
\begin{equation}
\label{eq:static_screened_interaction_with_rho}
\begin{aligned}
W_{\vb{kk'k-qk'+q}}^{nn'mm'}(t) = \sum_{{\vb{G,G'}}} & W_{\vb{GG'}}(\vb{q},t) \times \\
    & \rho_{n'm'}^{\vb{G'}}(\vb{k',q})\left(\rho^{\vb{G}}_{mn}(\vb{k-q},\vb{q})\right)^*,
\end{aligned}
\end{equation}
where the time-dependent Coulomb interaction matrix is built as explained in Supplementary Section 4.

The electron-phonon matrix elements in the phonon normal-mode basis are given by
\begin{equation}
\label{eq:elph_matrix_normal_modes}
    g_{mn}^{\nu}(\vb{k,q}) = \sum_{a\alpha} {d}^{a\alpha}_{mn}(\vb{k,q}) \dfrac{{e}^{a\alpha}_{\vb{q}\nu}}{\sqrt{2M_a\omega^0_{\vb{q}\nu}}}.
\end{equation}
The anharmonic interaction in the same basis reads
\begin{align}
\label{eq:anharmonic_matrix_normal_modes}
\Phi^{(3)}_{\nu_1\nu_2\nu_3}(\vb{q}_1,\vb{q}_2,\vb{q}_3) = \sum_{\substack{a,b,c \\ \alpha,\beta,\gamma}} & \dfrac{\Phi^{(3)}_{a\alpha b\beta c \gamma}(\vb{q}_1,\vb{q}_2,\vb{q}_3)}{\sqrt{\omega^0_{\vb{q}_1\nu_1}\omega^0_{\vb{q}_2\nu_2}\omega^0_{\vb{q}_3\nu_3}}} \nonumber \\
& \times \dfrac{{e}^{a\alpha}_{\vb{q}_1\nu_1}{e}^{b\beta}_{\vb{q}_2\nu_2}{e}^{c\gamma}_{\vb{q}_3\nu_3}}{\sqrt{8M_aM_bM_c}}.
\end{align}
Finally, equilibrium electron and phonon distributions, before the action of the perturbation ($t < 0$), are initialized to Fermi-Dirac and Bose-Einstein distributions at temperature $T$, while polarization, atomic displacements, and momenta are set to zero. With all ingredients in place, time propagation of the EOMs can begin.

\subsection{Time evolution of the nonequilibrium state}
\label{subsec:Time evolution of the nonequilibrium state}

The evolution of the nonequilibrium state is governed by the time evolution of the system’s dynamical variables, including electron and phonon occupations, microscopic polarizations, and atomic positions and momenta. The simulation proceeds through three main steps:

\begin{enumerate}[(i)]
    \item At time $t = t_i$, self-energy corrections are computed (see Supplementary Section 3). These are used to evaluate the renormalized electron and phonon energies and Rabi-frequency corrections (Eqs.~\eqref{eq:renorm_eig2},~\eqref{eq:renorm_Rabi_freq},~\eqref{eq:renormalized_phonon_frequencies2}). The rates, i.e., the time derivatives of the dynamical variables, are then calculated (Eqs.~\eqref{eq:EOM_el_polarizations}-\eqref{eq:EOM_el_occupations},~\eqref{eq:EOM_ph_occupations}, and \eqref{eq:EOM_atomic_displacement}).
    
    \item The rates are used to update the dynamical variables at $t_{i+1} = t_i + \Delta t$ via an explicit time-integration algorithm.
    
    \item The updated variables at $t_{i+1}$ are used to compute and store selected physical observables.
\end{enumerate}

These steps are repeated until the system reaches a quasi-equilibrium state.

Step (i) involves evaluating momentum averages over the first Brillouin zone. However, the presence of $\delta$ functions and the Cauchy principal part $\mathcal{P}$ makes these expressions well defined only in the continuum limit. To regularize them on finite grids, we introduce the regularized forms:
\begin{equation}
\label{eq:regularized_delta_principal_part}
    \delta(x) = \dfrac{1}{\sqrt{2\pi\sigma^2}} e^{-\dfrac{x^2}{2\sigma^2}}, \quad \mathcal{P}\left(\dfrac{1}{x}\right) = \dfrac{x}{x^2 + \sigma^2},
\end{equation}
with $\sigma$ a small positive smearing parameter.

In step (ii), the dynamical variables are propagated to $t_{i+1}$ using their values and derivatives at $t_i$. We employ explicit Runge-Kutta (RK) schemes, including second-order (RK2), fourth-order (RK4), and the adaptive fifth-fourth-order Dormand-Prince method (RKDP54)~\cite{Dormand1980}. Unlike RK2 and RK4, RKDP54 adjusts the timestep $\Delta t_i$ dynamically to maintain the local error below a specified threshold, estimated from the difference between fourth- and fifth-order solutions. Implementation details are provided in Supplementary Section 8 (see also Table S1).

In step (iii), relevant physical observables are computed. For example, the macroscopic polarization is
\begin{equation}
\label{eq:macroscopic_polarization}
    \vb{P}(t) = \dfrac{2}{N}\Re \left[\sum_{\vb{k},n,m} p_{\vb{k}nm}(t)(\mathbfcal{D}_{nm}(\vb{k}))^*\right],
\end{equation}
which captures linear and nonlinear optical transitions at zero momentum. The number of photoexcited electrons is
\begin{equation}
    n_{e}(t) = \dfrac{1}{N}\sum_{\vb{k},c} f_{\vb{k}c}(t),
\end{equation}
where $c$ denotes conduction-band indices. As the total number of electrons $n$ is conserved, the number of valence electrons is $n_v(t) = n - n_e(t)$, and the number of holes is $n_h(t) = n_e(t)$.

The nonequilibrium momentum-resolved phonon temperature is defined as
\begin{equation}
\label{eq:ph_mode_temperature}
    T_{\vb{q}\nu}(t) = \dfrac{{\omega}_{\vb{q}\nu}(t)}{\ln[1/n_{\vb{q}\nu}(t) + 1]},
\end{equation}
while the average lattice temperature is estimated as
\begin{equation}
\label{eq:average_lattice_temperature}
    T_{ph}(t) = \dfrac{1}{N N_{m}}\sum_{\vb{q},\nu} T_{\vb{q}\nu}(t),
\end{equation}
$N_m$ being the number of phonon modes. The average lattice temperature provides insight into photoinduced lattice heating.

The forces and displacements of atom $a$ are given by
\begin{align}
    \vb{F}_{a}(t) &= \dfrac{1}{\sqrt{N}}\sum_{\nu} F_{\nu}(t) \vb{e}^{a}_{\vb{0}\nu} \sqrt{\dfrac{M_a}{\omega_{\vb{0}\nu}}}, \label{eq:td_force_real_space}\\
    \vb{Q}_{a}(t) &= \dfrac{1}{\sqrt{N}}\sum_{\nu} Q_{\nu}(t) \dfrac{\vb{e}_{\vb{0}\nu}^{a}}{\sqrt{\omega_{\vb{0}\nu}M_a}},
\end{align}
capturing photoinduced structural distortions.

Additional observables, such as quasiparticle energies, linewidths, and occupations, can also be extracted from the time-dependent dynamical variables. In the next following, the computational optimizations and performances of the implementation are presented.

\subsection{Optimization and computational performance}

The explicit time integration of the EOMs can become significantly computationally demanding. The most time-consuming part of the algorithm is the evaluation of the collision integrals, which involve reciprocal-space integrations over the BZ and summations over Wannier bands and phonon branches.

Each carrier-carrier collision integral requires two BZ integrations and three summations over Wannier bands. This must be performed $N \times N_{w}$ times to obtain the collision rates for every electronic state. Consequently, each term in the summation must be evaluated and accumulated a total of $N^3 \times N_{w}^4$ times.

The carrier-phonon collision integrals, by contrast, involve a single BZ integration and two summations over Wannier bands (Wannier bands and phonon branches). Evaluating these rates requires summing $2N^2 \times N_{w}^2 \times N_{m}$ contributions.

Finally, phonon-phonon collision integrals involve two BZ integrations and two summations over phonon modes. However, due to the presence of a Kronecker $\delta$ in the definition of the third-order anharmonic matrix (Eq.~\eqref{eq:thirdorder_matrix_cartesian}), the number of BZ integrations is effectively reduced to one. As a result, computing these rates involves summing $N^2 \times N_{m}^3$ terms.

The computational cost can become substantial when using fine reciprocal-space grids or a large number of Wannier bands or phonon modes. Therefore, it is essential to adopt strategies that reduce the number of floating-point operations, thereby minimizing computational time and improving the efficiency of time propagation.

The most immediate optimization is to discard all terms where the Dirac $\delta$ function is violated significantly. Specifically, we ignore contributions for which the modulus of the Dirac delta argument exceeds $3\sigma$.

Further efficiency is obtained by neglecting contributions where the scattering amplitude, set by the squared modulus of the static Coulomb screening, electron-phonon interaction, or third-order anharmonic matrix elements, is below a chosen threshold. Another optimization leverages symmetries in the definition of the collision integrals.

For example, the computational cost of carrier-carrier collision integrals can be reduced (assuming time reversal symmetry) by exploiting their invariance under the transformation $\vb{k} \leftrightarrow \vb{k}'$, $n \leftrightarrow n'$, $m \leftrightarrow m'$, and $\vb{q} \to -\vb{q}$. This symmetry allows us to limit the summation over $n'$ to $n' \leq n$ and restrict the indices of $\vb{k}'$ to those less than or equal to the corresponding indices of $\vb{k}$.

Similar symmetries hold for carrier- and phonon-phonon collision integrals. However, since these are significantly less expensive than carrier-carrier contributions, we do not apply these optimizations to them.

Crystalline symmetries can also be used to restrict computations to the irreducible wedge of the BZ. However, the presence of an external electric field may break some spatial symmetries, preventing their application to collision-integral evaluations. For this reason, we do not exploit spatial symmetries and instead compute rates explicitly for all $\vb{k}$ and $\vb{q}$ points.

To enhance performance, we parallelize our implementation using the Message Passing Interface (MPI), distributing both computational workload and memory across $n$ MPI tasks. A natural choice is to distribute over reciprocal-space points in the BZ. By analyzing the structure of the matrix elements defining the scattering amplitudes in Eqs.~\eqref{eq:static_screened_interaction_with_rho}-\eqref{eq:anharmonic_matrix_normal_modes}, we find that distributing memory over $\vb{q}$ points is more efficient than that on $\vb{k}$ points, as it reduces communication overhead. As a result, matrix elements corresponding to electron-phonon interactions, oscillator strengths, and anharmonic matrices, interpolated on a double $\vb{k}$-$\vb{q}$ grid of dimension $N \times N$, are distributed among the $n$ MPI tasks in $\vb{q}$-chunks. Thus, each process handles a fraction of matrix elements on a double-grid of dimension $N \times N/n$. This ensures that the total memory required to store matrix elements remains independent of the number of MPI tasks, improving scalability.

\begin{figure}
\centering
\includegraphics[width=0.95\linewidth]{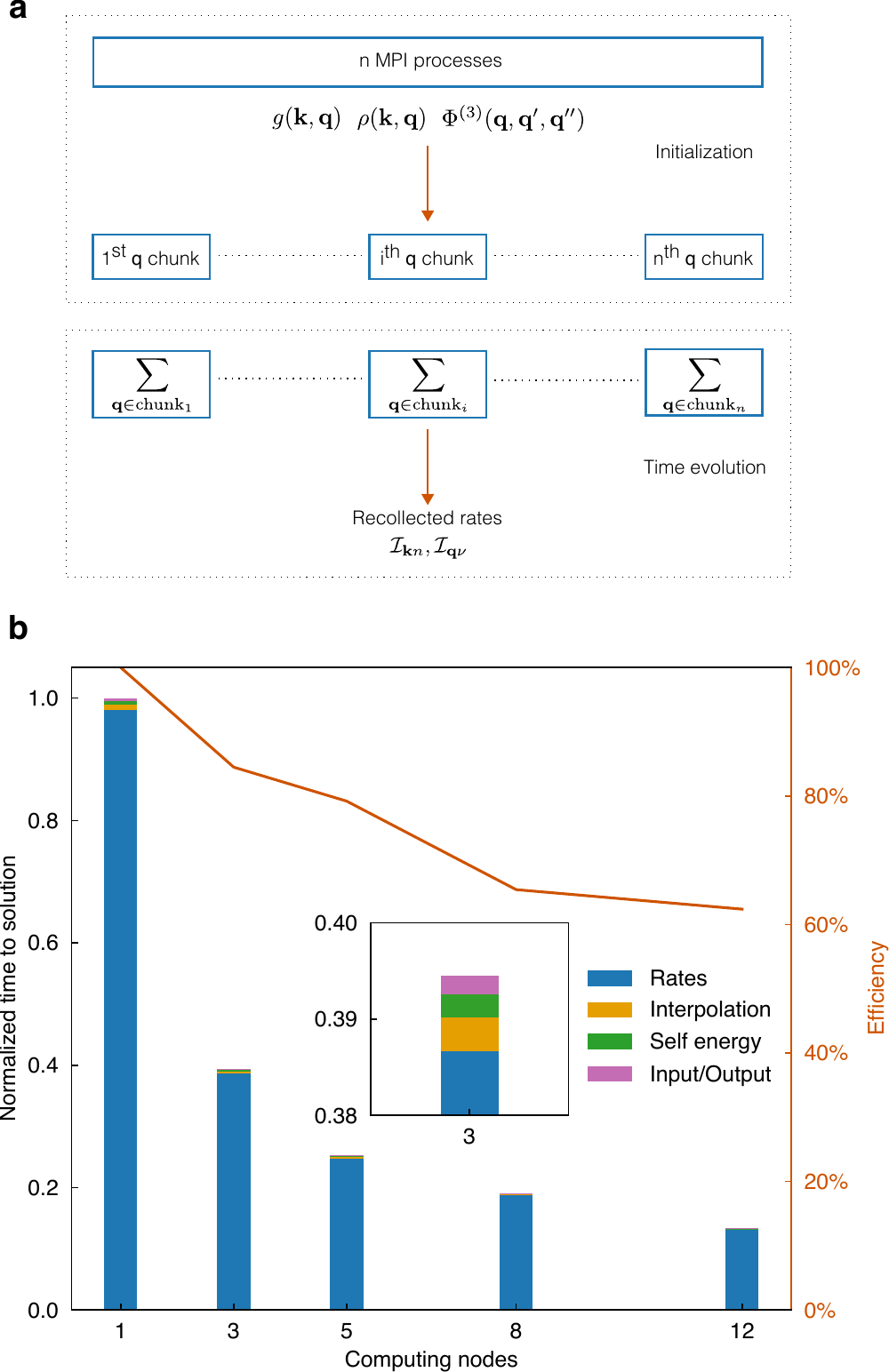}
\caption{\textbf{Parallelization and efficiency.} (a) Parallelization strategy for matrix-element storage and collision-integral computation. (b) Normalized time to solution against the number of computing nodes for the propagation of the EOMs in monolayer MoS$_2$.}
\label{fig:parallel_efficiency}
\end{figure}

For rate calculations, all summations over $\vb{q}$ are performed in chunks by each task, with the final result obtained via a synchronous MPI reduction. Conversely, summations over $\vb{k}$, which yield a $\vb{q}$-dependent quantity, are computed by each task and collected using a synchronous MPI gather. The current parallelization strategy is schematically depicted in Fig.~\ref{fig:parallel_efficiency}(a). Systematic improvements are possible, for instance, through multiple MPI layers or a hybrid MPI+OpenMP approach. These modifications would be straightforward to implement and represent a promising future development.

To test scalability, we perform strong-scaling tests by propagating the EOMs in Fig.~\ref{fig:EOM} for 90 fs in monolayer MoS$_2$; computational details are given later. This is done on the LEONARDO supercomputer at CINECA (Italy), where each compute node features two Intel Xeon Platinum 8480+ processors with 56 physical cores each. To quantify the parallelization efficiency, we plot the normalized time to solution, i.e., the wall time divided by the wall time of one computing node, as a function of the number of nodes employed. The results, shown in Fig.~\ref{fig:parallel_efficiency}(b), confirm that the dominant computational workload arises from the evaluation of collision rates, particularly those corresponding to carrier-carrier scattering. The code demonstrates good scalability, achieving a parallelization efficiency exceeding 60\% with 12 nodes.

Next, we analyze how the choice of EOMs integration algorithm affects computational workload and precision. As discussed above, the primary cost arises from rate evaluations. Thus, it is crucial to develop strategies that allow for large timesteps while maintaining high accuracy. In this context, adaptive-timestep algorithms offer a favorable trade-off between efficiency and precision. To validate this, we compare the performance of the three RK schemes we implemented (see Supplementary Section 8). These schemes are tested by propagating the EOMs for 90 fs in monolayer MoS$_2$.

\begin{figure} 
\centering 
\includegraphics[width=0.95\linewidth]{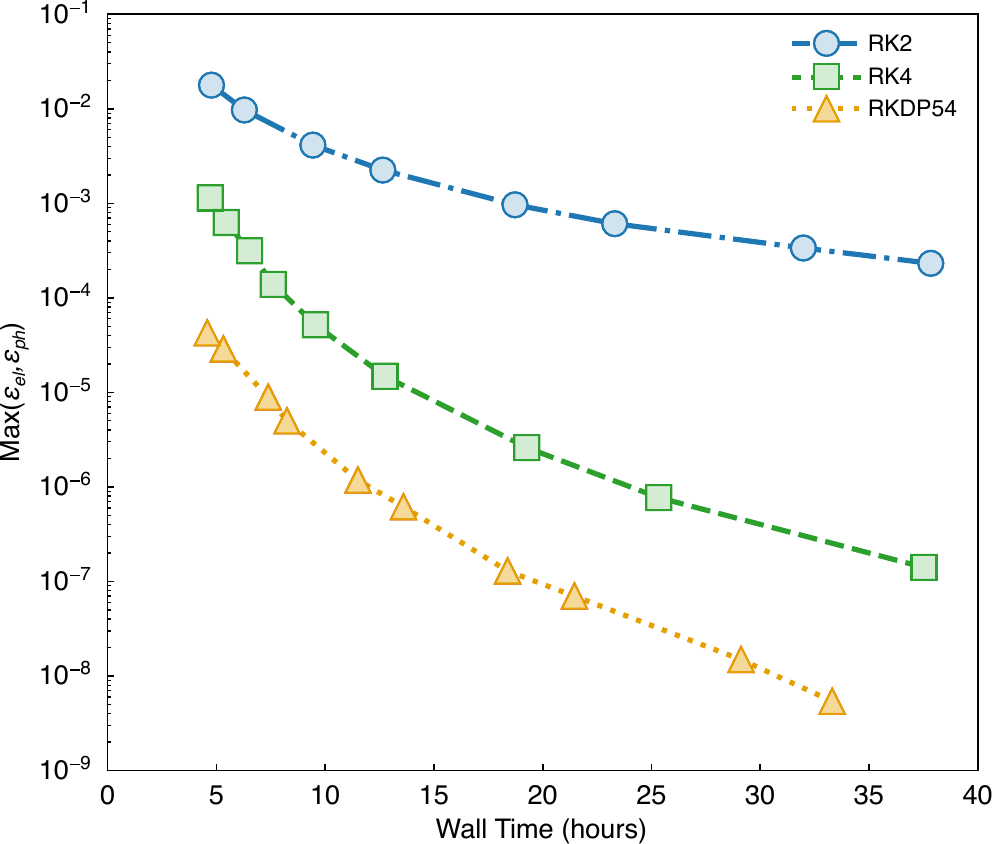} 
\caption{\textbf{Maximum error in the final electron and phonon occupation numbers as a function of computational time for three integration algorithms: RK2, RK4, and RKDP54.} The results show that the adaptive-stepsize RKDP54 algorithm substantially reduces computational time while maintaining high accuracy.} 
\label{fig:error_vs_cputime} 
\end{figure}

We evaluate accuracy by comparing results to a reference set of occupations at 90 fs, $f^{r}_{\vb{k}n}$ and $n^{r}_{\vb{q}\nu}$, obtained using the RKDP54 algorithm with $\epsilon_{\text{tol}} = 10^{-11}$; see Supplementary Section 8. The errors on the occupations at 90 fs, $f_{\vb{k}n}$ and $n_{\vb{q}\nu}$, obtained with a given method, are then
\begin{equation} 
\varepsilon_{el} = \max_{\vb{k},n} |f_{\vb{k}n} - f^{r}_{\vb{k}n}|, \quad \varepsilon_{ph} = \max_{\vb{q},\nu} |n_{\vb{q}\nu} - n^{r}_{\vb{q}\nu}|. 
\end{equation}
The overall error is defined as $\max(\varepsilon_{el}, \varepsilon_{ph})$. In Fig.~\ref{fig:error_vs_cputime}, we show the error corresponding to each integration method as a function of wall time. The adaptive-stepsize RKDP54 method significantly outperforms both RK2 and RK4, ensuring higher accuracy at lower computational cost.

Finally, we note that the stepsize, whether fixed or adaptive, is primarily determined by the equation for electronic polarization (see Eq.~\eqref{eq:EOM_el_polarizations}). The polarization exhibits a rapidly oscillating component due to the term proportional to the difference between electronic eigenvalues. Moreover, in the presence of an oscillating electric field, the Rabi term oscillates at frequencies on the order of fractions of a petahertz. To resolve this fast oscillatory behavior, the timestep must be sufficiently small to keep the propagation error below a given threshold. Conversely, in the absence of an oscillating polarization, the timestep would be determined by the much slower timescale associated with the collision integrals. As a result, when an oscillating polarization is present, the collision integrals must be evaluated far more frequently than would otherwise be required, substantially increasing computational overhead.

This represents a typical example of a problem involving a nonstiff (slow) component, determined by the scattering integrals, and a stiff (fast) component, arising from the rapidly oscillating polarization. Methods designed to disentangle these two timescales are commonly referred to as multirate infinitesimal (MRI) methods \cite{Schlegel2009, Schlegel2012}. In such frameworks, the fast component of the equation is integrated with a small timestep, while the slow component is treated through an effective equation using a much larger timestep.

This approach significantly reduces the number of collision-integral evaluations while preserving an accurate evolution of the fast oscillating polarization. MRI methods have already been applied to real-time electron and phonon dynamics in a semiclassical Boltzmann framework \cite{Yao2025}, demonstrating clear advantages over standard RK methods. Implementing MRI methods thus represents a promising avenue for future developments.

\subsection{Technical details}
\label{subsec:Technical details}

DFT and DFPT calculations are performed with \textsc{Quantum ESPRESSO} \cite{Giannozzi2009, Giannozzi2017}, using the local-density approximation (LDA) \cite{Perdew1981} for the exchange-correlation functional. A kinetic-energy cutoff of 80~Ry is adopted for the wave functions, and optimized norm-conserving Vanderbilt (ONCV) pseudopotentials \cite{Hamann2013} are employed. Spin-orbit coupling (SOC) is included for MoS$_2$. A vacuum spacing of 20~\AA\ prevents spurious out-of-plane interactions. The in-plane lattice constants are fixed to 2.496~\AA\ for h-BN and 3.16~\AA\ for MoS$_2$.

Electronic densities are computed on $\Gamma$-centered Monkhorst-Pack \cite{Monkhorst1976} $\vb{k}$-grids of $18 \times 18 \times 1$ for h-BN and $14 \times 14 \times 1$ for MoS$_2$. Force constants, electron-phonon couplings, and oscillator-strength matrices are obtained on $\Gamma$-centered $\vb{k}$- and $\vb{q}$-grids of $12 \times 12 \times 1$ and $8 \times 8 \times 1$ for h-BN and MoS$_2$, respectively. The variation of the KS potential and density is evaluated on a shifted $36 \times 36 \times 1$ $\vb{k}$-grid. The long-range polar contribution to the deformation potential, arising from the macroscopic field of Born effective charges, is removed following Refs.~\cite{Sohier2016, Sio2022,Sjakste_2015}.

Maximally localized Wannier functions are computed with \textsc{Wannier90} \cite{Pizzi2020}, yielding 9 (22) Wannierized bands from 11 (34) KS bands for h-BN (MoS$_2$) via disentanglement \cite{Marzari2012}.

Third-order anharmonic force constants are obtained with the \textsc{D3}q code \cite{Paulatto2013} on $6 \times 6 \times 1$ $\vb{q}$-grids, using a shifted $36 \times 36 \times 1$ $\vb{k}$-grid for the linear-response calculations.

Equilibrium quasiparticle eigenvalues are computed at the $G_0W_0$ level using \textsc{Yambo} \cite{Marini2009, Sangalli2019} within the plasmon-pole approximation (PPA) on the same $\vb{k}$-grid employed for Wannierization. The density-density response includes 500 states with a cutoff of 18~Ry. The correlation part of the GW self-energy is computed using 200 states and a $24 \times 24 \times 1$ $\vb{k}$-grid, combined with stochastic integration of the screened interaction \cite{Guandalini2023}.

In real-time simulations, the equilibrium Coulomb interaction is described by a model dielectric function for two-dimensional semiconductors \cite{Trolle2017}, see Supplementary Section 9. The time-dependent part $\Delta \varepsilon (\vb{q},t)$ is then calculated at each timestep as explained in Supplementary Section 4. The long-wavelength divergence is handled via the random-integration method~\cite{Marini2009}. We neglect local-field effects on the photoinduced renormalization of the dielectric function and Hartree self-energy.

For the real-time dynamics, $\Gamma$-centered $\vb{k}$- and $\vb{q}$-grids of $64 \times 64 \times 1$ and $36 \times 36 \times 1$ are used for h-BN and MoS$_2$, respectively. A Gaussian smearing of 20~meV, 4~meV, and 2~\text{meV} is employed for carrier-carrier, carrier-phonon, and phonon-phonon scattering integrals. Further details about convergence are provided in Supplementary Section 10. For the rate calculations, we restrict the basis to the two topmost valence bands and the two lowest conduction bands for both materials. This implies that we neglect nonradiative recombination processes (e.g., Auger) involving high-energy states, as well as radiative recombination; both will be addressed in future work.

The real-time dynamics is obtained by propagating the EOMs in Fig.~\ref{fig:EOM} using the explicit RKDP54 method introduced in before and detailed in Supplementary Section 8. The error tolerance is set to $\varepsilon_{tol} = 10^{-8}$.

In \nameref{sec3}, we apply the theoretical and computational framework to the coupled electron-phonon dynamics of photoexcited two-dimensional semiconductors. Its capabilities are showcased via case studies on monolayer molybdenum disulfide (MoS$_2$) and hexagonal boron nitride (h-BN) following above-gap excitation.

\section{Results}
\label{sec3}

\subsection{Real-time dynamics in monolayer MoS\texorpdfstring{$_2$}{2\texttwosuperior}}
\label{subsec:Real-time dynamics in monolayer MoS2}

We analyze the real-time dynamics of electrons and phonons in single-layer MoS$_2$ following an above-gap optical excitation. We focus on how different interactions govern the nonequilibrium evolution of the system and show that an accurate description of light-driven materials requires including all of them. Unless otherwise stated, we neglect phonon-frequency renormalization due to anharmonic effects and set the initial lattice temperature to 300 K.

The system's drive is a quasi-monochromatic electric-field pulse,
\begin{equation}
\label{eq:gaussian_pulse}
    \vb{E}(t)=\vb{E}_0 \exp\!\left[-\frac{(t-t_0)^2}{2\Delta t^2}\right]\sin(\omega t),
\end{equation}
where $t_0$ is the pulse center, $\Delta t$ the pulse width, and $\omega$ the field frequency. We model typical pump-probe conditions with a pump photon energy exceeding the bandgap ($\approx 2.4$ eV) and a duration of a few tens of femtoseconds. Specifically, we set $\hbar\omega=2.6$~eV and $\Delta t=5$~fs. The simulation starts at $t=0$~fs, with the pulse center at $t_0=3\Delta t=15$~fs.

We consider a linearly polarized field in the $xy$ plane with amplitude $E_0$. The field intensity uniquely determines the fluence $F$, defined as the electromagnetic energy per unit area. For the pulse in Eq.~\eqref{eq:gaussian_pulse}, the relation between $F$ and $E_0$ (see Supplementary Section 11) is
\begin{equation}
    F \simeq \frac{\sqrt{\pi}}{4}\,\varepsilon_0 c\,\Delta t\,E_0^2.
\end{equation}
With the above parameters, $F$ ranges from tens of $\mu$J/cm$^2$ to a few mJ/cm$^2$ as $E_0$ varies from $10^8$ to $10^9$~V/m.

\begin{figure*}
    \centering
    \includegraphics[width=0.95\linewidth]{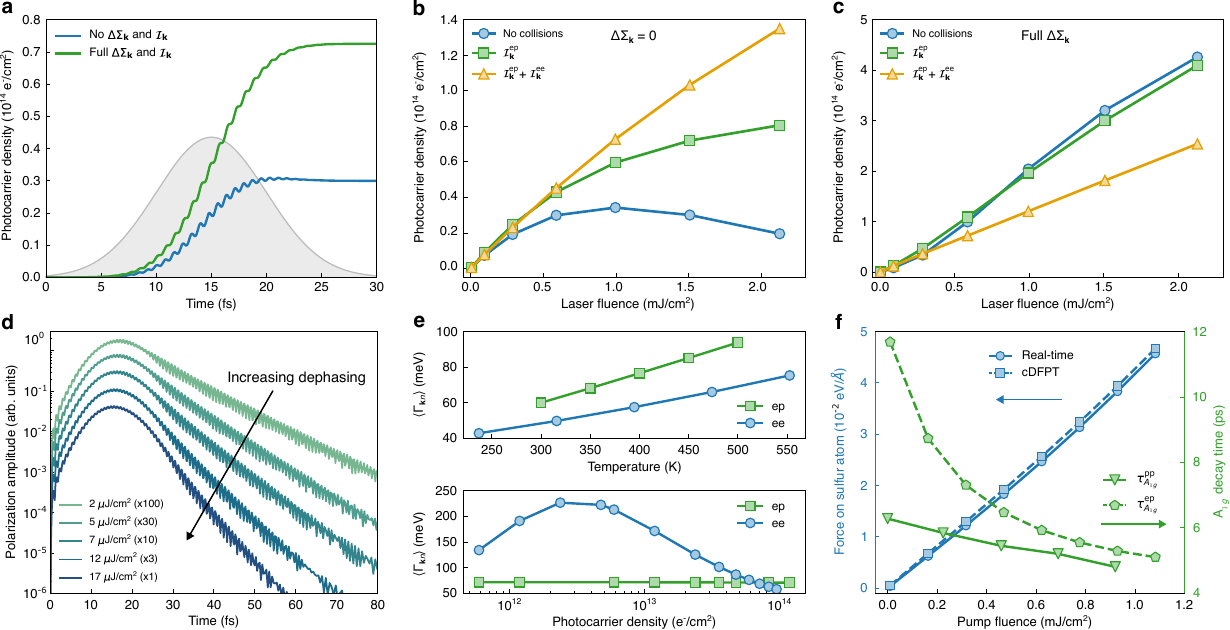}
    \caption{\textbf{MoS$_2$: role of quasiparticle renormalization and scattering in the ultrafast response.} (a) Time-dependent photocarrier density with (green) and without (blue) quasiparticle renormalization $\Delta\Sigma_{\mathbf{k}}$ and collision integrals $I_{\mathbf{k}}$. The gray curve is the Gaussian pulse envelope. (b,c) Photocarrier density vs fluence with (b) and without (c) quasiparticle renormalization. Symbols denote dynamics with no (blue circles), carrier-phonon only (green squares), and carrier-phonon + carrier-carrier (orange triangles) collision integrals. (d) Time-dependent macroscopic polarization amplitude vs fluence. The rescaling factor used for plotting purposes is reported in the legend. (e) Average electron-phonon (green squares) and electron-electron (blue circles) linewidth $\langle \Gamma_{\vb{k}n}\rangle $ vs temperature and photocarrier concentration. (f) Real-time force on S atoms (blue circles) compared with cDFPT (blue squares) vs fluence. Decay time of the coherent $A_{1g}$ amplitude due to electron-phonon (green pentagons) and phonon-phonon (green triangles) interactions vs fluence.}
    \label{fig:MoS2_fig1}
\end{figure*}

We begin by evaluating selected electronic observables from the propagation of the EOMs under different approximations. Specifically, we solve the dynamics with and without electronic quasiparticle renormalizations (Eqs.~\eqref{eq:renorm_eig2} and \eqref{eq:renorm_Rabi_freq}) and with carrier-phonon carrier-carrier collision integrals toggled on or off. In this initial analysis, we study the evolution up to 80 fs after irradiation and thus neglect the role of phonon-phonon collision integrals, as they contribute mostly to phonon and coherent dynamics at longer times.

Figure~\ref{fig:MoS2_fig1}(a) shows the time-resolved photocarrier density induced by the pulse at a fluence of $0.6$~mJ/cm$^2$. 
The gray-filled curve represents the Gaussian pulse envelope, the green one corresponds to the EOMs solution including both quasiparticle renormalizations and collision integrals, and the blue curve to the EOMs solution neglecting both. In the latter case, the number of photoexcited electrons is strongly underestimated.

This trend is clearer when we examine the laser-induced photocarrier density against fluence at 30 fs across various approximations to band structure renormalization and scattering. Figures~\ref{fig:MoS2_fig1}(b)-\ref{fig:MoS2_fig1}(c) display results without and with quasiparticle renormalizations, respectively. In each panel, the blue, green, and orange symbols denote no collisions, carrier-phonon only, and carrier-phonon + carrier-carrier collisions.

If quasiparticle renormalizations are neglected (Fig.~\ref{fig:MoS2_fig1}(b)), the interpretation is straightforward: adding scattering channels allows excited carriers to relax, hence reducing Pauli blocking. This frees states for additional carriers promoted by the field. As expected, the effect is most pronounced at large fluence. In the absence of scattering (blue circles), phase-space filling (Pauli blocking) quickly limits further excitation after the first few femtoseconds, and stimulated emission reduces the number of photocarriers, leading to a counterintuitive photocarrier density decrease with increasing fluence. 

Even with all scattering channels included, the carrier density exhibits a nonlinear increase at high fluence. In this regime, states are not freed rapidly enough to accommodate newly excited carriers, despite rapid carrier-carrier interactions, leading to saturation.

When quasiparticle renormalizations are included (Fig.~\ref{fig:MoS2_fig1}(c)), the picture changes substantially. Band structure renormalization, most notably the single-particle bandgap reduction, provides additional phase space for photoexcited carriers throughout the pulse, regardless of scattering channels. As a result, the no-collision curve becomes qualitatively similar to the carrier-phonon-only case, while both differ markedly from the full-scattering result. This separation originates from different polarization dephasing rates, see Eq.~\eqref{eq:EOM_el_polarizations}. Large dephasing drives a rapid decay of the polarization and hinders carrier promotion, see Eq.~\eqref{eq:EOM_el_occupations}. 

Specifically, in the carrier-phonon-only regime, the quasiparticle linewidths of the dipole-active states near the valence-band maximum and conduction-band minimum remain small, whereas carrier-carrier contributions are non-negligible. This explains the remarkable difference between the green and yellow symbols. Lastly, across the explored fluence range, we do not observe pronounced nonlinearities in the induced photocarrier density, underscoring that band structure renormalization counteracts phase-space filling and mitigates saturation effects.

We next analyze the real-time evolution of the macroscopic polarization following optical excitation at different fluences. We focus on the polarization amplitude,
\begin{equation}
    P_{A}(t)=\left(\frac{2}{N}\sum_{\vb{k},n,m}\big|\,p_{\vb{k}nm}(t)\,(\mathbfcal{D}_{nm}(\vb{k}))^*\,\big|^2\right)^{1/2},
\end{equation}
which differs from the macroscopic polarization in Eq.~\eqref{eq:macroscopic_polarization} as $P_A(t)$ is positive definite and encodes the microscopic polarization amplitude.

Figure~\ref{fig:MoS2_fig1}(d) shows $P_A(t)$ for six fluences between $2$ and $17~\mu$J/cm$^2$. The long-time dephasing rate increases with fluence, indicating an increase in the electronic linewidth for dipole-active states over this range. This trend agrees qualitatively with previous reports at comparable fluences~\cite{Perfetto2023, Mieck2000}. We stress that each curve is multiplied by a different arbitrary constant for visual clarity, see the legend. 

To disentangle electron-phonon and electron-electron contributions to polarization decay, we evaluate the average electronic linewidths $\langle \Gamma_{\vb{k}n} \rangle$ (Eqs.~\eqref{eq:elel_dephasing_rate} and \eqref{eq:elph_dephasing_rate}) within a 0.5~eV window around the valence-band maximum and conduction-band minimum. Linewidths are computed while varying the electronic and lattice temperature and the photocarrier density, using a Fermi-Dirac distribution with two distinct chemical potentials, see Eq.~\eqref{eq:fermi_distribution}.

In Fig.~\ref{fig:MoS2_fig1}(e), we report the electron-phonon (green squares) and electron-electron (blue circles) average linewidth versus temperature (top panel) and versus photocarrier density (bottom panel). We study the temperature dependence of the average electronic linewidth by fixing a photocarrier density of $2 \times 10^{13}$ e$^{-}/$cm$^2$. Both contributions grow monotonically with temperature in the explored range. 

This behavior originates from two different mechanisms. On one hand, the increase of the lattice temperature enhances carrier-phonon scattering channels due to an increase in the phonon occupations, see Eq.\eqref{eq:elph_dephasing_rate}. On the other hand, the increase of electronic temperature yields an amplification of the carrier-carrier scattering probability due to Pauli-blocking weakening, see Eq.\eqref{eq:elel_dephasing_rate}. 

\begin{figure*}
    \centering
    \includegraphics[width=0.95\linewidth]{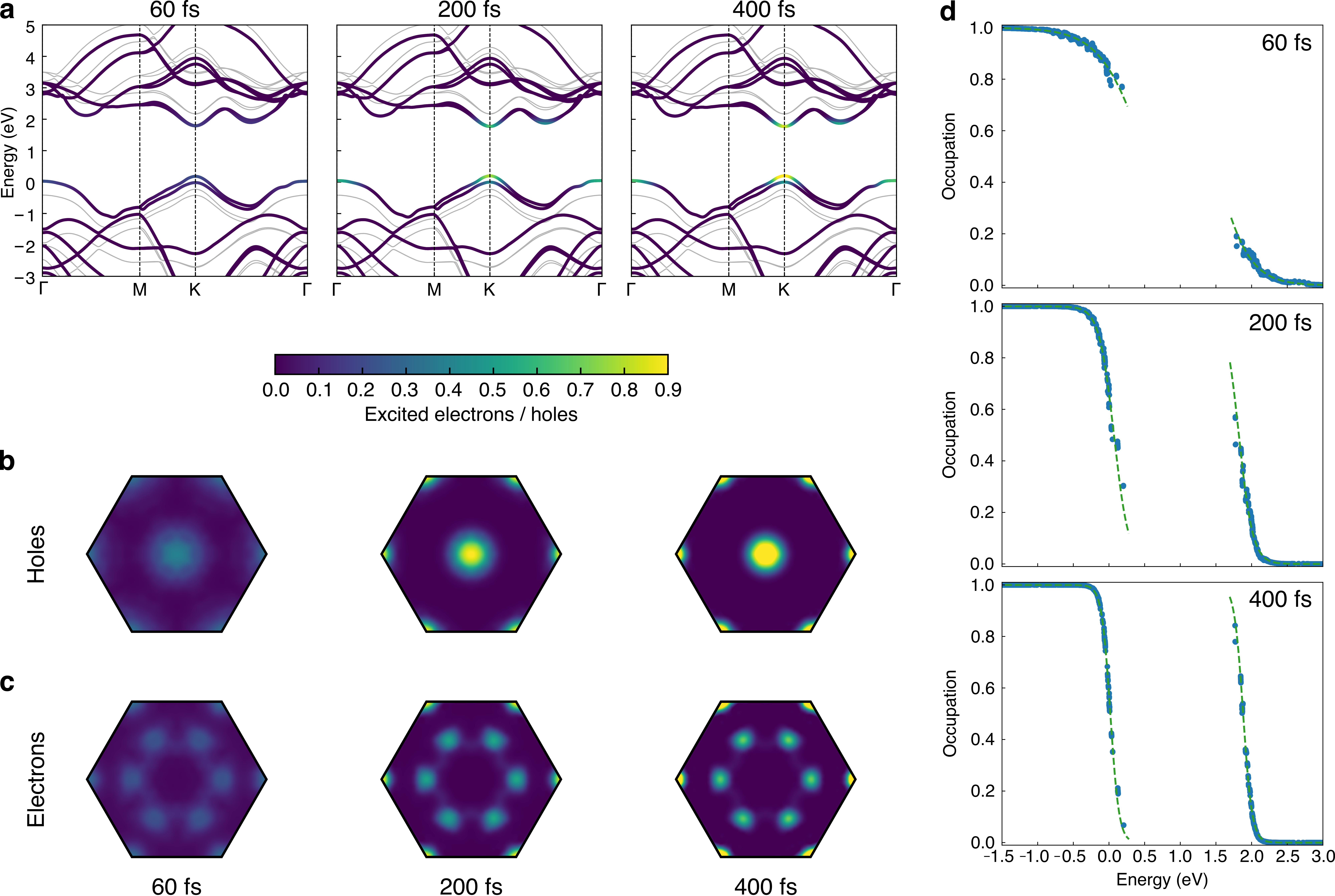}
    \caption{\textbf{Real-time evolution of electronic observables in monolayer MoS$_2$ after above-gap excitation.} (a) Band structure and occupations along a high-symmetry path at selected snapshots (60, 200, and 400 fs). The gray curves represent the equilibrium bands. (b,c) Momentum-resolved occupations of photoexcited electrons (b) and holes (c) at selected times. (d) Energy-resolved electronic occupations at selected times (blue circles) with fits to a two-chemical-potential Fermi-Dirac distribution (green dashed line).}
    \label{fig:MoS2_fig2}
\end{figure*}

To study the photocarrier density dependence of the average electronic linewidth, we instead fix the electronic and lattice temperature to 1000 and 300 K, respectively. The electron-electron linewidth exhibits a non-monotonic dependence on carrier density. Conversely, we find that the electron-phonon contribution is essentially independent of photocarrier density. Its weak photocarrier density dependence largely reflects the neglect of screening-induced renormalizations of the electron-phonon coupling (see the related discussion in \nameref{sec2}). This effect will be studied in future work.

Two regimes emerge for the electron-electron linewidth. At low photocarrier densities, it increases due to occupation changes that open additional scattering channels. At high densities, enhanced electronic screening reduces the linewidth. The competition between phase-space opening and screening produces a crossover, with a maximum near $\sim 3\times 10^{12}$~e$^{-}$/cm$^2$.

For the ionic dynamics, we compute the photoinduced forces (Eqs.~\eqref{eq:ep_ionic_force_normal_modes} and \eqref{eq:td_force_real_space}) and the decay time of the coherent ionic motion due to electron-phonon and phonon-phonon interactions (Eqs.~\eqref{eq:gamma_nu_elph} and \eqref{eq:gamma_nu_phph}). The only nonzero normal-mode component of the force is along the symmetry-preserving $A_{1g}$ mode, consistent with a coherent displacive excitation mechanism~\cite{Cheng1991, Zeiger1992}. This force acts on the S atoms along the $\hat{z}$ axis. Accordingly, the relevant decay time for the ionic dynamics is that of the $A_{1g}$ mode.

We evaluate the forces and the $A_{1g}$ decay time as functions of pump fluence after the initial carrier relaxation at 100 fs. Real-time forces are compared against cDFPT calculations~\cite{Marini2021} at the same photocarrier density and electronic temperature.

The results are shown in Fig.~\ref{fig:MoS2_fig1}(f). The real-time forces (blue circles) agree very well with cDFPT (blue squares) and increase steadily with fluence. The remarkable agreement confirms that the linear expansion of the density, see Eqs. \eqref{eq:tot_density_variation}-\eqref{eq:ionic_force_real_space_final}, is, in practice, an excellent approximation for photoexcited semiconductors.
The $A_{1g}$ decay times due to both electron-phonon and phonon-phonon processes decrease with fluence. For electron-phonon (green pentagons), the reduction reflects weakened Pauli blocking as the photocarrier population grows. For phonon-phonon (green triangles), the fluence dependence arises from changes in the average nonequilibrium lattice temperature induced by the energy exchange between electrons and phonons. The two contributions are comparable in magnitude, particularly near $F\!\approx\!1$~mJ/cm$^2$, underscoring the need to account for all interactions to accurately capture ionic dynamics after illumination. Overall, Fig.~\ref{fig:MoS2_fig1} shows that an accurate real-time description of photoexcited MoS$_2$ requires all relevant interactions. 

Guided by these results, we now focus on a specific laser-induced dynamics by fixing the pump fluence to a typical experimental value, namely $0.85$~mJ/cm$^2$, inducing an excited electron and hole density of $\approx 10^{14}$ e$^-$/cm$^2$, see Fig.~\ref{fig:MoS2_fig1}(c). We study the evolution of electronic and lattice observables with the full inclusion of quasiparticle renormalizations and scattering integrals.

We begin with electronic observables and track carrier relaxation after excitation. Figure~\ref{fig:MoS2_fig2}(a) shows the band structure and nonequilibrium electron and hole occupations along a high-symmetry path at $t=60$, $200$, and $400$~fs. Dot colors encode the occupations, and gray lines indicate the equilibrium band structure. The bands exhibit substantial renormalization with a pronounced gap reduction, yielding a single-particle bandgap of $\sim1.8$~eV, consistent with time-resolved ARPES reporting a saturated gap of $\sim1.9$~eV at a carrier density of $5\times10^{13}$~cm$^{-2}$~\cite{Liu2019}. This remarkable band structure renormalization stems from the change in screening induced by excited carriers, see \nameref{sec2} for a detailed discussion.

Within the first few tens of femtoseconds, carriers rapidly accumulate near the band extrema. Ultrafast carrier-carrier scattering provides efficient intra- and interband redistribution, while carrier-phonon scattering enables energy transfer to the lattice. Without the former, carriers remain comparatively delocalized in $\vb{k}$ space (see Fig.~\ref{fig:sumup}(a) for $K$-valley dynamics). Over the following hundreds of femtoseconds, the carrier distribution cools and sharpens: by $t=400$~fs, most excess energy has been transferred to the lattice and electrons (holes) are strongly concentrated near the dominant conduction- (valence-) band valleys. Specifically, electrons predominantly populate $K$ and $Q$ points, while holes occupy $K$ and $\Gamma$.

The dynamics exhibits a two-stage behavior: a carrier-carrier-dominated redistribution on the tens-of-femtoseconds timescale, followed by carrier-phonon-assisted cooling/thermalization over a few hundred femtoseconds. This two-timescale trend is consistent with time-resolved spectroscopy measurements in the same density/fluence range, which report a fast (tens of fs) and a slower (hundreds of fs) component~\cite{Nie2014, DalConte2015}, and with pump-probe photoconductivity/photoluminescence reporting sub-ps relaxation at comparable densities~\cite{Docherty2014}. We emphasize that including carrier-carrier scattering is essential; omitting it leads to multi-ps relaxation and the loss of the observed two-stage character, see Fig.~\ref{fig:sumup}(a).

\begin{figure*}
    \centering
    \includegraphics[width=0.95\linewidth]{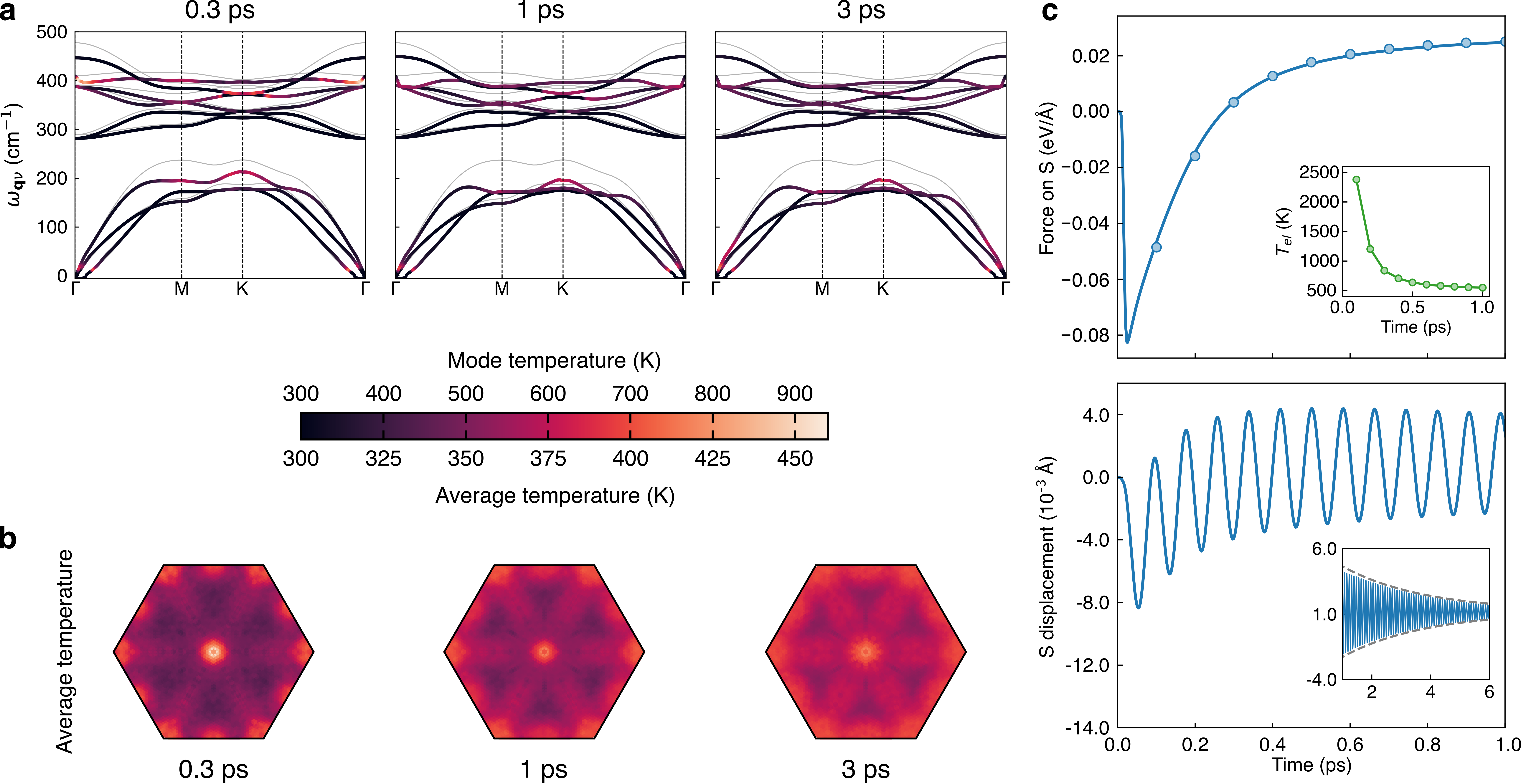}
    \caption{\textbf{Real-time evolution of phonon observables in monolayer MoS$_2$ after above-gap excitation.} (a) Phonon dispersion and mode temperatures along a high-symmetry path at selected snapshots (0.3, 1, and 3 ps); the gray curves represent equilibrium phonon bands in the absence of photocarriers. (b) Momentum-resolved average lattice temperature at selected times. (c) Top: time-dependent force on S atoms (blue curve) compared with two-chemical-potential Fermi-Dirac forces (blue circles) evaluated at the fitted electronic real-time temperature shown in the inset (green circles). Bottom: time-dependent displacement of the S atoms. The inset shows the long-time decay of the coherent amplitude; the gray line represents an exponential-envelope fit.}
    \label{fig:MoS2_fig3}
\end{figure*}

The photocarrier relaxation pathway is even more evident in the momentum-resolved electron and hole densities (see Eq.~\eqref{eq:electron_hole_momentum_densities}) of Figs.~\ref{fig:MoS2_fig2}(b)-\ref{fig:MoS2_fig2}(c), which highlight intervalley scattering among the main valleys driven by large-$\vb{q}$ carrier-phonon scattering. 

Finally, Fig.~\ref{fig:MoS2_fig2}(d) reports the energy-resolved occupations (blue symbols) together with fits to a two-chemical-potential Fermi-Dirac distribution (green dashed lines). Already at $t=60$~fs the real-time occupations are well described by this distribution, albeit with an electronic temperature of several thousand Kelvin. Over the next few hundred femtoseconds, the electron-hole plasma cools as energy flows into the phonon degrees of freedom, yielding a near-thermal distribution with an electronic temperature of a few hundred Kelvin.

These observations support the use of constrained electronic occupations in first-principles calculations~\cite{Marini2021} to describe time-dependent electronic properties even shortly after photoexcitation, and suggest that, to leading order, energy transfer to the lattice can be modeled simply by changing the electronic temperature.

Next, we examine time-resolved phonon and lattice observables. Figure~\ref{fig:MoS2_fig3}(a) shows the phonon dispersion together with the mode-resolved temperatures (Eq.~\eqref{eq:ph_mode_temperature}) at $t=0.3$, $1$, and $3$~ps. Dot color encodes the mode temperature, while gray lines represent the equilibrium phonon dispersions.

The phonon spectrum is markedly renormalized by the adiabatic electron-hole self-energy (Eq.~\eqref{eq:renormalized_phonon_frequencies}). Unlike the electronic bands, the phonon renormalization depends strongly on the instantaneous electronic temperature: the dispersion evolves noticeably between $0.3$ and $1$~ps as the electronic temperature $T_e$ drops, reflecting the progressive weakening of Pauli blocking.

The mode temperatures reveal the pathway of energy flow. Within the first few hundred femtoseconds, excited carriers scatter primarily with a limited set of strongly coupled modes, most prominently the zone-center $A_{1g}$ and modes near $M$ and $K$, so the lattice energy is stored in localized BZ hot spots. Over the subsequent picosecond, anharmonic scattering redistributes this energy across the BZ, driving the phonon population toward a quasi-thermal distribution by $t=3$~ps.

The role of scattering channels is even clearer when analyzing the $A_{1g}$-mode dynamics with and without phonon-phonon scattering (Fig.~\ref{fig:sumup}(b)). With carrier-phonon scattering alone, the phonon mode temperature saturates within $\sim 2$~ps, indicating incomplete lattice thermalization. By contrast, introducing anharmonic channels yields a pronounced long-time tail, with a several-picosecond drift toward equilibrium. This behavior is consistent with picosecond-scale anharmonic lattice relaxation observed by time-resolved thermal diffuse scattering at comparable photocarrier densities~\cite{Pan2025}. Thus, including phonon-phonon scattering is essential to capture the long-time hot-lattice relaxation.

Figure~\ref{fig:MoS2_fig3}(b) reports the momentum-resolved temperature averaged over all branches, see Eq.\eqref{eq:average_lattice_temperature}, at the same snapshots. Consistent with the mode-resolved figure, the early-time distribution is highly nonthermal with peaks at $\Gamma$, $M$, and $K$. By $3$~ps the lattice exhibits a spatially smoother hot-phonon state with an average temperature of $\sim 400$~K.

Figure~\ref{fig:MoS2_fig3}(c) shows the ionic response. The upper panel represents the real-time force on S atoms (blue line) compared with two-Fermi-Dirac forces (blue circles) evaluated using the real-time electronic temperature extracted from two-chemical-potential Fermi-Dirac fits (inset, green symbols). The initial rapid force build-up reflects ultrafast carrier promotion and carrier-carrier thermalization, while the subsequent decay follows the decrease of $T_e$, in excellent agreement with the two-chemical-potential Fermi-Dirac benchmark.

The lower panel shows the S displacement along $z$ driven by the nonequilibrium force. The motion is harmonic at the $A_{1g}$ frequency ($\approx 12.3$~THz) and superimposed on a displacive shift of the equilibrium position. The oscillation amplitude decays with an approximately exponential envelope (gray dashed line in the inset), yielding a coherent damping time of $\approx 2.8$~ps, consistent with Fig.~\ref{fig:MoS2_fig2}(f). Including phonon-phonon on top of electron-phonon effects further reduces the damping time, bringing it closer to the $\sim\!1.7$~ps reported by transient absorption~\cite{Trovatello2020}, see Fig.~\ref{fig:sumup}(c). Importantly, experiments resolve the selective activation of the $A_{1g}$ mode in pump-probe traces, with no detectable $E'$ oscillations, consistent with our nonequilibrium forces result.

In summary, photoexcitation rapidly produces a quasi-thermal electronic distribution via carrier-carrier scattering, while carrier-phonon collisions allow for energy transfer to a restricted set of phonon modes, generating strong nonequilibrium hot spots. On the picosecond timescale, combined carrier-phonon and phonon-phonon processes redistribute this energy, leading to lattice thermalization. The associated ionic forces exhibit a fast rise due to carrier heating and a slower relaxation governed by the electronic temperature decrease, and they coherently drive the $A_{1g}$ oscillations with damping set by both electron-phonon and phonon-phonon interactions.

\subsection{Fluence-dependent properties in monolayer h-BN}
\label{subsec:fluence-dependent properties in monolayer h-BN at quasiequilbrium}

In this part, we focus on the light-induced modification of screening, optical absorption, and phonon spectrum in monolayer h-BN following above-gap photoexcitation after carrier thermalization.
Thus, we analyze the quasi-equilibrium state forming within tens of femtoseconds after carrier thermalization and before electron-hole recombination, occurring on much longer timescales. 

As demonstrated previously, this state is well described by a Fermi-Dirac distribution with two distinct chemical potentials, one for valence holes and one for conduction electrons. Accordingly, the electronic occupations are initialized as
\begin{equation}
\label{eq:fermi_distribution}
    f_{\vb{k}n} = \dfrac{1}{1 + \exp\left[\beta (\varepsilon_{\vb{k}n} - \mu_{n})\right]}, 
\end{equation}
where $\varepsilon_{\vb{k}n}$ are the electronic eigenvalues of the photoexcited system, $\beta$ is the inverse carrier temperature, and the chemical potential $\mu_n$ takes one of two values depending on whether $n$ belongs to the valence or conduction manifold. For a target photocarrier concentration $n_{\text{exc}}$, the chemical potentials $\mu_v$ and $\mu_c$ are obtained by enforcing
\begin{equation}
\label{eq:electrons_holes_number}
    n_{\text{exc}} = \dfrac{1}{N} \sum_{\vb{k}} n_h(\vb{k}) = \dfrac{1}{N}\sum_{\vb{k}} n_e(\vb{k}),
\end{equation}
with momentum-resolved electron and hole densities defined as
\begin{equation}
\label{eq:electron_hole_momentum_densities}
    n_h(\vb{k}) = \sum_{v} (1 - f_{\vb{k}v}), \quad n_e(\vb{k}) = \sum_c f_{\vb{k}c},
\end{equation}
where $v$ and $c$ denote valence and conduction states, respectively. Since the eigenvalues in Eq.~\eqref{eq:fermi_distribution} depend on occupations through Eq.~\eqref{eq:renorm_eig2}, occupations, eigenvalues, and chemical potentials are determined self-consistently.

\begin{figure*}[t]
    \centering
    \includegraphics[width=0.95\linewidth]{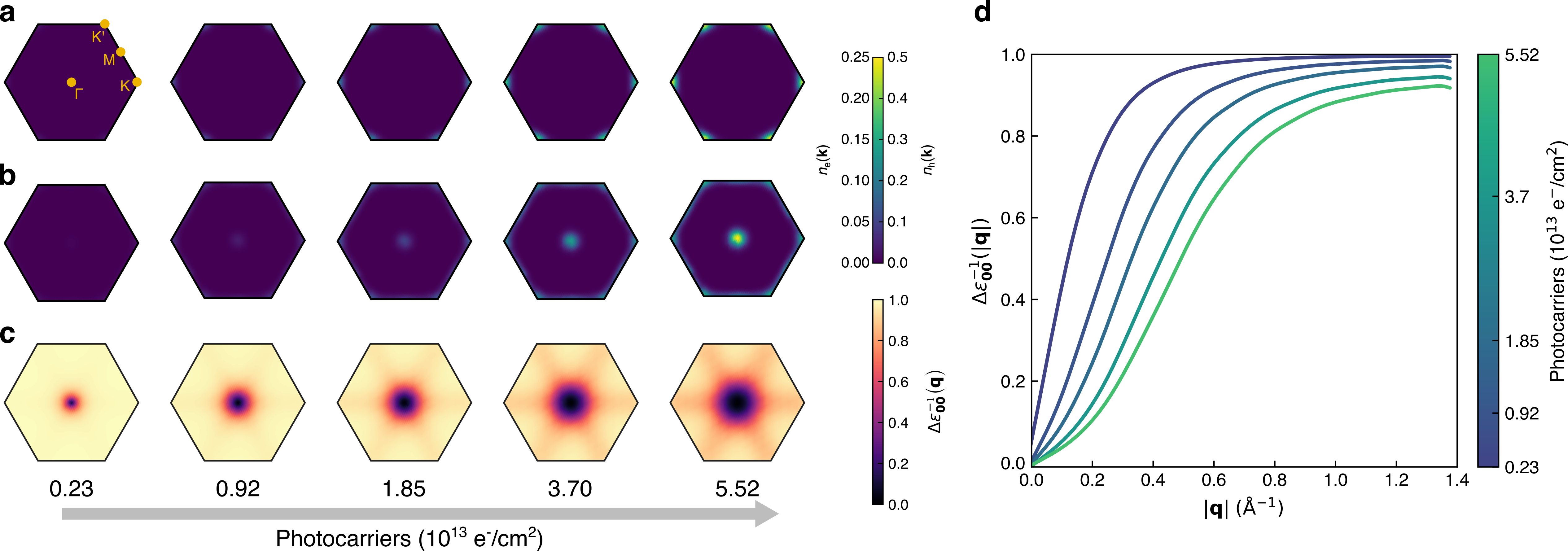}
    \caption{\textbf{h-BN: carrier distributions and photoinduced screening.} Momentum-resolved distributions of valence holes (a) and conduction electrons (b) in the BZ as functions of photocarrier concentration. High-symmetry points are indicated for clarity. (c,d) Momentum-resolved (c) and isotropically averaged (d) photoinduced contributions to the inverse dielectric function for selected photocarrier concentrations.}
    \label{fig:h-BN_fig1}
\end{figure*}

\begin{figure*}
    \centering
    \includegraphics[width=0.95\linewidth]{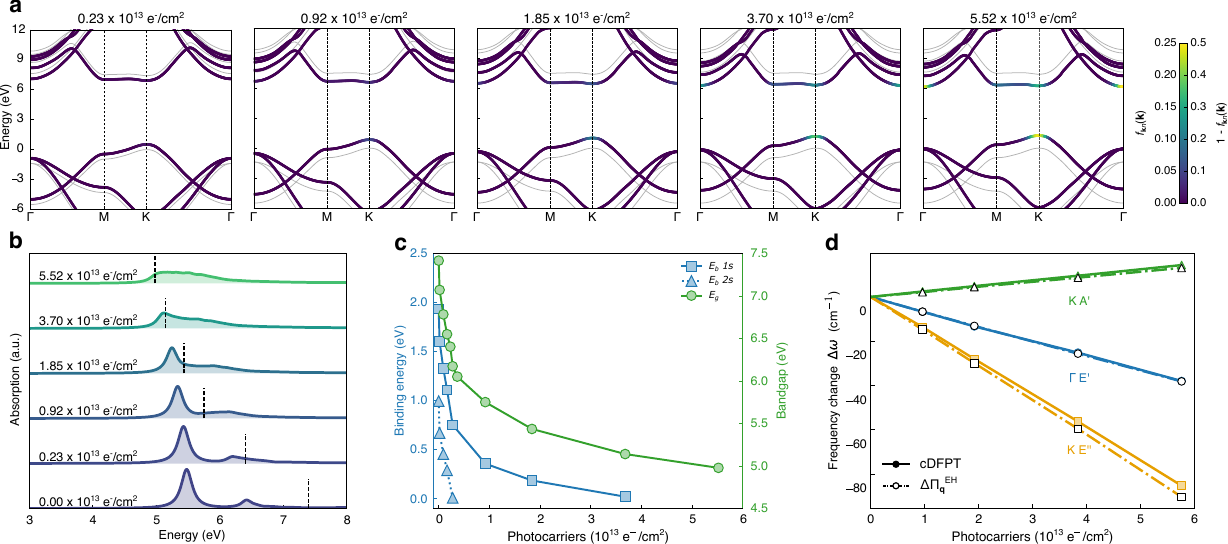}
    \caption{\textbf{Photodoping dependence of electronic and phononic observables in monolayer h-BN.} (a) Band structure and occupations along a high-symmetry path; gray curves are equilibrium bands without photocarriers. (b) Linear optical absorption. Black dashed lines indicate the direct single-particle gap. (c) Binding energies of $1s$ and $2s$ excitons (blue squares/triangles) and the single-particle gap (green circles). (d) Phonon frequency shifts for three modes at $\Gamma$ and $K$; filled symbols correspond to cDFPT~\cite{Marini2021}, open symbols to adiabatic electron-hole self-energy renormalization.}
    \label{fig:h-BN_fig2}
\end{figure*}

To describe the state immediately after carrier thermalization, we fix the Fermi-Dirac temperature to 1000~K, representing a hot electron-hole plasma before electron-lattice thermalization (occurring in hundreds of femtoseconds). Quasi-equilibrium occupations and eigenvalues are then computed for several photodoping levels within the range typically achieved by optical pumping \cite{Cunningham2017, Liu2019, Hofmann2025}. Figures~\ref{fig:h-BN_fig1}(a-b) display the momentum-resolved carrier densities from Eq.~\eqref{eq:electron_hole_momentum_densities} for different photocarrier concentrations. Holes accumulate near $K$ and $K'$, while electrons populate regions close to $\Gamma$ and along $K$-$K'$. These regions coincide with the valence-band maxima and conduction-band minima, respectively (see Fig.~\ref{fig:h-BN_fig2}(a)).

As discussed in \nameref{sec2}, a key consequence of photoexcitation is the modification of screening by free carriers, see also Supplementary Section 4. At equilibrium, the inverse static dielectric function remains finite as $|\vb{q}| \to 0$, implying a long-ranged weakly screened Coulomb interaction. In the photoexcited system, photocarriers enable charge rearrangements that fully screen the long-range component. The change in screening is quantified by the photoinduced change of the inverse dielectric function, $\Delta \epsilon^{-1}_{\vb{00}}(\vb{q})$, which equals one in the absence of photocarriers and drops below unity when carriers are present (Eqs.(S34)--(S40) in the Supplementary Information). Here we consider only its homogeneous part, neglecting local-field effects.

Figure~\ref{fig:h-BN_fig1}(c) shows momentum-resolved $\Delta \epsilon^{-1}_{\vb{00}}(\vb{q})$ for selected photocarrier concentrations. Even modest densities produce a marked reduction at long wavelengths, with $\Delta \epsilon^{-1}_{\vb{00}}(\vb{q}) \to 0$ as $|\vb{q}| \to 0$, while changes near the BZ boundary are weaker. To highlight this, Fig.~\ref{fig:h-BN_fig1}(d) reports the isotropic average
\begin{equation}
    \Delta \epsilon^{-1}_{\vb{00}}(|\vb{q}|) = \dfrac{1}{2\pi}\int_{0}^{2\pi} d\theta \, \Delta \epsilon^{-1}_{\vb{00}}(\vb{q}),
\end{equation}
with $\vb{q} = |\vb{q}|(\cos\theta,\sin\theta)$. The average shows that the inverse dielectric function vanishes linearly as $|\vb{q}| \to 0$ for all concentrations considered, i.e., the total static inverse dielectric function also vanishes linearly at long wavelength, consistent with metallic behavior in two dimensions. Increasing photocarrier density strengthens this trend and enhances anisotropy away from the zone center (Fig.~\ref{fig:h-BN_fig1}(c)), supporting the picture of a hot electron-hole plasma that efficiently screens the long-range Coulomb interaction.

The resulting changes in electronic density and screening have direct consequences for electronic properties and lattice dynamics. We first assess band-structure changes induced by photoexcited carriers by solving Eqs.~\eqref{eq:renorm_eig2} and \eqref{eq:fermi_distribution}--\eqref{eq:electron_hole_momentum_densities} self-consistently. In the quasi-equilibrium state with time-independent occupations, the quasiparticle renormalization is also time independent, and only the Fan-Migdal and COHSEX terms contribute (Eqs.(S28) and (S29) in the Supplementary Information). The dominant effect arises from COHSEX, which reflects the modified screening and leads to significant band-structure renormalization.

Figure~\ref{fig:h-BN_fig2}(a) shows the electronic band structure along a high-symmetry path for increasing photocarrier densities. The line color encodes the momentum-resolved carrier density, while gray lines indicate the equilibrium bands. Consistent with Fig.~\ref{fig:h-BN_fig1}(a), electrons and holes accumulate near the conduction-band minimum and valence-band maximum. The most striking consequence of photodoping is the bandgap reduction, as valence and conduction manifolds shift closer together. This renormalization stems from the buildup of metallic screening and can be viewed as a fingerprint of mutual electron-hole attraction. The green curve in Fig.~\ref{fig:h-BN_fig2}(c) illustrates the progressive bandgap reduction with photocarrier density, which becomes substantial at high doping. Comparable renormalizations have been obtained in full-frequency GW under electron-only \cite{Liang2015, Gao2017} and combined electron-hole doping \cite{Meckbach2018}, supporting our static-GW treatment of the electron-hole plasma. Large photoinduced bandgap renormalizations have also been observed in monolayer MoS$_2$ \cite{Liu2019, Pogna2016} and WS$_2$ \cite{Chernikov2015, Cunningham2017, Hofmann2025}, underscoring the central role of photocarriers in reshaping electronic and optical properties.

We next examine the optical response of h-BN in the presence of photocarriers. We solve the EOMs (Fig.~\ref{fig:EOM}) by initializing the population in the quasi-equilibrium state of Eq.~\eqref{eq:fermi_distribution}, and perturb the system with a weak, delta-like in-plane electric field, $\vb{E}(t)=\vb{E}_0\,\delta(t-t_0)$. Since the field is weak, we neglect collision integrals, phonons, and atomic displacements, and assume occupations remain fixed during the response. The Fourier transform of the in-plane macroscopic polarization (Eq.~\eqref{eq:macroscopic_polarization}) yields the linear susceptibility
\begin{equation}
    \chi(\omega) = \frac{P(\omega)}{E(\omega)}.
\end{equation}
The delta-like perturbation excites all frequencies, providing the full optical spectrum. An exponential damping is applied to the time-domain polarization to mimic finite linewidths of optical excitations. The absorption spectrum, $\mathrm{Im}\,\chi(\omega)$, is then evaluated for different photocarrier concentrations; see Fig.~\ref{fig:h-BN_fig2}(b).

In the absence of photocarriers, the spectrum exhibits two dominant peaks corresponding to the $1s$ and $2s$ excitons, with binding energies of approximately 2.0~eV and 1.0~eV, respectively. With increasing photocarrier density, the peaks broaden and lose intensity, consistent with the picture proposed in Ref.~\cite{Mahan1967}. As the single-particle bandgap (black dashed line) is strongly renormalized, the exciton binding energies are markedly reduced. For densities exceeding $\sim 4 \times 10^{13}$~cm$^{-2}$, no excitonic peaks remain, yielding an independent-particle-like spectrum with no excitations below the single-particle gap.

The progressive disappearance of excitonic features with photodoping (exciton melting) has been predicted in transition-metal dichalcogenides \cite{Steinhoff2014, Steinhoff2017, Erben2018, Meckbach2018, Erben2022} and observed via transient absorption \cite{Chernikov2015, Cunningham2017}. Figure~\ref{fig:h-BN_fig2}(c) reports the binding energies of the $1s$ and $2s$ excitons, together with the single-particle bandgap, as functions of photocarrier concentration. The $2s$ exciton vanishes at $\sim 2.5 \times 10^{12}$~cm$^{-2}$, while the $1s$ peak persists up to $\sim 3.8 \times 10^{13}$~cm$^{-2}$.

Finally, we assess how lattice dynamics is affected by photoexcited carriers. Phonon frequency changes are evaluated from Eq.~\eqref{eq:renormalized_phonon_frequencies2}, neglecting anharmonic effects, for several photodoping levels. Specifically, we compute the adiabatic phonon renormalization due to the electron-phonon contribution to the self-energy, $\Delta \Pi^{\text{EH}}_{\vb{q}}$. To benchmark our predictions, we compare with constrained density-functional perturbation theory (cDFPT) \cite{Tangney1999, Marini2021}, which yields phonon frequencies in the presence of an excited electron-hole plasma and has shown excellent agreement with experiments \cite{Tangney2002, Mocatti2023}. 

We analyze frequency shifts of the $E'$ mode at $\Gamma$ and the $A'$ and $E'$ modes at $K$ for different photocarrier concentrations. The results in Fig.~\ref{fig:h-BN_fig2}(d) (solid lines and filled markers) are compared with cDFPT predictions (dashed lines and open markers). The shifts from the self-energy approach closely match those from cDFPT, with deviations of only a few cm$^{-1}$. This validates the self-energy method for computing phonon-frequency changes induced by excited carriers within the explored doping range and provides an efficient route to evaluate phonon frequencies in photoexcited materials directly from the ground-state dynamical matrix, without additional linear-response calculations.

\section{Discussion}\label{sec: Conclusion and perspectives}

In this work, we developed a first-principles many-body framework to describe the coupled dynamics of photocarriers, phonons, and ions in semiconductors following ultrafast excitation. Our approach incorporates explicit light-matter coupling together with \textit{ab initio} carrier-carrier, carrier-phonon, and phonon-phonon collision integrals, thereby treating electron and phonon interactions on the same footing in real time.

The time evolution of the density matrix is performed in a maximally localized Wannier basis, ensuring gauge-consistent evaluation of scattering integrals and enabling efficient, ultradense sampling of electron and phonon momenta over long propagation times. This implementation allows for direct, parameter-free comparison with realistic pump–probe experimental conditions. Furthermore, the method can be coupled to constrained density-functional theory to access light-induced structural phase transitions at longer times after the light pulse.

Key advances relative to previous approaches include: (i) the equal-footing treatment of carrier-carrier, carrier-phonon, and anharmonic phonon-phonon scattering channels; (ii) the inclusion of time-dependent electronic screening in the nonequilibrium dynamics; (iii) the incorporation of quasiparticle-energy and Rabi-frequency renormalizations in the real-time evolution; and (iv) the first-principles calculation of coherent, damped nuclear oscillations driven by displacive forces, going beyond state-of-the-art Ehrenfest dynamics in which electronic orbitals are typically kept fixed.

We showcased the capabilities and predictive power of this framework on MoS$_2$ and h-BN monolayers. For MoS$_2$, we resolved photoinduced renormalizations of electronic and lattice properties, ultrafast carrier relaxation, hot-phonon dynamics, and displacive coherent phonon motion, highlighting the crucial role of scattering channels in setting the relevant time scales. For h-BN, we quantify photoinduced changes in the electronic, optical, and lattice responses in quasi-equilibrium after photoexcitation, demonstrating a fluence-dependent enhancement of screening and melting of excitonic features. Taken together, these results establish our approach as a versatile and quantitatively predictive computational platform for interpreting and guiding ultrafast experiments in semiconductors and related quantum materials, enabling first-principles investigations of nonequilibrium phenomena across a broad class of periodic systems.

\subsubsection{Data availability}
The datasets generated and/or analyzed during the current study are not publicly available during peer review because the Zenodo repository is not yet public, but are available from the corresponding author on reasonable request. The datasets will be made publicly available upon publication at \href{https://doi.org/10.5281/zenodo.18985083}{10.5281/zenodo.18985083}.

\subsubsection{Code availability}
The code described and used in this work is publicly available as the open-source package \textsc{EPIq}, released under the GNU General Public License v3.0. The source code and documentation are available at \href{https://the-epiq-team.gitlab.io/epiq-site/}{the-epiq-team.gitlab.io/epiq-site/}.

\subsubsection{Acknowledgements}
Funded by the European Union (ERC, DELIGHT, 101052708). Views and opinions expressed are however those of the authors only and do not necessarily reflect those of the European Union or the European Research Council. Neither the European Union nor the granting authority can be held responsible for them.

We acknowledge the CINECA award under the ISCRA initiative (projects IscrC\_Uf--DynFP and IscrB\_EPhoCS), for the availability of high performance computing resources and support. We acknowledge EUROHPC (project 465000468) for the availability of high performance computing resources and support. 

We are grateful for the fruitful discussion with Prof. Enrico Perfetto and Prof. Gianluca Stefanucci.

\subsubsection{Author contributions}

The computational framework was implemented by S.M. with contributions from G.M. and G.V.
The theoretical framework and the corresponding approximations were devised by S.M. with contributions from G.M., P.C. and M.C.
Numerical simulations were performed by S.M.
Data analysis was performed by S.M. with contributions from G.M. and M.C.
Manuscript preparation was done by S.M. with contributions from G.M. and M.C.
All authors participated in the revision and correction of the manuscript.  
M.C.  supervised the project.

All authors have read and approved the final version of the manuscript.

\subsubsection{Competing interests}
The authors declare no competing financial or non-financial interests.

\bibliography{bibliography}

\end{document}